\documentclass[12pt]{article}
\usepackage[a4paper, total={6in, 9in}]{geometry}
\usepackage{cite}
\usepackage{amsthm}
\usepackage{amsmath,amssymb,amsfonts}
\usepackage{xcolor}

\definecolor{reviewblue}{RGB}{0,70,150}
\newenvironment{reviewnote}[2]{\par\medskip\begingroup\color{reviewblue}\small\normalfont\noindent\textbf{Review #1 --- #2.}\ }{\par\endgroup\medskip}
\usepackage{graphicx}
\usepackage{hyperref}
\hypersetup{hidelinks}
\usepackage{textcomp}
\usepackage{graphicx} % Required for inserting images
\usepackage{mathtools}
\usepackage{amsfonts}

\usepackage{comment}
\usepackage{caption}
\usepackage{subcaption}

\usepackage{mathtools}
\DeclarePairedDelimiter\bra{\langle}{\rvert}
\DeclarePairedDelimiter\ket{\lvert}{\rangle}
\DeclarePairedDelimiterX\braket[2]{\langle}{\rangle}{#1\,\delimsize\vert\,\mathopen{}#2}

\DeclareMathOperator{\Tr}{Tr}

\newtheorem{theorem}{Theorem}[section]
\newtheorem{remark}[theorem]{Remark}

\newtheorem{lemma}[theorem]{Lemma}
\newtheorem{prop}[theorem]{Proposition}
\title{Draft cat qubit}

\title{Exponential suppression of bit-flip errors in dissipative cat qubits}
\usepackage{authblk}

\author{Ruikang Liang}
\author{R\'emi Robin}

\affil{Laboratoire de Physique de l'\'Ecole Normale Sup\'erieure,
Inria, Mines Paris, CNRS, ENS-PSL, Sorbonne Universit\'e,
PSL Research University, Paris, France}
\date{October 2026}
\begin{document}

\maketitle
\maketitle
\begin{abstract}
    We study the infinite-dimensional Lindblad generator $\mathcal L_\epsilon=D[a^2-\alpha^2]+\epsilon D[a]$, describing a dissipative cat qubit subject to single-photon loss. For sufficiently large $\alpha$ and sufficiently small $\epsilon>0$, we identify a two-dimensional spectral subspace near zero and establish an upper bound on the slow decay rate that decreases exponentially with $\alpha^2$, uniformly in $\epsilon$. This proves exponential suppression of bit-flip errors.
\end{abstract}
\section*{Introduction}

We study the Lindblad master equation
\[
\frac{\mathrm d\rho}{\mathrm dt}=\mathcal L_\epsilon(\rho),\qquad
\mathcal L_\epsilon=D[a^2-\alpha^2]+\epsilon D[a],
\]
where $\alpha>0$ and $\epsilon\geq0$. This equation describes a bosonic mode subject to engineered two-photon dissipation and a coherent two-photon drive (whose intensity is proportional to $\alpha^2$), together with single-photon loss. Time is measured in units of the inverse two-photon dissipation rate $\kappa_2$, and $\epsilon=\kappa_1/\kappa_2$ is the ratio of the single-photon and two-photon loss rates. Here $a$ and $a^\dagger$ denote the annihilation and creation operators, and $\rho$ is a positive trace-class operator of unit trace on the bosonic Fock space $\mathcal H=\ell^2(\mathbb N)$, unitarily identified with $L^2(\mathbb R)$ through the Hermite basis. Precise mathematical definitions are given in Section~\ref{sec:functional_space}.

The two-dimensional subspace
\[
\mathcal C_\alpha=\mathrm{span}\{\ket\alpha,\ket{-\alpha}\}
=\ker(a^2-\alpha^2)
\]
provides an encoding of a qubit in a single oscillator. In the absence of single-photon loss, every operator supported on $\mathcal C_\alpha$ is stationary, yielding a four-dimensional kernel of $\mathcal L_0$. Well-posedness and convergence to this stationary manifold have been established in an adapted trace-class framework by Azouit, Sarlette and Rouchon~\cite{rouchonazouit}. Using Costa Rico, Gondolf and M\"obus instantaneous Sobolev regularization for dissipative bosonic dynamics~\cite{costarico2025}, we can easily deduces the exponential convergence of any state to the stationary manifold in the trace norm.

Nevertheless, the main interest of this encoding relies on its robustness to various sources of physical noise~\cite{mirrahimi2014,leghtas2015}, including the single-photon loss that we consider here. For $\epsilon>0$, the code space is no longer invariant, but the population imbalance between its two coherent components is expected to decay on a time scale exponentially long in $\alpha^2$. This meta-stability property, often called bit-flip suppression, has been observed experimentally in~\cite{lescanne2020,berdou2023,reglade2024,putterman2025}. This noise bias motivates error-correction architectures that predominantly correct the remaining phase errors~\cite{guillaud2019,guillaud2023}.

From a spectral viewpoint, this behavior is associated with a separation between the slow population-imbalance mode and the faster coherence modes, in the spirit of the spectral description of metastability~\cite{macieszczak2016}. Perturbative calculations predict a stationary eigenvalue $0$, a slow eigenvalue $-\lambda$, and two coherence-decay eigenvalues with leading scale $-2\epsilon\alpha^2$~\cite{azouit2015,dubovitskii}. In the physics litterature, the physical bit-flip rate is defined as $\Gamma_{\mathrm{bf}}=\kappa_2\lambda/2$.

Several theoretical works have addressed the problem of exponential suppression of bit-flip errors~\cite{gautier2022,thompson2022,dubovitskii}. A remarkable contribution is that of Dubovitskii~\cite{dubovitskii}, who performs a second-order perturbative calculation in $\epsilon$ to compute the slow non-zero eigenvalue of $\mathcal L_\epsilon$. The first- and second-order contributions indeed scale as $e^{-4\alpha^2}$ and $e^{-2\alpha^2}$, respectively, up to polynomial factors in $\alpha$. However, it is unclear how this approach could prove that the remainder beyond second order in $\epsilon$ is also exponentially small in $\alpha^2$ as $\alpha\to\infty$ at fixed small $\epsilon$.

% Editorial follow-up: reconcile Theorem~\ref{theorem:main_3} with the stronger
% bound in Eq.~\eqref{eq:improved_eigenvalue_bound}; resolve the existing
% domain and semigroup review notes before finalizing the result claims.
In this work, we establish a uniform exponential upper bound on the slow decay rate of the perturbed infinite-dimensional generator.  In Section~2, we construct the resolvent of a modified generator using the decomposition into the cat-code subspace and its orthogonal complement. Its spectral structure, together with relative perturbation estimates, allows us to isolate a two-dimensional spectral subspace of the original Lindbladian $\mathcal L_\epsilon$ near zero. In Section~3, we construct a basis of this subspace whose two elements are localized near the coherent-state projections $|\alpha\rangle\langle\alpha|$ and $|-\alpha\rangle\langle-\alpha|$, respectively. After applying the respective displacements that send $\ket{\pm\alpha}$ to the vacuum state $\ket{0}$, these basis elements are close to the vacuum projection $\ket{0}\bra{0}$ and have exponentially small tails beyond a Fock cutoff of order $\alpha^2$. These localization estimates follow from bounds on the truncated resolvent obtained with an exponential Lyapunov weight. Finally, parity symmetry identifies the stationary and slow modes as the symmetric and antisymmetric combinations of these basis elements, and the truncation estimates yield the desired bound on the slow eigenvalue.
\section{Functional spaces}
\label{sec:functional_space}
Let $\mathcal{H}=L^2(\mathbb{R})$, and denote by $a^\dagger$ and $a$ the creation and annihilation operators, respectively, and by $N=a^\dagger a$ the number operator. For $\alpha\in\mathbb{R}$, define
\begin{equation*}
L=a^2-\alpha^2,
\qquad
L^\dagger=(a^\dagger)^2-\alpha^2.
\end{equation*}

% MAGENTA-CONVENTION-BEGIN
{We use the standard closed realizations of the bosonic operators and interpret inequalities involving unbounded operators in the sense of quadratic forms.\par}
% MAGENTA-CONVENTION-END

Denote by $\mathcal{B}(\mathcal{H})$ the space of bounded linear operators on $\mathcal{H}$, and define the space of trace-class operators by
\begin{equation*}
    \mathcal{K}_{1}=\left\{\rho\in B(\mathcal{H})\mid \Tr(|\rho|)<\infty\right\}
\end{equation*}
with $|\rho|=\sqrt{\rho^{\dagger}\rho}$. Equipped with the trace norm
\[\|\rho\|_{1}=\Tr(|\rho|),\]
$\mathcal{K}_{1}$ is a Banach space. For $\alpha>0$, let $P$ denote the orthogonal projection onto the two-dimensional subspace $\operatorname{span}\{\ket\alpha,\ket{-\alpha}\}$, and set $Q=I_{\mathcal H}-P$, where $I_{\mathcal H}$ is the identity operator on $\mathcal H$. The dependence of $P$ and $Q$ on $\alpha$ is suppressed in the notation. Define the positive operator
\[
S=(L^\dagger L+{P})^{1/2}.
\]
\begin{lemma}
\label{lemma:gap_1}
     For all $\alpha>0$,
    \[LL^{\dagger}\geq\max\left(2\alpha^2-\frac{1}{4},2\right),\quad L^{\dagger}L\geq\max\left(2\alpha^2-\frac{1}{4},2\right){Q}.\]
\end{lemma}
\begin{proof}
    By definition of $L$, we have
    \[LL^{\dagger}=\left(a^2-\alpha^2\right)\left((a^{\dagger})^2-\alpha^2\right)=(N+1)(N+2)-\alpha^2\left(a^2+(a^{\dagger})^2\right)+\alpha^4.\]
    Since $-a^2-(a^{\dagger})^2=(a-a^{\dagger})(a^{\dagger}-a)-2N-1\geq -2N-1$, we have
    \begin{equation*}
        \begin{aligned}
            LL^{\dagger}&\geq (N+1)(N+2)-\alpha^2(2N+1)+\alpha^4=N^2-(2\alpha^2-3)N+\alpha^4-\alpha^2+2\\
            &=\left(N-\left(\alpha^2-\frac{3}{2}\right)\right)^2+\alpha^4-\alpha^2+2-\left(\alpha^2-\frac{3}{2}\right)^2\geq 2\alpha^2-\frac{1}{4}.
        \end{aligned}
    \end{equation*}
Moreover,  $LL^{\dagger}=L^{\dagger}L+4N+2\geq 2$, hence
\begin{equation*}
    LL^{\dagger}\geq \max\left(2\alpha^2-\frac{1}{4},2\right).
\end{equation*}
For $\psi=\sum_{n\geq0}c_n\ket n\in D(L)$, the equation $L\psi=0$ is equivalent to
\[
\sqrt{(n+1)(n+2)}\,c_{n+2}=\alpha^2c_n,
\qquad n\in\mathbb N.
\]
The even and odd solutions are proportional to $\ket\alpha+\ket{-\alpha}$ and $\ket\alpha-\ket{-\alpha}$, respectively. Hence
\[
\ker L=\operatorname{span}\{\ket\alpha,\ket{-\alpha}\}.
\]
By the polar decomposition of $L$, the preceding bound on $LL^\dagger$ yields
\[
L^\dagger L\geq\max\left(2\alpha^2-\frac14,2\right){Q},
\]
which proves the claim.
\end{proof}

By Lemma~\ref{lemma:gap_1}, $S\geq I_{\mathcal H}$, so $S^{-1}\in\mathcal B(\mathcal H)$. We define
\[
\mathcal K_L
=\left\{S^{-1}\tau S^{-1}:\tau\in\mathcal K_1\right\},
\qquad
\|S^{-1}\tau S^{-1}\|_L=\|\tau\|_1.
\]
The map $\tau\mapsto S^{-1}\tau S^{-1}$ is a linear isometric isomorphism from $\mathcal K_1$ onto $\mathcal K_L$. Consequently, $(\mathcal K_L,\|\cdot\|_L)$ is a Banach space.
For $\rho=S^{-1}\tau S^{-1}\in\mathcal K_L$, we have
\[
\|\rho\|_L=\|S\rho S\|_1.
\]
\par
% MAGENTA-SECTION1-END

Throughout the paper, $\|\cdot\|_{\mathrm{op},1}$ and $\|\cdot\|_{\mathrm{op},L}$ denote the operator norms of bounded superoperators on $\mathcal{K}_1$ and $\mathcal{K}_L$, respectively, while $\|\cdot\|_{\infty}$ denotes the operator norm of bounded operators on $\mathcal{H}$. We denote by $I_{\mathcal{H}}$ the identity operator on the Hilbert space $\mathcal{H}$ and by $\mathbf{I}$ the identity superoperator on spaces of operators over $\mathcal{H}$. As usual, scalar multiples of $I_{\mathcal{H}}$ are identified with the corresponding scalars whenever no ambiguity arises.

Notice that
\[L{Q}=L,\quad {Q}L^{\dagger}=L^{\dagger},\quad {Q}L^{\dagger}L {Q}=L^{\dagger}L.\]

\begin{lemma}
\label{lemma:gap_2}
    For all $\lambda>0$, there exists $\alpha_0>0$ such that for all $\alpha\geq\alpha_0$,
    \[3L^{\dagger}L\geq {Q}(N+\lambda){Q}.\]
\end{lemma}
\begin{proof}
    Let us take $\lambda>0$. By the definition of $L^{\dagger}L$, we have
    \begin{equation*}
    \begin{aligned}
        L^{\dagger}L-(N+\lambda)&={\left((a^{\dagger})^2-\alpha^2\right)\left(a^2-\alpha^2\right)}-(N+\lambda)\\&=N^2-2N-
        \alpha^2(a^2+(a^{\dagger})^2)+\alpha^4-\lambda.
    \end{aligned}
    \end{equation*}
    Using again $-a^2-(a^{\dagger})^2=(a-a^{\dagger})(a^{\dagger}-a)-2N-1\geq -2N-1$, we obtain
    \begin{equation*}
    \begin{aligned}
        L^{\dagger}L-(N+\lambda)&\geq N^2-(2\alpha^2+2)N+\alpha^4-\alpha^2-\lambda\\&=(N-(\alpha^2+1))^2+\alpha^4-\alpha^2-\lambda-(\alpha^2+1)^2\\
        &\geq -3\alpha^2-\lambda-1.
    \end{aligned}
    \end{equation*}
    Consequently
    \[(3\alpha^2+\lambda+1){Q}\geq {Q}(N+\lambda){Q}-L^{\dagger}L.\]
    By Lemma~\ref{lemma:gap_1}, there exists $\alpha_0>0$ such that, for every $\alpha\geq\alpha_0$,
\begin{equation*}
L^\dagger L
\geq
\left(
\frac32\alpha^2+\frac12(\lambda+1)
\right)
{Q}.
\end{equation*}
It follows that
\begin{equation*}
\begin{aligned}
3L^\dagger L
\geq
L^\dagger L
+(3\alpha^2+\lambda+1){Q}\
\geq
{Q}
(N+\lambda)
{Q},
\end{aligned}
\end{equation*}
which proves the claim.
\end{proof}

\section{Perturbed Lindbladian $\mathcal{L}_{\epsilon}$\texorpdfstring{\\}{ }and {modified operator} $\tilde{\mathcal{L}}_{\epsilon}$}For $\alpha>0$ and $\epsilon\geq 0$, consider the perturbed Lindbladian $\mathcal{L}_{\epsilon}$ on $\mathcal{K}_{L}$,
\[\mathcal{L}_{\epsilon}(\rho)=D[L](\rho)+\epsilon D[a](\rho)=L\rho L^{\dagger}+\epsilon a\rho a^{\dagger}-\frac{1}{2}(L^{\dagger}L+\epsilon N)\rho-\frac{1}{2}\rho(L^{\dagger}L+\epsilon N).\]
{
{
Let $\mathcal D_\infty=\bigcap_{k\geq0}D(N^k)$.
This dense subspace is invariant under $a$, $a^\dagger$, $P$ and $Q$.
We interpret $\mathcal L_\epsilon$ as its maximal weak realization
on $\mathcal K_L$, namely
\[
D(\mathcal L_\epsilon)
=\left\{\rho\in\mathcal K_L:
\begin{array}{l}
\text{there exists }f\in\mathcal K_L\text{ such that}\\
\langle u,fv\rangle
=\langle u,\mathcal L_\epsilon^{\mathrm{form}}(\rho)v\rangle
\quad\text{for all }u,v\in\mathcal D_\infty
\end{array}\right\},
\qquad
\mathcal L_\epsilon(\rho)=f.
\]
Here $\mathcal L_\epsilon^{\mathrm{form}}$ denotes the formal expression, evaluated weakly by transferring the
unbounded factors to the test vectors; for example,
$\langle u,L\rho L^\dagger v\rangle
=\langle L^\dagger u,\rho L^\dagger v\rangle$. We use the same convention for $\tilde{\mathcal L}_\epsilon$
on $\mathcal K_L$, and for $\hat{\mathcal L}_\epsilon$
on $\hat{\mathcal K}_L$, using
$\hat{\mathcal D}_\infty
=Q\mathcal D_\infty
=\mathcal D_\infty\cap\hat{\mathcal H}$
as the test space in the latter case.}
}
\begin{prop}[Trace preserving]
    Let \(\alpha>0\), \(\epsilon\ge0\), and let \(\mathcal L_\epsilon\) be the maximal weak realization on \(\mathcal K_L\) defined above. Then
\[
\operatorname{Tr}\bigl(\mathcal L_\epsilon(\rho)\bigr)=0,
\qquad
\text{for all }\rho\in D(\mathcal L_\epsilon).
\]
\end{prop}
\begin{proof}
    First, note that \[D(S)=D(L)=D(a^2)=D(N).\]
By closed graph theorem, we can deduce that 
\[B=(N+1)S^{-1}\]
is bounded. Then for all $\rho\in\mathcal{K}_{L}$, 
\[\Theta=(N+1)\rho(N+1)=B(S\rho S)B^{\dagger}\]
is trace class. For $m,n\in\mathbb{N}$, let us write
\[\rho_{mn}=\bra{m}\rho\ket{n},\quad\Theta_{mn}=\bra{m}\Theta\ket{n}=(m+1)(n+1)\rho_{mn}.\]
Since $\Theta$ is in trace class, We can deduce that when $n$ tends to infinity
\[
n(n-1)\rho_{nn}\longrightarrow0,
\qquad
n\rho_{nn}\longrightarrow0,\]
and
\[
\sqrt{n(n-1)}\,\rho_{n,n-2}\longrightarrow0,
\qquad
\sqrt{n(n-1)}\,\rho_{n-2,n}\longrightarrow0.
\]
Now take \(\rho\in D(\mathcal L_\epsilon)\), set
\[
f=\mathcal L_\epsilon(\rho)\in\mathcal K_L,
\qquad
P_M=\sum_{n=0}^{M}|n\rangle\langle n|,
\]
and let \(M\ge2\). We have
\[
\begin{aligned}
\Tr(P_Mf)
=\Tr(\mathcal{L}_{\epsilon}^{*}(P_M)\rho)=&\sum_{n=M+1}^{M+2}n(n-1)\rho_{nn}
+\epsilon(M+1)\rho_{M+1,M+1}\\
&-\frac{\alpha^2}{2}
\sum_{n=M+1}^{M+2}\sqrt{n(n-1)}
\bigl(\rho_{n,n-2}+\rho_{n-2,n}\bigr),
\end{aligned}
\]
where $\mathcal{L}_{\epsilon}^{*}$ denotes $\mathcal{L}_{\epsilon}$ transferred to the finite diagonal Fock matrix $P_M$. When $M$ tends to infinity, all terms on the right tend to zero by the estimates above. Since \(f\) is trace class and \(P_M\) converges strongly to \(I_{\mathcal H}\), we obtain
\[
\operatorname{Tr}(f)
=\lim_{M\to\infty}\operatorname{Tr}(P_Mf)=0.
\]
This concludes the proof.
\end{proof}
\subsection{Main results}
\begin{figure}[!h]
\centering
\includegraphics[width=0.4\linewidth]{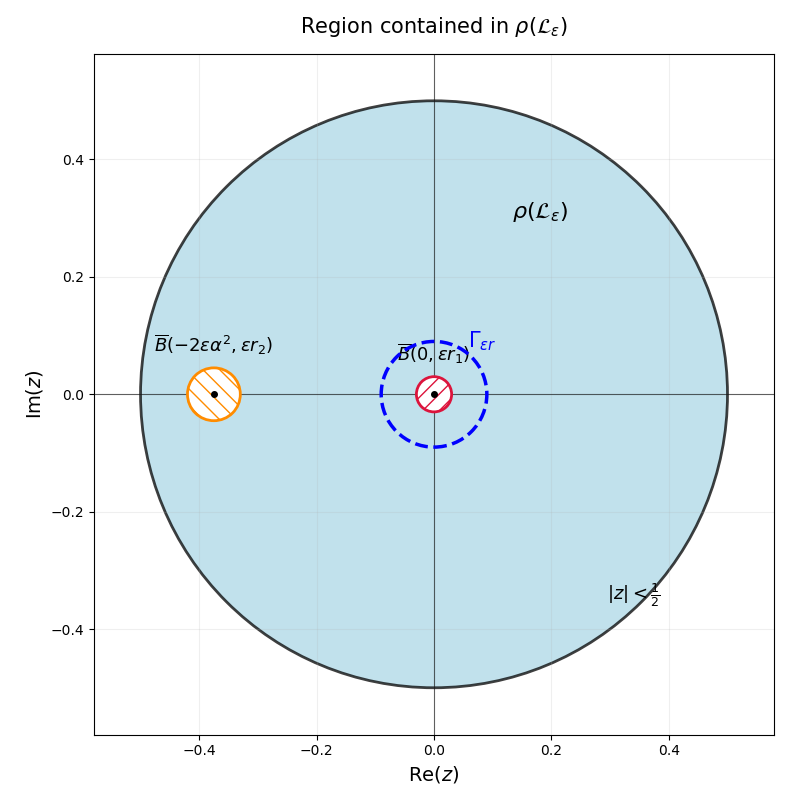}
   \caption{Illustration of the resolvent region of $\mathcal{L}_{\epsilon}$. The shaded region corresponds to the subset of the resolvent set $D_{\frac{1}{2}}\setminus\left(\bar{B}(0,\epsilon r_1)\cup \bar{B}(-2\epsilon\alpha^2,\epsilon r_2)\right)$, where the two excluded closed disks contain the eigenvalues near $0$ and $-2\epsilon\alpha^2$, respectively.}
    \label{fig:example_1}
\end{figure}

In this section, we present our main results concerning the spectrum of the perturbed Lindbladian. For $r>0$, let $\Gamma_r$ and $D_r$ denote, respectively, the circle and the closed disk of radius $r$ centered at the origin:
\begin{equation*}
\Gamma_r
=
\left\{
\lambda\in\mathbb{C}
:
|\lambda|=r
\right\},
\qquad
D_r
=
\left\{
\lambda\in\mathbb{C}
:
|\lambda|\leq r
\right\}.
\end{equation*}
{Throughout the paper, every contour $\Gamma_r$ is oriented counterclockwise.}
More generally, for $p\in\mathbb{C}$ and $r>0$, we denote by
\begin{equation*}
\overline{B}(p,r)
=
\left\{
\lambda\in\mathbb{C}
:
|\lambda-p|\leq r
\right\}
\end{equation*}
the closed disk of radius $r$ centered at $p$.
\begin{theorem}[Resolvent set of $\mathcal{L}_{\epsilon}$]
\label{theorem:main_1}
    There exist $\alpha_0> 0$, $\epsilon_0>0$ and $r_1,r_2>0$ such that for all $\alpha\geq \alpha_0$ and $\epsilon\in(0,\epsilon_0]$, 
    \begin{equation}
    \label{eq:theorem_1}
        D_{\frac{1}{2}}\setminus\left(\bar{B}(0,\epsilon r_1)\cup \bar{B}(-2\epsilon\alpha^2,\epsilon r_2)\right)
    \end{equation}
    is contained in the resolvent set of $\mathcal{L}_{\epsilon}$ as an operator on $\mathcal{K}_{L}$.
\end{theorem}
See an illustrative example of Theorem~\ref{theorem:main_1} in Fig.\ref{fig:example_1}.

\begin{theorem}[Spectral projection associated with the spectrum near $0$]
\label{theorem:main_2}
    There exist $\alpha_0> 0$, $\epsilon_0>0$, and $r>0$ such that, for every $\alpha\geq\alpha_0$ and $\epsilon\in(0,\epsilon_0]$, the contour $\Gamma_{\epsilon r}$ is contained in the resolvent set of $\mathcal{L}_{\epsilon}$ as an operator on $\mathcal{K}_{L}$ and separates the two excluded disks in~\eqref{eq:theorem_1}. Moreover, the spectral projection
    \begin{equation}
    \label{eq:spectral_projection}
        \mathcal{P}_{\epsilon,\alpha}=\frac{1}{2\pi i}\oint_{\Gamma_{\epsilon r}}(\xi\mathbf{I}-\mathcal{L}_{\epsilon})^{-1}\text{d}\xi,
    \end{equation}
    is well defined and has rank $2$.
\end{theorem}

See an illustrative example of Theorem~\ref{theorem:main_1} and Theorem~\ref{theorem:main_2} in Figure.\ref{fig:example_1}.

In order to prove the results stated above, we first introduce some perturbations to $\mathcal{L}_{\epsilon}$ and define a {modified operator} $\tilde{\mathcal{L}}_{\epsilon}$.
The resolvent of $\tilde{\mathcal{L}}_{\epsilon}$ as well as its spectrum are easier to solve (see Proposition~\ref{prop:resolvent_tilde}). Finally we show that the spectra of $\mathcal{L}_{\epsilon}$ and $\tilde{\mathcal{L}}_{\epsilon}$ are $\epsilon$-close to each other in the neighborhood of $0$.
\subsection{{Modified operator} $\tilde{\mathcal{L}}_{\epsilon}$}
For $\alpha>0$ and $\epsilon\geq 0$, define the {modified operator} on $\mathcal{K}_{L}$:
\begin{equation*}
\begin{aligned}
\tilde{\mathcal{L}}_{\epsilon}(\rho)=&L\rho L^{\dagger}+\epsilon (a-{P}a{Q})\rho (a^{\dagger}-{Q}a^{\dagger}{P})\\&-\frac{1}{2}\left(L^{\dagger}L+\epsilon {Q}N{Q}+\epsilon\alpha^2{P}\right)\rho-\frac{1}{2}\rho\left(L^{\dagger}L+\epsilon {Q}N{Q}+\epsilon\alpha^2 {P}\right).
\end{aligned}
\end{equation*}

{This auxiliary operator is not generally trace-preserving.}
Take $\rho\in D(\tilde{\mathcal{L}}_{\epsilon})$ and $\lambda>0$. Define $f=(\lambda\mathbf I-\tilde{\mathcal L}_\epsilon)(\rho)$ and decompose $\rho$ and $f$ into blocks:
\begin{equation*}
    \begin{aligned}
         &\hat{\rho}={Q}\rho {Q},\quad \rho_1={Q}\rho {P},\quad\rho_2={P}\rho {Q},\quad \rho_3={P}\rho {P},\\
    &\hat{f}={Q}f {Q},\quad f_1={Q}f {P},\quad f_2={P}f {Q},\quad f_3={P}f {P}.
    \end{aligned}
\end{equation*}
By the definition of $\tilde{\mathcal{L}}_{\epsilon}$, we can obtain the following equations
\begin{equation}
\label{eq:0}
\begin{aligned}
    \hat{f}=&-{Q}L{Q}\hat{\rho}{Q}L^{\dagger}{Q}-\epsilon {Q}a{Q}\hat{\rho}{Q}a^{\dagger}{Q}\\
    &+\frac{1}{2}(L^{\dagger}L+\epsilon {Q}N{Q}+\lambda)\hat{\rho}+\frac{1}{2}\hat{\rho}(L^{\dagger}L+\epsilon {Q}N{Q}+\lambda),\\
    f_1=&-{Q}L\hat{\rho}L^{\dagger}{P}-\epsilon {Q}a {Q}\rho_1{P} a^{\dagger} {P}\\&+\frac{1}{2}(L^{\dagger}L+\epsilon {Q}N{Q}+\lambda)\rho_1+\frac{1}{2}(\epsilon\alpha^2+\lambda)\rho_1,\\
    f_2=&-{P}L\hat{\rho}L^{\dagger}{Q}-\epsilon {P}a {P}\rho_2{Q}a^{\dagger}{Q}\\
    &+\frac{1}{2}\rho_2(L^{\dagger}L+\epsilon {Q}N{Q}+\lambda)+\frac{1}{2}(\epsilon\alpha^2+\lambda)\rho_2, \\
    f_3=& -{P}L\hat{\rho}L^{\dagger}{P}-\epsilon {P}a{P}\rho_3{P}a^{\dagger}{P}+(\epsilon \alpha^2+\lambda)\rho_3.
\end{aligned}
\end{equation}

% MAGENTA-REDUCED-SPACE-BEGIN
{
Define $\hat{\mathcal H}=Q\mathcal H$. Let $I_{\hat{\mathcal H}}$ denote the identity operator on $\hat{\mathcal H}$ and $\hat{\mathbf I}$ the identity superoperator on spaces of operators on $\hat{\mathcal H}$. As before, scalar multiples of $I_{\hat{\mathcal H}}$ are identified with the corresponding scalars whenever no ambiguity arises. Let $\hat{\mathcal K}_1$ denote the trace-class operators on $\hat{\mathcal H}$.

Identifying operators on $\hat{\mathcal H}$ with their zero extensions to $\mathcal H$, we define
\[
\hat{\mathcal K}_L
=\{\rho\in\mathcal K_L:\rho=Q\rho Q\}
=\{Q\rho Q:\rho\in\mathcal K_L\},
\]
with the norm inherited from $\mathcal K_L$. Since $Q$ commutes with $S$, the map $\rho\mapsto Q\rho Q$ is a bounded projection on $\mathcal K_L$, and
\[
\|Q\rho Q\|_L=\|Q(S\rho S)Q\|_1\leq\|\rho\|_L,
\qquad \rho\in\mathcal K_L.
\]
Thus $\hat{\mathcal K}_L$ is a closed subspace of $\mathcal K_L$ and is a Banach space.

Let $\hat S=S|_{D(S)\cap\hat{\mathcal H}}$. By Lemma~\ref{lemma:gap_1}, $\hat S\geq\sqrt2\,I_{\hat{\mathcal H}}$, so its inverse $\hat S^{-1}$ is bounded on $\hat{\mathcal H}$. For $\hat\rho\in\hat{\mathcal K}_L$,
\[
\|\hat\rho\|_L=\|\hat S\hat\rho\hat S\|_1.
\]
In the following, $\|\cdot\|_{\mathrm{op},1}$ and $\|\cdot\|_{\mathrm{op},L}$ also denote the operator norms of bounded superoperators on $\hat{\mathcal K}_1$ and $\hat{\mathcal K}_L$, respectively, while $\|\cdot\|_\infty$ also denotes the operator norm of bounded operators on $\hat{\mathcal H}$.
\par}
% MAGENTA-REDUCED-SPACE-END

Consider the superoperator $\hat{\mathcal{L}}_{\epsilon}$ on $\hat{\mathcal{K}}_{L}$:
\begin{equation}
\begin{aligned}
    \hat{\mathcal{L}}_{\epsilon}(\hat{\rho})=&{Q}L{Q}\hat{\rho}{Q}L^{\dagger}{Q}+\epsilon {Q}a{Q}\hat{\rho}{Q}a^{\dagger}{Q}\\
    &-\frac{1}{2}(L^{\dagger}L+\epsilon {Q}N{Q})\hat{\rho}-\frac{1}{2}\hat{\rho}(L^{\dagger}L+\epsilon {Q}N{Q}).
\end{aligned}
\end{equation}
Using equation~\eqref{eq:0}, we obtain
\begin{equation}
\label{eq:1}
(\lambda\hat{\mathbf{I}}-\hat{\mathcal{L}}_{\epsilon})(\hat{\rho})=\hat{f}.
\end{equation}
\begin{remark}
\label{remark:decomposition}
For every $\alpha>0$, the coherent states $\ket{\alpha}$ and $\ket{-\alpha}$ satisfy
\begin{equation*}
\braket{\alpha}{\alpha}
=
\braket{-\alpha}{-\alpha}
=
1,
\qquad
\braket{\alpha}{-\alpha}
=
e^{-2\alpha^2}.
\end{equation*}
Define 
\begin{equation*}
\begin{aligned}
    P_{\alpha}&=\frac{1}{1-e^{-4\alpha^2}}\ket{\alpha}\bra{\alpha}-\frac{e^{-2\alpha^2}}{1-e^{-4\alpha^2}}\ket{\alpha}\bra{-\alpha},\\ P_{-\alpha}&=-\frac{e^{-2\alpha^2}}{1-e^{-4\alpha^2}}\ket{-\alpha}\bra{\alpha}+\frac{1}{1-e^{-4\alpha^2}}\ket{-\alpha}\bra{-\alpha}.
\end{aligned}
\end{equation*}
A direct computation shows that $P_{\alpha}$ and $P_{-\alpha}$ are projections and satisfy
\begin{equation*}
P_{\alpha}^2=P_{\alpha},
\qquad
P_{-\alpha}^2=P_{-\alpha},
\qquad
P_{\alpha}P_{-\alpha}
=
P_{-\alpha}P_{\alpha}
=
0,
\end{equation*}
and that
\begin{equation*}
{P}=P_{\alpha}+P_{-\alpha}.
\end{equation*}
$P_{\alpha}$ and $P_{-\alpha}$ are not orthogonal projections, and hence $P_{\alpha}\neq P_{\alpha}^{\dagger}$ and $P_{-\alpha}\neq P_{-\alpha}^{\dagger}$.
Nevertheless, since ${P}$ is orthogonal ${P}=P_{\alpha}+P_{-\alpha}=P_{\alpha}^{\dagger}+P_{-\alpha}^{\dagger}$.
For all $\alpha>0$, we have
\begin{equation*}
\|P_{\alpha}\|_{\infty}\leq\frac{1}{1-e^{-2\alpha^2}},\quad\|P_{-\alpha}\|_{\infty}\leq\frac{1}{1-e^{-2\alpha^2}}.    
\end{equation*}
\end{remark}

Using $aP_{\alpha}=\alpha P_{\alpha}$, $aP_{-\alpha}=-\alpha P_{-\alpha}$, and the decomposition ${P}=P_{\alpha}^{\dagger}+P_{-\alpha}^{\dagger}$, we obtain
\begin{equation*}
    \begin{aligned}
        &f_1 P_{\alpha}^{\dagger}=-{Q}L\hat{\rho}L^{\dagger}P_{\alpha}^{\dagger}+\left(\frac{1}{2}(L^{\dagger}L+\epsilon {Q}N{Q}+\epsilon\alpha^2)-\epsilon \alpha{Q}a{Q}+\lambda\right)\rho_1P_{\alpha}^{\dagger},\\
         &f_1 P_{-\alpha}^{\dagger}
        =-{Q}L\hat{\rho}L^{\dagger}P_{-\alpha}^{\dagger}+\left(\frac{1}{2}(L^{\dagger}L+\epsilon {Q}N{Q}+\epsilon\alpha^2)+\epsilon \alpha {Q}a{Q}+\lambda\right)\rho_1P_{-\alpha}^{\dagger},
    \end{aligned}
\end{equation*}
and
\begin{equation}
\label{eq:2}
     f_3=-{P}L\hat{\rho}L^{\dagger}{P}+\lambda(P_{\alpha}\rho_3 P_{\alpha}^{\dagger}+P_{-\alpha}\rho_3 P_{-\alpha}^{\dagger})+(\lambda+2\epsilon\alpha^2)(P_{\alpha}\rho_3 P_{-\alpha}^{\dagger}+P_{-\alpha}\rho_3 P_{\alpha}^{\dagger}).
\end{equation}
Define the operators $\hat{V}_{\alpha}$ and $\hat{V}_{-\alpha}$ on $\hat{\mathcal{H}}$:
\begin{equation}
    \begin{aligned}
        \hat{V}_{\alpha}&=\frac{1}{2}L^{\dagger}L+\frac{\epsilon}{2}{Q}N{Q}-\epsilon\alpha {Q}a {Q}+\frac{\epsilon\alpha^2}{2}\\
        \hat{V}_{-\alpha}&=\frac{1}{2}L^{\dagger}L+\frac{\epsilon}{2}{Q}N{Q}+\epsilon\alpha {Q}a {Q}+\frac{\epsilon\alpha^2}{2}
    \end{aligned}
\end{equation}
Therefore, we have
\begin{equation}
\label{eq:3}
 \begin{aligned}   f_1P_{\alpha}^{\dagger}+{Q}L\hat{\rho}L^{\dagger} P_{\alpha}^{\dagger}&=(\lambda+\hat{V}_{\alpha})\rho_1 P_{\alpha}^{\dagger},\\
 f_1P_{-\alpha}^{\dagger}+{Q}L\hat{\rho}L^{\dagger}P_{-\alpha}^{\dagger}&=(\lambda+\hat{V}_{-\alpha})\rho_1P_{-\alpha}^{\dagger},
\end{aligned} 
\end{equation}
and similarly
\begin{equation}
\label{eq:4}
 \begin{aligned}   P_{\alpha}f_2+P_{\alpha}L\hat{\rho}L^{\dagger} {Q}&=P_{\alpha}\rho_2 (\lambda+\hat{V}_{\alpha}^{\dagger}),\\
 P_{-\alpha}f_2+P_{-\alpha}L\hat{\rho}L^{\dagger}{Q}&=P_{-\alpha}\rho_2(\lambda+\hat{V}_{-\alpha}^{\dagger}).
\end{aligned} 
\end{equation}
We can solve the equations in~\eqref{eq:0} block by block, provided that $\lambda\hat{\mathbf{I}}-\hat{\mathcal{L}}_{\epsilon}$ is invertible on $\hat{\mathcal{K}}_{L}$, that $\lambda+\hat{V}_{\alpha}$ and $\lambda+\hat{V}_{-\alpha}$ are invertible on $\hat{\mathcal{H}}$, and that $\lambda\notin\{0,-2\epsilon\alpha^2\}$. We now analyze these conditions one by one.

\subsubsection{Inverse of $\lambda\hat{\mathbf{I}}-\hat{\mathcal{L}}_{\epsilon}$}
  Take $\alpha>0$, $\epsilon\geq 0$ and $\lambda\in\mathbb{R}$ such that $L^{\dagger}L+\epsilon {Q}N{Q}+\lambda$ is a strictly positive operator on $\hat{\mathcal{H}}$. Set $C=L^\dagger L+\epsilon QNQ+\lambda I_{\hat{\mathcal H}}$, which is self-adjoint and satisfies $C>cI_{\hat{\mathcal{H}}}$ for some $c>0$. Then for $\hat{\rho}\in\hat{\mathcal{K}}_1$ (respectively, $\hat{\rho}\in\mathcal{B}(\hat{\mathcal{H}})$), let us define
\begin{equation}
    \hat{\Pi}_{\lambda}^{\epsilon}(\hat{\rho})=\int_{0}^{\infty}e^{-\frac{s}{2}C}\hat{\rho}e^{-\frac{s}{2}C}\text{d}s.
\end{equation}
On \(\mathcal X=\hat{\mathcal K}_1\), (respectively \(\mathcal X=\mathcal B(\hat{\mathcal H})\)), we consider the maximal weak realization \(\mathcal A_C\) of the map \(\rho\mapsto\frac12(C\rho+\rho C)\). Its domain consists of all \(\rho\in\mathcal X\) for which there exists \(f\in\mathcal X\) satisfying
\[
\frac12\bigl(\langle Cu,\rho v\rangle+\langle u,\rho Cv\rangle\bigr)
=\langle u,fv\rangle,
\qquad\text{for all }u,v\in D(C),
\]
and we set \(\mathcal A_C(\rho)=f\). This realization is bijective, and its inverse is precisely
\[
\mathcal A_C^{-1}(f)
=\hat\Pi_\lambda^\epsilon(f)
=\int_0^\infty e^{-tC/2}f\,e^{-tC/2}\,\mathrm dt,
\]
where the integral converges in trace norm (respectively operator norm).
% REVIEW-END 14
\begin{remark}
Let $A$ and $B$ be positive self-adjoint operators on $\mathcal{H}$ (respectively, on $\hat{\mathcal{H}}$), and assume that $A^{1/2}B^{1/2}$ extends to a bounded operator. If $A^{1/2}BA^{1/2}$ is trace class, then $B^{1/2}AB^{1/2}$ is also trace class and
\begin{equation*}
\Tr\left(A^{1/2}BA^{1/2}\right)
=
\Tr\left(B^{1/2}AB^{1/2}\right).
\end{equation*}
The operator product $AB$ itself need not be trace class. With a slight abuse of notation, we therefore use the convention
\begin{equation*}
\Tr(AB)
:=
\Tr\left(A^{1/2}BA^{1/2}\right),
\qquad
\Tr(BA)
:=
\Tr\left(B^{1/2}AB^{1/2}\right).
\end{equation*}
With this convention,
\begin{equation*}
\Tr(AB)=\Tr(BA).
\end{equation*}
In the following, whenever we write $\Tr(AB)<\infty$, it is understood in the above sense; namely,
\begin{equation*}
\Tr\left(A^{1/2}BA^{1/2}\right)<\infty.
\end{equation*}
\end{remark}

\begin{remark}
\label{remark:Pi_lambda_epsilon}
Let
\begin{equation*}
\hat{S}_{\epsilon}
=
\left(
L^{\dagger}L
+
\epsilon
{Q}N{Q}
\right)^{1/2},
\end{equation*}
which is a strictly positive operator on $\hat{\mathcal{H}}$. {For $k\in\mathbb{R}$, consider a positive operator $\hat{\rho}_k$ such that there exists a positive $\hat{\rho}_0\in\hat{\mathcal{K}_1}$ (respectively in $\mathcal{B}(\hat{\mathcal{H}})$) satisfying
\[\hat{\rho}_k=\hat{S}_{\epsilon}^{k}\hat{\rho}_0\hat{S}_{\epsilon}^k.\]}
We extend the definition of $\hat{\Pi}_{\lambda}^{\epsilon}$ to such operators by setting
\begin{equation*}
\begin{aligned}
\hat{\Pi}_{\lambda}^{\epsilon}(\hat{\rho}_k)
=
\hat{S}_{\epsilon}^{k}
\bigg(
\int_0^\infty
&e^{-\frac{s}{2}(\hat{S}_{\epsilon}^2+\lambda)}
\hat{\rho}_0
e^{-\frac{s}{2}(\hat{S}_{\epsilon}^2+\lambda)}
\mathrm{d}s
\bigg)
\hat{S}_{\epsilon}^{k}.
\end{aligned}
\end{equation*}
Since $\hat{S}_{\epsilon}$ commutes with
$e^{-\frac{s}{2}(\hat{S}_{\epsilon}^2+\lambda)}$, this extension satisfies
\begin{equation*}
\begin{aligned}
\frac12(\hat{S}_{\epsilon}^2+\lambda)
\hat{\Pi}_{\lambda}^{\epsilon}(\hat{\rho}_k)
+
\frac12
\hat{\Pi}_{\lambda}^{\epsilon}(\hat{\rho}_k)
(\hat{S}_{\epsilon}^2+\lambda)
=
\hat{\rho}_k.
\end{aligned}
\end{equation*}

Let now $\hat{\rho}\in\hat{\mathcal{K}}_1$ be positive and let $\hat{A}$ be a positive self-adjoint operator on $\hat{\mathcal{H}}$. Assume that there exists $k>0$ {and a positive $\hat{A}_0\in\mathcal{B}(\hat{\mathcal{H}})$ such that
\begin{equation*}
\hat{A}=\hat{S}_{\epsilon}^k\hat{A}_0\hat{S}_{\epsilon}^{k},
\end{equation*}}
so that $\hat{\Pi}_{\lambda}^{\epsilon}(\hat{A})$ is defined by the above extension. If
\begin{equation*}
\Tr\left(
\hat{A}
\hat{\Pi}_{\lambda}^{\epsilon}(\hat{\rho})
\right)
<\infty,
\end{equation*}
where the trace is understood in the sense of the convention introduced above, then
\begin{equation*}
\Tr\left(
\hat{A}
\hat{\Pi}_{\lambda}^{\epsilon}(\hat{\rho})
\right)
=
\Tr\left(
\hat{\Pi}_{\lambda}^{\epsilon}(\hat{A})
\hat{\rho}
\right).
\end{equation*}
Indeed, setting
\begin{equation*}
T(s)
=
e^{-\frac{s}{2}(\hat{S}_{\epsilon}^2+\lambda)},
\end{equation*}
the positivity of $\hat{\rho}$ and $\hat{A}$ and Tonelli's theorem yield
\begin{equation*}
\begin{aligned}
\Tr\left(
\hat{A}
\hat{\Pi}_{\lambda}^{\epsilon}(\hat{\rho})
\right)
&=
\int_0^\infty
\Tr\left(
\hat{\rho}^{1/2}
T(s)\hat{A}T(s)
\hat{\rho}^{1/2}
\right)\mathrm{d}s.
\end{aligned}
\end{equation*}
Since $T(s)$ commutes with $\hat{S}_{\epsilon}$,
\begin{equation*}
\int_0^\infty
\Tr\left(
\hat{\rho}^{\frac{1}{2}}
T(s)\hat{A}T(s)
\hat{\rho}^{\frac{1}{2}}
\right)\mathrm{d}s
=\int_0^{\infty}\Tr\left(\hat{\rho}^{\frac{1}{2}}
\hat{S}_{\epsilon}^{k}
T(s)
\hat{A}_0
T(s)
\hat{S}_{\epsilon}^{k}\hat{\rho}^{\frac{1}{2}}\right)\mathrm{d}s.
\end{equation*}
 Hence, applying Tonelli's theorem once more,
\begin{equation*}
\Tr\left(
\hat{\rho}^{1/2}
\hat{\Pi}_{\lambda}^{\epsilon}(\hat{A})
\hat{\rho}^{1/2}
\right)
=
\Tr\left(
\hat{\Pi}_{\lambda}^{\epsilon}(\hat{A})
\hat{\rho}
\right),
\end{equation*}
where the last equality is understood according to the trace convention introduced above.
\end{remark}
On $\hat{\mathcal{K}}_1$, let us define
{\[
\hat Q_\lambda^\epsilon(\hat\rho)
=QLQ\,\hat\Pi_\lambda^\epsilon(\hat\rho)\,QL^\dagger Q
+\epsilon QaQ\,\hat\Pi_\lambda^\epsilon(\hat\rho)\,Qa^\dagger Q.
\]}

\begin{comment}
\begin{lemma}
\label{lemma:bound_4}
    There exists $\alpha_0>0$ such that for all $\alpha\geq\alpha_0$, we have 
    \begin{equation*}
        L^{\dagger}LN{Q}+{Q}NL^{\dagger}L\geq-8L^{\dagger}L.
    \end{equation*}
\end{lemma}
    \begin{proof}
        Since $LN=NL+2a^2$, we have
        \begin{equation*}
            \begin{aligned}
                L^{\dagger}LN{Q}+{Q}NL^{\dagger}L&=L^{\dagger}(NL+2a^2){Q}+{Q}(L^{\dagger}N+2(a^{\dagger})^2)L\\
                &=2L^{\dagger}NL+2L^{\dagger}L+2L^{\dagger}a^2{Q}+2{Q}(a^{\dagger})^2L.
            \end{aligned}
        \end{equation*}
        Since $(L^{\dagger}a+{Q}a^{\dagger})(a^{\dagger}L+a{Q})\ge 0$, we have
        \[{Q}(a^{\dagger})^2L+L^{\dagger}a^2{Q}\geq -L^{\dagger}(N+1)L-{Q}N {Q}.\]
        By Lemma~\ref{lemma:gap_2}, there exists $\alpha_0>0$ such that for all $\alpha\geq \alpha_0$, we have $3L^{\dagger}L\geq {Q}N{Q}$. Therefore, for all $\alpha\geq \alpha_0$,
        \begin{equation*}
\begin{aligned}L^{\dagger}LN{Q}+{Q}NL^{\dagger}L&=2L^{\dagger}NL+2L^{\dagger}L+2L^{\dagger}a^2{Q}+2{Q}(a^{\dagger})^2L\\
&\geq 2L^{\dagger}NL-2L^{\dagger}(N+1)L-2{Q}N{Q}\geq-8L^{\dagger}L.
            \end{aligned}
        \end{equation*}
    \end{proof}
\end{comment}
\begin{lemma}
\label{lemma:inverse_1}
    For all $\alpha>0$, $\epsilon\ge 0$ and $\lambda>0$, we have \[\|\hat{Q}_{\lambda}^{\epsilon}\|_{\mathrm{op},1}\leq 1.\]
\end{lemma}
\begin{proof}
    Take $\alpha>0$, $\epsilon\geq 0$ and $\lambda>0$. By the complete positivity of $\hat{Q}_{\lambda}^{\epsilon}$ , it suffices to prove  that for all positive $\hat{\rho}\in\hat{\mathcal{K}}_1$, we have $\Tr(\hat{Q}_{\lambda}^{\epsilon}(\hat{\rho}))\leq\Tr(\hat{\rho})$. Notice that
    \[L^{\dagger}{Q}L+\epsilon {Q}a^{\dagger}{Q}a{Q}\leq L^{\dagger}L+\epsilon {Q}N{Q}+\lambda.\]
    Thus
    \[\hat{\Pi}_{\lambda}^{\epsilon}(L^{\dagger}{Q}L+\epsilon {Q}a^{\dagger}{Q}a{Q})\leq \hat{\Pi}_{\lambda}^{\epsilon}(L^{\dagger}L+\epsilon {Q}N{Q}+\lambda)= I_{\hat{\mathcal{H}}}.\]
    By Remark~\ref{remark:Pi_lambda_epsilon}, we can conclude that for all positive $\hat{\rho}\in\hat{\mathcal{K}}_{1}$,
    \begin{equation*}
        \begin{aligned}
            \Tr(\hat{Q}_{\lambda}^{\epsilon}(\hat{\rho}))&=\Tr((L^{\dagger}{Q}L+\epsilon {Q}a^{\dagger}{Q}a{Q})\hat{\Pi}_{\lambda}^{\epsilon}(\hat{\rho}))\\
            &=\Tr(\hat{\Pi}_{\lambda}^{\epsilon}(L^{\dagger}{Q}L+\epsilon {Q}a^{\dagger}{Q}a{Q})\hat{\rho})\leq\Tr(\hat{\rho}).
        \end{aligned}
    \end{equation*}
\end{proof}
{Lemma~\ref{lemma:inverse_1} ensures that for all $r\in(0,1)$
and $\rho\in \hat{\Pi}_{\lambda}^{\epsilon}(\hat{\mathcal{K}}_1)$ with $\lambda>0$, we can define
\begin{equation}
\label{eq:hat_L_epsilon_r}
\begin{aligned}
    \hat{\mathcal{L}}^{r}_{\epsilon}(\hat{\rho})=&r{Q}L{Q}\hat{\rho}{Q}L^{\dagger}{Q}+\epsilon r{Q}a{Q}\hat{\rho}{Q}a^{\dagger}{Q}\\
    &-\frac{1}{2}(L^{\dagger}L+\epsilon {Q}N{Q})\hat{\rho}-\frac{1}{2}\hat{\rho}(L^{\dagger}L+\epsilon {Q}N{Q}).
\end{aligned}
\end{equation}
Notice that here $D(\hat{\mathcal{L}}_{\epsilon}^r)=\hat{\Pi}_{\lambda}^{\epsilon}(\hat{\mathcal{K}}_{1})$ and the later is independent of $\lambda>0$.}
\begin{lemma}
\label{lemma:inverse_K_1}
    For all $\alpha>0$, $\epsilon\geq 0$, $\lambda>0$ and $r\in(0,1)$, $\lambda\hat{\mathbf{I}}-\hat{\mathcal{L}}_{\epsilon}^{r}$ is invertible on $\hat{\mathcal{K}}_1$ and 
    \[\|(\lambda\hat{\mathbf{I}}-\hat{\mathcal{L}}_{\epsilon}^{r})^{-1}\|_{\mathrm{op},1}\leq \frac{1}{\lambda}.\]
\end{lemma}
\begin{proof}
    Take $\alpha>0$, $\epsilon\geq 0$, $\lambda>0$ and $r\in(0,1)$. By Lemma~\ref{lemma:inverse_1}, it can be easily verified that $\lambda\hat{\mathbf{I}}-\hat{\mathcal{L}}_{\epsilon}^{r}$ is invertible on $\hat{\mathcal{K}}_1$ and its inverse is given by 
    \[(\lambda\hat{\mathbf{I}}-\hat{\mathcal{L}}_{\epsilon}^{r})^{-1}=\sum_{n=0}^{\infty}r^n\hat{\Pi}_{\lambda}^{\epsilon}\circ\left(\hat{Q}_{\lambda}^{\epsilon}\right)^{n}.\]
    Since $(\lambda\mathbf{I}-\hat{\mathcal{L}}_{\epsilon}^{r})^{-1}$ is a completely positive operator on $\hat{\mathcal{K}}_1$, to conclude the proof, it suffices to prove that all positive $\hat{f}\in\hat{\mathcal{K}}_1$,
    \[\Tr((\lambda\hat{\mathbf{I}}-\hat{\mathcal{L}}^{r}_{\epsilon})^{-1}(\hat{f}))\leq\frac{1}{\lambda}\Tr(\hat{f}).\]
Take a positive $\hat{f}\in\hat{\mathcal{K}}_1$ and define the following sequence of positive operators in $\hat{\mathcal{K}}_1$
\[\hat{\xi}_0=\hat{f},\quad \hat{\xi}_{n+1}=\hat{Q}_{\lambda}^{\epsilon}(\hat{\xi}_{n}),\quad\forall n\in\mathbb{N}.\]
By the definition of the sequence $\{\hat{\xi}_{n}\}_{n\in\mathbb{N}}$ and Remark~\ref{remark:Pi_lambda_epsilon}, for all $n\in\mathbb{N}$, we have
\begin{equation*}
    \begin{aligned}
        \Tr(\hat{\xi}_{n+1})&=\Tr(\hat{Q}_{\lambda}^{\epsilon}(\hat{\xi}_{n}))=\Tr((L^{\dagger}{Q}L+\epsilon {Q}a^{\dagger}{Q}a{Q})\hat{\Pi}_{\lambda}^{\epsilon}(\hat{\xi}_{n}))\\
        &\leq\Tr((L^{\dagger}L+\epsilon {Q}N{Q}+\lambda)\hat{\Pi}_{\lambda}^{\epsilon}(\hat{\xi}_{n}))-\lambda\Tr(\hat{\Pi}_{\lambda}^{\epsilon}(\hat{\xi}_{n}))\\
        &=\Tr(\hat{\xi}_{n})-\lambda\Tr(\hat{\Pi}_{\lambda}^{\epsilon}(\hat{\xi}_{n}))
    \end{aligned}
\end{equation*}
We can deduce that for all $n\in\mathbb{N}$, we have 
\[\Tr(\hat{\Pi}_{\lambda}^{\epsilon}(\hat\xi_{n}))\leq\frac{1}{\lambda}(\Tr(\hat{\xi}_{n})-\Tr(\hat{\xi}_{n+1})).\]
By adding this inequality with $n$ from 0 to $N_0\in\mathbb{N}$, we can obtain
\begin{equation*}
    \Tr\left(\sum_{n=0}^{N_0}\hat{\Pi}_{\lambda}^{\epsilon}(\hat{\xi}_{n})\right)\leq \frac{1}{\lambda}\Tr(\hat{\xi}_0)-\frac{1}{\lambda}\Tr(\hat{\xi}_{N_0+1})\leq\frac{1}{\lambda}\Tr(\hat{\xi}_0)=\frac{1}{\lambda}\Tr(\hat{f}).
\end{equation*}
Tend $N_0$ to infinity, and we can obtain
\[\Tr\left(\sum_{n=0}^{\infty}\hat{\Pi}_{\lambda}^{\epsilon}(\hat{\xi}_{n})\right)\leq\frac{1}{\lambda}\Tr(\hat{f})<\infty.\]
Since $\{\hat{\Pi}_{\lambda}^{\epsilon}(\hat{\xi}_{n})\}_{n}$ is a sequence of positive operators in $\hat{\mathcal{K}}_1$, we have
\begin{equation*}
\begin{aligned}
    \Tr\left((\lambda\hat{\mathbf{I}}-\hat{\mathcal{L}}_{\epsilon}^{r})^{-1}(\hat{f})\right)=\Tr\left(\sum_{n=0}^{\infty}r^n\hat{\Pi}_{\lambda}^{\epsilon}(\hat{\xi}_{n})\right)\leq \Tr\left(\sum_{n=0}^{\infty}\hat{\Pi}_{\lambda}^{\epsilon}(\hat{\xi}_{n})\right)\leq\frac{1}{\lambda}\Tr(\hat{f}).
\end{aligned}
\end{equation*}
\end{proof}

\begin{prop}
\label{prop:resolvent_hat_L_epsilon}
    There exists $\alpha_{0}> 0$ such that for all $\alpha\geq\alpha_{0}$, $\epsilon\in[0,4]$ and $\lambda\in(-2+4\epsilon,+\infty)$, $\lambda\hat{\mathbf{I}}-\hat{\mathcal{L}}_{\epsilon}$ is invertible on $\hat{\mathcal{K}}_{L}$. Moreover, for all $\hat{f}\in\hat{\mathcal{K}}_{L}$, we have
    \[\|\Delta(\epsilon,\lambda)^{\frac{1}{2}}(\lambda\hat{\mathbf{I}}-\hat{\mathcal{L}_{\epsilon}})^{-1}(\hat{f})\Delta(\epsilon,\lambda)^{\frac{1}{2}}\|_1\leq \|\hat{f}\|_{L},\]
    with $\Delta(\epsilon,\lambda)=(4-\epsilon)L^{\dagger}NL+(2-4\epsilon+\lambda)L^{\dagger}L$.
\end{prop}
\begin{proof}
By Lemma~\ref{lemma:gap_1} and Lemma~\ref{lemma:gap_2}, there exists $\alpha_0>0$ such that for all $\alpha\geq \alpha_0$, $L^{\dagger}L\geq 3{Q}$ and $3L^{\dagger}L\geq {Q}N{Q}$. Take $\alpha\geq \alpha_0$, $\epsilon\in[0,4]$ and $\lambda\in(-2+4\epsilon,+\infty)$. 
Using $LN=NL+2a^2$, we obtain
        \begin{equation*}
            \begin{aligned}
                L^{\dagger}LN{Q}+{Q}NL^{\dagger}L&=L^{\dagger}(NL+2a^2){Q}+{Q}(L^{\dagger}N+2(a^{\dagger})^2)L\\
                &=2L^{\dagger}NL+2L^{\dagger}a^2{Q}+2{Q}(a^{\dagger})^2L.
            \end{aligned}
        \end{equation*}
        Since $(L^{\dagger}a+{Q}a^{\dagger})(a^{\dagger}L+a{Q})\ge 0$
        \begin{equation*}\begin{aligned}{Q}(a^{\dagger})^2L+L^{\dagger}a^2{Q}&\geq -L^{\dagger}(N+1)L-{Q}N {Q}\\
        &\geq -L^{\dagger}(N+1)L-3L^{\dagger}L.\end{aligned}\end{equation*}
We can deduce
\begin{equation}
  \label{eq:prop_hat_L_epsilon_1}\begin{aligned}L^{\dagger}LN{Q}+{Q}NL^{\dagger}L&=2L^{\dagger}NL+2L^{\dagger}a^2{Q}+2{Q}(a^{\dagger})^2L\\
  &\geq2L^{\dagger}NL-2L^{\dagger}(N+1)L-6L^{\dagger}L\geq -8L^{\dagger}L.
  \end{aligned}
\end{equation}
Let us define \begin{equation*}\Delta(\epsilon,\lambda)=(4-\epsilon)L^{\dagger}NL+(2-4\epsilon+\lambda)L^{\dagger}L.\end{equation*}
Notice that $L^{\dagger}L+\epsilon {Q}N{Q}+\lambda>I_{\hat{\mathcal{H}}}$ and $\hat{\Pi}_{\lambda}^{\epsilon}(\Delta(\epsilon,\lambda))$ is well defined and positive. Moreover, using equation~\eqref{eq:prop_hat_L_epsilon_1} and $LL^{\dagger}=L^{\dagger}L+4N+2$ we obtain
\begin{equation*}
    \begin{aligned}
        &\frac{1}{2}\left(L^{\dagger}L+\epsilon {Q}N{Q}+\lambda\right)L^{\dagger}L+\frac{1}{2}L^{\dagger}L(L^{\dagger}L+\epsilon {Q}N {Q}+\lambda)-\Delta(\epsilon,\lambda)\\
        &=L^{\dagger}LL^{\dagger}L+\lambda L^{\dagger}L+\frac{\epsilon}{2}{Q}NL^{\dagger}L+\frac{\epsilon}{2}L^{\dagger}LN{Q}-\Delta(\epsilon,\lambda)\\
        &\geq L^{\dagger}LL^{\dagger}L+\lambda L^{\dagger}L-4\epsilon L^{\dagger}L-\Delta(\epsilon,\lambda)\\&=L^{\dagger}(L^{\dagger}L+4N+2)L+(\lambda-4\epsilon)L^{\dagger}L-\Delta(\epsilon,\lambda)=L^{\dagger}L^{\dagger}LL+\epsilon L^{\dagger}NL.
        \end{aligned}
\end{equation*}
 Applying $\hat{\Pi}_{\lambda}^{\epsilon}$ on both sides of the equation above, we obtain
\begin{equation}
    L^{\dagger}L-\hat{\Pi}_{\lambda}^{\epsilon}(\Delta(\epsilon,\lambda))\geq \hat{\Pi}_{\lambda}^{\epsilon}(L^{\dagger}L^{\dagger}LL+\epsilon L^{\dagger}NL)\geq 0.
\end{equation}
Take a positive operator $\hat{f}\in\hat{\mathcal{K}}_{L}$ and define the following sequence of positive operators:
\[\hat{\xi}_0=\hat{f},\quad \hat{\xi}_{n+1}=\hat{Q}_{\lambda}^{\epsilon}(\hat{\xi}_{n}),\quad\forall n\in\mathbb{N}.\]
Then, by the construction of the sequence $\{\hat{\xi}_{n}\}_{n\in\mathbb{N}}$ and Remark~\ref{remark:Pi_lambda_epsilon}, for all $n\in\mathbb{N}$, we have
\begin{equation}
\begin{aligned}\Tr(L^{\dagger}L\hat{\xi}_{n+1})&=\Tr(L^{\dagger}L\hat{Q}_{\lambda}^{\epsilon}(\hat{\xi}_{n}))=\Tr((L^{\dagger}L^{\dagger}LL+\epsilon {Q}a^{\dagger}L^{\dagger}La{Q})\hat{\Pi}_{\lambda}^{\epsilon}(\hat{\xi}_{n}))\\&=\Tr((L^{\dagger}L^{\dagger}LL+\epsilon L^{\dagger}NL)\hat{\Pi}_{\lambda}^{\epsilon}(\hat{\xi}_{n}))=\Tr(\hat{\Pi}_{\lambda}^{\epsilon}(L^{\dagger}L^{\dagger}LL+\epsilon L^{\dagger}NL)\hat{\xi}_{n})\\&\leq \Tr((L^{\dagger}L-\hat{\Pi}_{\lambda}^{\epsilon}(\Delta(\epsilon,\lambda)))\hat{\xi}_{n}).
    \end{aligned}
\end{equation}
By Remark~\ref{remark:Pi_lambda_epsilon}, we can deduce that for all $n\in\mathbb{N}$, we have
\begin{equation*}
    \Tr(\Delta(\epsilon,\lambda)\hat{\Pi}_{\lambda}^{\epsilon
    }(\hat{\xi}_{n}))=\Tr(\hat{\Pi}_{\lambda}^{\epsilon}(\Delta(\epsilon,\lambda))\hat{\xi}_{n})\leq \Tr(L^{\dagger}L\hat{\xi}_{n})-\Tr(L^{\dagger}L\hat{\xi}_{n+1}).
\end{equation*}

By adding this inequality with $n$ from $0$ to $N_0\in\mathbb{N}$, we can obtain
\begin{equation*}
\label{eq:}
\begin{aligned}\Tr\left(\Delta(\epsilon,\lambda)\sum_{n=0}^{N_0}\hat{\Pi}_{\lambda}^{\epsilon}(\hat{\xi}_{n})\right)&\leq \Tr(L^{\dagger}L\hat{\xi}_0)-\Tr(L^{\dagger}L\hat{\xi}_{N_0+1}) \leq\Tr(L^{\dagger}L\hat{\xi}_0)=\Tr(L^{\dagger}L\hat{f}).\end{aligned}\end{equation*}
Tend $N_0$ to infinity and we can obtain
\begin{equation}
\label{eq:prop_hat_L_epsilon_2}
(2-4\epsilon+\lambda)\Tr\left(L^{\dagger}L\sum_{n=0}^{\infty}\hat{\Pi}_{\lambda}^{\epsilon}(\hat{\xi}_n)\right)\leq\Tr\left(\Delta(\epsilon,\lambda)\sum_{n=0}^{\infty}\hat{\Pi}_{\lambda}^{\epsilon}(\hat{\xi}_n)\right)\leq\Tr(L^{\dagger}L\hat{f})\leq \infty.
\end{equation}
    We can then well define in $\hat{\mathcal{K}}_{L}$
\[\hat{R}_{\lambda}^{\epsilon}(\hat{f}):=\sum_{n=0}^{\infty}\hat{\Pi}_{\lambda}^{\epsilon}(\hat{\xi}_n).\]
We can extend the definition of $\hat{R}_{\lambda}^{\epsilon}$ to all $\hat{\mathcal{K}}_{L}$ by observing that every element in $\hat{\mathcal{K}}_{L}$ can be written as a complex linear combination of positive operators in $\hat{\mathcal{K}}_{L}$. $\hat{R}_{\lambda}^{\epsilon}$ is a bounded and completely positive operator on $\hat{\mathcal{K}}_{L}$. {Moreover, we can verify that $\hat{R}_{\lambda}^{\epsilon}(\hat{\mathcal{K}}_{L})\subset D(\hat{\mathcal L}_\epsilon)$ under maximal weak realization and  $(\lambda\hat{\mathbf{I}}-\hat{\mathcal{L}}_{\epsilon})\circ\hat{R}_{\lambda}^{\epsilon}=\hat{\mathbf{I}}$.} Therefore $\lambda\hat{\mathbf{I}}-\hat{\mathcal{L}}_{\epsilon}$ is surjective. In order to conclude that $\hat{R}_{\lambda}^{\epsilon}$ is the inverse of $\lambda\hat{\mathbf{I}}-\hat{\mathcal{L}}_{\epsilon}$, we still need to prove the injectivity. Take $\hat{\rho}_0\in\hat{\mathcal{K}}_{L}$ such that $(\lambda\hat{\mathbf{I}}-\hat{\mathcal{L}}_{\epsilon})(\hat{\rho}_0)=0$. 
If $\lambda>0$, then for all $r\in(0,1)$, by the definition of $\hat{\mathcal{L}}_{\epsilon}^{r}$ in equation~\eqref{eq:hat_L_epsilon_r} we have
\begin{equation*}
(\lambda\hat{\mathbf{I}}-\hat{\mathcal{L}}_{\epsilon}^{r})(\hat{\rho}_0)=(1-r)\hat{f}_0,
\end{equation*}
with \[\hat{f}_0={Q}L\hat{\rho}_0L^{\dagger} {Q}+\epsilon {Q}a{Q}\hat{\rho}_0{Q}a^{\dagger}{Q}.\]
Since $\hat{\rho}_0$ is in $\hat{\mathcal{K}}_L$, it can be easily verified that $\hat{f}_0$ is in $\hat{\mathcal{K}}_1$. By Lemma~\ref{lemma:inverse_K_1}, we can deduce that $\hat{\rho}_0=(1-r)(\lambda\hat{\mathbf{I}}-\hat{\mathcal{L}}_{\epsilon}^{r})^{-1}(\hat{f}_0)$
and
\[\|\hat{\rho}_0\|_1=\|(1-r)(\lambda\hat{\mathbf{I}}-\hat{\mathcal{L}}_{\epsilon}^r)^{-1}(\hat{f_0})\|_1\le \frac{1-r}{\lambda}\|\hat{f}_0\|_1,\]
for all $r\in(0,1)$. Thus $\|\hat{\rho}_0\|_1=0$, $\hat{\rho}_0=0$ and  $\lambda\hat{\mathbf{I}}-\hat{\mathcal{L}}_{\epsilon}$ is injective on $\hat{\mathcal{K}}_L$. We can conclude that $\lambda>0$, $\lambda\hat{\mathbf{I}}-\hat{\mathcal{L}}_{\epsilon}$ is invertible on $\hat{\mathcal{K}}_{L}$ and
\[(\lambda\hat{\mathbf{I}}-\hat{\mathcal{L}}_{\epsilon})^{-1}=\hat{R}_{\lambda}^{\epsilon}.\]
If $-2+4\epsilon<\lambda\leq 0$, let us take $\lambda_0>0$ and we can obtain by the reasoning above that $\lambda_0\hat{\mathbf{I}}-\hat{\mathcal{L}}_{\epsilon}$ is invertible on $\hat{\mathcal{K}}_L$. Furthermore, by the complete positivity of its inverse and equation~\eqref{eq:prop_hat_L_epsilon_2}, we can deduce that
\[\|(\lambda-\lambda_0)(\lambda_0\hat{\mathbf{I}}-\hat{\mathcal{L}_\epsilon})^{-1}\|_{\mathrm{op},L}\leq\frac{\lambda_0-\lambda}{2-4\epsilon+\lambda_0}<1.\]
Since $\lambda\hat{\mathbf{I}}-\hat{\mathcal{L}}_{\epsilon}=(\hat{\mathbf{I}}+(\lambda-\lambda_0)(\lambda_0\hat{\mathbf{I}}-\hat{\mathcal{L}_{\epsilon}})^{-1})\circ(\lambda_0\hat{\mathbf{I}}-\hat{\mathcal{L}}_{\epsilon})$, $\lambda\hat{\mathbf{I}}-\hat{\mathcal{L}}_{\epsilon}$ is invertible on $\hat{\mathcal{K}}_{L}$ and thus bijective. We finally conclude that for all $\alpha\geq \alpha_0$, $\epsilon\in[0,4]$ and $\lambda\in(-2+4\epsilon,\infty)$, $\lambda\hat{\mathbf{I}}-\hat{\mathcal{L}}_{\epsilon}$ is invertible on $\hat{\mathcal{K}}_{L}$ and as its right inverse, $\hat{R}_{\lambda}^{\epsilon}$ coincides with its inverse:
\[(\lambda\hat{\mathbf{I}}-\hat{\mathcal{L}}_{\epsilon})^{-1}=\hat{R}_{\lambda}^{\epsilon}.\]
Furthermore, by the complete positivity of $\hat{R}_{\lambda}^{\epsilon}$ and equation~\eqref{eq:prop_hat_L_epsilon_2}, for all $\hat{f}\in\hat{\mathcal{K}}_{L}$, we have
    \[\|\Delta(\epsilon,\lambda)^{\frac{1}{2}}(\lambda\hat{\mathbf{I}}-\hat{\mathcal{L}_{\epsilon}})^{-1}(\hat{f})\Delta(\epsilon,\lambda)^{\frac{1}{2}}\|_1\leq \|\hat{f}\|_{L}.\]
\end{proof}

\begin{comment}
\begin{remark}
    Since $\hat{\mathcal{L}}_{\epsilon}$ is a completely positive map, for all $\hat{f}\in\hat{\mathcal{K}}_{L}$, we have
    \[\|\Delta(\epsilon,\lambda)^{\frac{1}{2}}(\lambda\mathbf{I}-\hat{\mathcal{L}_{\epsilon}})^{-1}(\hat{f})\Delta(\epsilon,\lambda)^{\frac{1}{2}}\|_1\leq \|\hat{f}\|_{L}.\]
\end{remark}
\end{comment}
\begin{lemma}
\label{lemma:hat_L}
    There exist $\alpha_{0}> 0$ and $\epsilon_0>0$ such that for all $\alpha\geq \alpha_{0}$, $\epsilon\in[0,\epsilon_0]$ and $\lambda\in D_{\frac{1}{2}}$, $\lambda\hat{\mathbf{I}}-\hat{\mathcal{L}}_{\epsilon}$ is invertible on $\hat{\mathcal{K}}_{L}$ and for all $\hat{f}$ in $\hat{\mathcal{K}}_{L}$, we have 
\begin{equation*}
\begin{aligned}
&\|(N+1)^{\frac{1}{2}}L(\lambda\hat{\mathbf{I}}-\hat{\mathcal{L}}_{\epsilon})^{-1}(\hat{f})L^{\dagger}(N+1)^{\frac{1}{2}}\|_1\leq \|\hat{f}\|_{L},\\
&\|(L^{\dagger}L+1)^{\frac{1}{2}}a(\lambda\hat{\mathbf{I}}-\hat{\mathcal{L}}_{\epsilon})^{-1}(\hat{f})a^{\dagger}(L^{\dagger}L+1)^{\frac{1}{2}}\|_1\leq3\|\hat{f}\|_{L},\\
&\|\hat{S}(\lambda\hat{\mathbf{I}}-\hat{\mathcal{L}}_{
\epsilon
})^{-1}(\hat{f})\hat{S}\|_1=\|(\lambda\hat{\mathbf{I}}-\hat{\mathcal{L}}_{\epsilon})^{-1}(\hat{f})\|_L\leq\|\hat{f}\|_{L}.
\end{aligned}
\end{equation*}
\begin{proof}
By Lemma~\ref{lemma:gap_2} and Proposition~\ref{prop:resolvent_hat_L_epsilon} there exist $\alpha_0>0$ and $\epsilon_0\in(0,\frac{1}{8}]$ such that, for all $\alpha\ge \alpha_0$ and $\epsilon\in[0,\epsilon_0]$, \[3L^{\dagger}L\geq {Q}N{Q},\] and for all $\lambda>-\frac{3}{2}$, the operator $\lambda\hat{\mathbf{I}}-\hat{\mathcal{L}}_{\epsilon}$ is invertible on $\hat{\mathcal{K}}_{L}$. Moreover, for all $\hat{f}\in\hat{\mathcal{K}}_L$,
\begin{equation*}
       (2-4\epsilon) \|\hat{\mathcal{L}}_{\epsilon}^{-1}(\hat{f})\|_{L}\leq\|\Delta(\epsilon,0)^{\frac{1}{2}}\hat{\mathcal{L}}_{\epsilon}^{-1}(\hat{f})\Delta(\epsilon,0)^{\frac{1}{2}}\|_{1}\leq \|\hat{f}\|_{L},
    \end{equation*}
 where $\Delta(\epsilon,0)=(4-\epsilon)L^{\dagger}NL+(2-4\epsilon)L^{\dagger}L$. Now take $\alpha\geq \alpha_0$, $\epsilon\in[0,\epsilon_0]$ and $\hat{f}\in\hat{\mathcal{K}}_{L}$. Then, we have \[\|\hat{\mathcal{L}}_{\epsilon}^{-1}\|_{\mathrm{op},L}\leq\frac{1}{2-4\epsilon}\leq\frac{1}{2-4\epsilon_0}\leq\frac{2}{3}.\]
    For all $\lambda\in D_{\frac{1}{2}}$,
        \[\lambda\hat{\mathbf{I}}-\hat{\mathcal{L}}_{\epsilon}=-(\hat{\mathbf{I}}-\lambda\hat{\mathcal{L}}_{\epsilon}^{-1})\circ\hat{\mathcal{L}}_{\epsilon}.\]
    Since $-\hat{\mathcal{L}}_{\epsilon}=0\hat{\mathbf{I}}-\hat{\mathcal{L}}_{\epsilon}$ is invertible on $\hat{\mathcal{K}}_{L}$ with \[\|\lambda\hat{\mathcal{L}}_{\epsilon}^{-1}\|_{\mathrm{op},L}\leq \frac{{|\lambda|}}{2-4\epsilon}\leq \frac{1}{2(2-4\epsilon_0)}{\leq}\frac{1}{3},\]
   $\hat{\mathbf{I}}-\lambda{\hat{\mathcal{L}}_{\epsilon}^{-1}}$ is invertible on $\hat{\mathcal{K}}_{L}$ with 
   \[\left\|{(\hat{\mathbf{I}}-\lambda\hat{\mathcal{L}}_{\epsilon}^{-1})^{-1}}\right\|_{\mathrm{op},L}\leq \left(1-\frac{1}{3}\right)^{-1}=\frac{3}{2}.\]
   Therefore $\lambda\hat{\mathbf{I}}-\hat{\mathcal{L}}_{\epsilon}$ is invertible on $\hat{\mathcal{K}}_{L}$ with
   \[(\lambda\hat{\mathbf{I}}-\hat{\mathcal{L}}_{\epsilon})^{-1}=-\hat{\mathcal{L}}_{\epsilon}^{-1}\circ (\hat{\mathbf{I}}-\lambda\hat{\mathcal{L}}_{\epsilon}^{-1})^{-1}.\]
   {Since $\epsilon\leq1/8$, we have
   \begin{equation*}
   \Delta(\epsilon,0)
   =(4-\epsilon)L^\dagger NL+(2-4\epsilon)L^\dagger L
   \geq\frac32L^\dagger(N+1)L.
   \end{equation*}} Then for all $\hat{f}\in\hat{\mathcal{K}}_L$, 
   \begin{equation*}
   \begin{aligned}
       &\left\|(N+1)^{\frac{1}{2}}L(\lambda\hat{\mathbf{I}}-\hat{\mathcal{L}_{\epsilon}})^{-1}(\hat{f})L^{\dagger}(N+1)^{\frac{1}{2}}\right\|_1\\&=\frac{2}{3}\left\|\left(\frac{3}{2}N+\frac{3}{2}\right)^{\frac{1}{2}}L(\lambda\hat{\mathbf{I}}-\hat{\mathcal{L}_{\epsilon}})^{-1}(\hat{f})L^{\dagger}\left(\frac{3}{2}N+
    \frac{3}{2}\right)^{\frac{1}{2}}\right\|_1\\&=\frac{2}{3}\left\|\left(\frac{3}{2}N+\frac{3}{2}\right)^{\frac{1}{2}}L\hat{\mathcal{L}}_{\epsilon}^{-1}\left((\hat{\mathbf{I}}-\lambda{\hat{\mathcal{L}}_{\epsilon}^{-1}})^{-1}(\hat{f})\right)L^{\dagger}\left(\frac{3}{2}N+\frac{3}{2}\right)^{\frac{1}{2}}\right\|_1\\
       &\leq\frac{2}{3}\left\|\Delta(\epsilon,0)^{\frac{1}{2}}\hat{\mathcal{L}}_{\epsilon}^{-1}\left((\hat{\mathbf{I}}-\lambda{\hat{\mathcal{L}}_{\epsilon}^{-1}})^{-1}(\hat{f})\right)\Delta(\epsilon,0)^{\frac{1}{2}}\right\|_1\\
       &\leq\frac{2}{3} \|{(\hat{\mathbf{I}}-\lambda\hat{\mathcal{L}}_{\epsilon}^{-1})^{-1}}(\hat{f})\|_{L}\leq \|\hat{f}\|_{L}.
    \end{aligned}
   \end{equation*}
   Then we also have
   \begin{equation*}
   \begin{aligned}
       \|(\lambda\hat{\mathbf{I}}-\hat{\mathcal{L}}_{\epsilon})^{-1}(\hat{f})\|_{L}&=\|\hat{S}(\lambda\hat{\mathbf{I}}-\hat{\mathcal{L}}_{\epsilon})^{-1}(\hat{f})\hat{S}\|_{1}=\|L(\lambda\hat{\mathbf{I}}-\hat{\mathcal{L}}_{\epsilon})^{-1}(\hat{f})L^{\dagger}\|_{1}\\&\leq\left\|(N+1)^{\frac{1}{2}}L(\lambda\hat{\mathbf{I}}-\hat{\mathcal{L}_{\epsilon}})^{-1}(\hat{f})L^{\dagger}(N+1)^{\frac{1}{2}}\right\|_1\leq\|\hat{f}\|_{L}.
    \end{aligned}
   \end{equation*}
   Since {$La=aL$ and}
   \begin{equation*}\begin{aligned}L^{\dagger}(N+1)L&{=} L^{\dagger}NL+L^{\dagger}L\geq \frac{1}{3}L^{\dagger}NL+\frac{1}{3}{Q}N{Q}=\frac{1}{3}{Q}a^{\dagger}(L^{\dagger}L+1)a{Q},\end{aligned}\end{equation*}
we obtain 
\begin{multline*}
    \left\|(L^{\dagger}L+1)^{\frac{1}{2}}a(\lambda\hat{\mathbf{I}}-\hat{\mathcal{L}}_{\epsilon})^{-1}(\hat{f})a^{\dagger}(L^{\dagger}L+1)^{\frac{1}{2}}\right\|_{1}\\\leq3\left\|(N+1)^{\frac{1}{2}}L(\lambda\hat{\mathbf{I}}-\hat{\mathcal{L}}_{\epsilon})^{-1}(\hat{f})L^{\dagger}(N+1)^{\frac{1}{2}}\right\|_1\leq3\|\hat{f}\|_{L}.
\end{multline*}
\end{proof}

\end{lemma}
\subsubsection{Inverses of $\lambda+\hat{V}_{\alpha}$ and $\lambda+\hat{V}_{-\alpha}$}
\begin{lemma}
\label{lemma:bound_1}
    There exist $\alpha_0> 0$, $\epsilon_0>0$ and $C_1,C_2>0$ such that for all $\alpha\geq \alpha_0$, $\epsilon\in[0,\epsilon_0]$ and $\lambda\in D_{\frac{1}{2}}$, $\lambda+\hat{V}_{\alpha}$ and $\lambda+\hat{V}_{-\alpha}$ are invertible on $\hat{\mathcal{H}}$ and we have
    \begin{equation*}
        \|(\lambda+\hat{V}_{\alpha})^{-1}\|_{\infty}\leq \frac{C_1}{\alpha^2},\quad \|(\lambda+\hat{V}_{-\alpha})^{-1}\|_{\infty}\leq \frac{C_1}{\alpha^2},
    \end{equation*}
and
\begin{equation*}
    \|\hat{S}(\lambda+\hat{V}_{\alpha})^{-1}\|_{\infty}\leq \frac{C_2}{\alpha},\quad \|\hat{S}(\lambda+\hat{V}_{-\alpha})^{-1}\|_{\infty}\leq \frac{C_2}{\alpha}.
\end{equation*}
\end{lemma}
\begin{proof}
For $\alpha> 0$ and $\epsilon\geq 0$, {define $\hat{V}$ as the self-adjoint operator on $\hat{\mathcal{H}}$}\[\hat{V}=\frac{1}{2}L^{\dagger}L+\frac{\epsilon}{2}{Q}N{Q}+\frac{\epsilon\alpha^2}{2}.\]
By Lemma~\ref{lemma:gap_1} and Lemma~\ref{lemma:gap_2}, there exists ${\alpha'_0\geq 1}$ such that for all $\alpha\geq\alpha'_0$ and $\epsilon\geq 0$, ${\hat{V}}$ is invertible on $\hat{\mathcal{H}}$ and we have
\begin{equation*}
    {\hat{V}\geq(\alpha^2-1/8)I_{\hat{\mathcal{H}}}\geq\frac{\alpha^2}{2}I_{\hat{\mathcal{H}}},\quad}\hat{V}^{-1}\leq {\frac{2}{\alpha^2}},\quad {Q}N {Q}\leq 3L^{\dagger}L\leq 6\hat{V}.
\end{equation*}
Then we have
\begin{equation*}
\begin{aligned}
\hat{V}^{-1}{Q}a^{\dagger}{Q}a{Q}\hat{V}^{-1}&\leq \hat{V}^{-1}{Q}N{Q}\hat{V}^{-1}\leq 6\hat{V}^{-1}\leq {\frac{12}{\alpha^2}},\\
\hat{V}^{-1}\hat{S}^2\hat{V}^{-1}&=\hat{V}^{-1}L^{\dagger}L\hat{V}^{-1}\leq 2\hat{V}^{-1}\leq {\frac{4}{\alpha^2}}.
    \end{aligned}
\end{equation*}
Therefore, we can deduce that $\|{Q}a {Q}\hat{V}^{-1}\|_{\infty}\leq{\frac{\sqrt{12}}{\alpha}}$ and $\|\hat{S}\hat{V}^{-1}\|{_{\infty}}\leq{\frac{2}{\alpha}}$. Moreover, for all $\lambda\in D_{\frac{1}{2}}$, we have $\|\lambda \hat{V}^{-1}\|_{\infty}\leq{\frac{1}{\alpha^2}}$. Notice that{, on $D(\hat{V})$,} 
\begin{equation*}
    \begin{aligned}
        \lambda+\hat{V}_{\alpha}&=\hat{V}-\epsilon{\alpha} {Q}a{Q}+\lambda=\left({I_{\hat{\mathcal{H}}}}-\epsilon{\alpha} {Q}a{Q}\hat{V}^{-1}+\lambda \hat{V}^{-1}\right)\hat{V}.
    \end{aligned}
\end{equation*}
Once $\epsilon{\alpha}\|{Q}a{Q}\hat{V}^{-1}\|_{\infty}+\|\lambda \hat{V}^{-1}\|_{\infty}<1$, $\lambda+\hat{V}_{\alpha}$ and $\lambda+\hat{V}_{-\alpha}$ are invertible, and 
\begin{equation*}
\begin{aligned}(\lambda+{\hat{V}}_{\alpha})^{-1}&=\hat{V}^{-1}\left(I_{\hat{\mathcal{H}}}-\epsilon{\alpha} {Q}a{Q}\hat{V}^{-1}+\lambda \hat{V}^{-1}\right)^{-1},\\
\hat{S}(\lambda+{\hat{V}}_{\alpha})^{-1}&=\hat{S}\hat{V}^{-1}\left(I_{\hat{\mathcal{H}}}-\epsilon{\alpha} {Q}a{Q}\hat{V}^{-1}+\lambda \hat{V}^{-1}\right)^{-1}.
\end{aligned}
\end{equation*}
{Choose $\alpha_0\geq\max\{\alpha_0',2\}$ and $0<\epsilon_0\leq1/(4\sqrt{12})$. Then} for all $\alpha\geq \alpha_0$, $\epsilon\in[0,\epsilon_0]$ and $\lambda\in D_{\frac{1}{2}}$, 
\[\epsilon{\alpha}\|{Q}a {Q}\hat{V}^{-1}\|_{\infty}+\|\lambda\hat{V}^{-1}\|_{\infty}\leq {\sqrt{12}\epsilon+\frac{2|\lambda|}{\alpha^2}\leq\sqrt{12}\epsilon_0+\frac{1}{\alpha_0^2}}\leq\frac{1}{2}.\]
Consequently,
  \[\left\|\left(I_{\hat{\mathcal{H}}}-\epsilon{\alpha} {Q}a{Q}\hat{V}^{-1}+\lambda \hat{V}^{-1}\right)^{-1}\right\|_{\infty}\leq 2,\]
and
\begin{equation*}
    \begin{aligned}
        \left\|(\lambda+\hat{V}_{\alpha})^{-1}\right\|_{\infty}&=\left\|\hat{V}^{-1}\left(I_{\hat{\mathcal{H}}}-\epsilon{\alpha} {Q}a{Q}\hat{V}^{-1}+\lambda \hat{V}^{-1}\right)^{-1}\right\|_{\infty}\\
        &\leq\left\|\hat{V}^{-1}\right\|_{\infty}\cdot\left\|\left(I_{\hat{\mathcal{H}}}-\epsilon{\alpha} {Q}a{Q}\hat{V}^{-1}+\lambda \hat{V}^{-1}\right)^{-1}\right\|_{\infty}\leq{\frac{4}{\alpha^2}},\\
        \left\|{\hat{S}}(\lambda+\hat{V}_{\alpha})^{-1}\right\|_{\infty}&=\left\|\hat{S}\hat{V}^{-1}\left(I_{\hat{\mathcal{H}}}-\epsilon{\alpha} {Q}a{Q}\hat{V}^{-1}+\lambda \hat{V}^{-1}\right)^{-1}\right\|_{\infty}\\
        &\leq\left\|\hat{S}\hat{V}^{-1}\right\|_{\infty}\cdot\left\|\left(I_{\hat{\mathcal{H}}}-\epsilon{\alpha} {Q}a{Q}\hat{V}^{-1}+\lambda \hat{V}^{-1}\right)^{-1}\right\|_{\infty}\leq{\frac{4}{\alpha}}.
    \end{aligned}
\end{equation*}
{Set $C_1=C_2=4$.}
Then for all $\alpha{\geq}\alpha_0$, $\epsilon\in[0,\epsilon_0]$ and $\lambda\in D_{\frac{1}{2}}$,
\begin{equation*}
    \|(\lambda+\hat{V}_{\alpha})^{-1}\|_\infty\leq \frac{C_1}{\alpha^2},\quad \|\hat{S}(\lambda+\hat{V}_{\alpha})^{-1}\|_{\infty}\leq\frac{C_2}{\alpha}.
\end{equation*}
{The same estimates hold for $\hat{V}_{-\alpha}$, with $-\epsilon\alpha QaQ$ replaced by $+\epsilon\alpha QaQ$, and we can conclude the proof.}
\end{proof}

\begin{prop}[Resolvent of $\tilde{\mathcal{L}}_{\epsilon}$]
\label{prop:resolvent_tilde}
    There exist $\alpha_0> 0$ and $\epsilon_0>0$ such that for all $\alpha\geq \alpha_0$, $\epsilon\in [0,\epsilon_0]$ and $\lambda\in D_{\frac{1}{2}}\setminus\{-2\epsilon\alpha^2,0\}$, $\lambda\hat{\mathbf{I}}-\hat{\mathcal{L}}_{\epsilon}$ is invertible on $\hat{\mathcal{K}}_{L}$, {$\lambda+\hat{V}_{\alpha}$, $\lambda+\hat{V}_{-\alpha}$, $\lambda+\hat{V}^{\dagger}_{\alpha}$ and $\lambda+\hat{V}_{-\alpha}^{\dagger}$ are invertible on $\hat{\mathcal{H}}$} and $\lambda\mathbf{I}-\tilde{\mathcal{L}}_{\epsilon}$ is invertible on $\mathcal{K}_{L}$. For $f\in\mathcal{K}_{L}$, let us note
    \[\hat{\rho}(\lambda)=(\lambda\hat{\mathbf{I}}-\hat{\mathcal{L}_{\epsilon}})^{-1}({Q}f {Q}),\]
    and define
    \begin{equation}
    \label{eq:resolvent_tilde}
        \begin{aligned}
            &\tilde{R}_{\lambda}^{\epsilon}(f)=\hat{\rho}(\lambda)+(\lambda+\hat{V}_{\alpha})^{-1}{Q}(L\hat{\rho}(\lambda)L^{\dagger}+f)P_{\alpha}^{\dagger}
            \\&\phantom{=}+(\lambda+\hat{V}_{-\alpha})^{-1}{Q}(L\hat{\rho}(\lambda)L^{\dagger}+f)P_{-\alpha}^{\dagger}+P_{\alpha}(L\hat{\rho}(\lambda)L^{\dagger}+f){Q}(\lambda+\hat{V}_{\alpha}^{\dagger})^{-1}\\&\phantom{=}+P_{-\alpha}(L\hat{\rho}(\lambda)L^{\dagger}+f){Q}(\lambda+\hat{V}_{-\alpha}^{\dagger})^{-1}\\&\phantom{=}+\frac{1}{\lambda}(P_{\alpha}(L\hat{\rho}(\lambda)L^{\dagger}+f)P_{\alpha}^{\dagger}+P_{-\alpha}(L\hat{\rho}(\lambda)L^{\dagger}+f)P_{-\alpha}^{\dagger})\\
            &\phantom{=}+\frac{1}{\lambda+2\epsilon\alpha^2}(P_{\alpha}(L\hat{\rho}(\lambda)L^{\dagger}+f)P_{-\alpha}^{\dagger}+P_{-\alpha}(L\hat{\rho}(\lambda)L^{\dagger}+f)P_{\alpha}^{\dagger}).
        \end{aligned}
    \end{equation}
    Then we have $(\lambda\mathbf{I}-\tilde{\mathcal{L}}_{\epsilon})^{-1}=\tilde{R}_{\lambda}^{\epsilon}.$
\end{prop}
\begin{proof}
    By Lemma~\ref{lemma:hat_L} and Lemma~\ref{lemma:bound_1}, there exist $\alpha_0>0$ and $\epsilon_0>0$ such that for all $\alpha\geq \alpha_0$, $\epsilon\in[0,\epsilon_0]$ and $\lambda\in D_{\frac{1}{2}}\setminus\{-2\epsilon\alpha^2,0\}$, $(\lambda\hat{\mathbf{I}}-\hat{\mathcal{L}}_{\epsilon})$ is invertible on $\hat{\mathcal{K}}_{L}$ and {$\lambda+\hat{V}_{\alpha}$, $\lambda+\hat{V}_{-\alpha}$, $\lambda+\hat{V}^{\dagger}_{\alpha}$ and $\lambda+\hat{V}_{-\alpha}^{\dagger}$ are invertible on $\hat{\mathcal{H}}$}. Moreover, $\tilde{R}_{\lambda}^{\epsilon}$ is a bounded operator on $\mathcal{K}_L$ (see Lemma~\ref{lemma:boundedness} for a detailed justification). A direct computation shows that $(\lambda\mathbf{I}-\tilde{\mathcal{L}}_{\epsilon})\circ \tilde{R}_{\lambda}^\epsilon=\mathbf{I}$. Now we still need to prove the injectivity of $\lambda\mathbf{I}-\tilde{\mathcal{L}}_{\epsilon}$. Take $\rho\in\mathcal{K}_{L}$ such that $(\lambda\mathbf{I}-\tilde{\mathcal{L}}_{\epsilon})(\rho)=0$. By decomposing $\rho$ into
    \[\hat{\rho}={Q}\rho {Q},\quad 
    \rho_1={Q}\rho {P},\quad \rho_2={P}\rho {Q},\quad \rho_3={P}\rho{P},\]
    and by equations~\eqref{eq:0}, \eqref{eq:1}, \eqref{eq:3} and \eqref{eq:4}, we can obtain that
    \begin{equation*}
        \begin{aligned}
            &(\lambda\mathbf{I}-\hat{\mathcal{L}}_{\epsilon})(\hat{\rho})=0\\
            &(\lambda+\hat{V}_{\alpha})\rho_1 P_{\alpha}^{\dagger}={Q}L\hat{\rho}L^{\dagger}P_{\alpha}^{\dagger},\quad(\lambda+\hat{V}_{-\alpha})\rho_1 P_{-\alpha}^{\dagger}={Q}L\hat{\rho}L^{\dagger}P_{-\alpha}^{\dagger},\\
            &P_{\alpha}\rho_2(\lambda+\hat{V}_{\alpha}^{\dagger})=P_{\alpha}L\hat{\rho}L^{\dagger}Q,\quad P_{-\alpha}\rho_2(\lambda+\hat{V}_{-\alpha}^{\dagger})=P_{-\alpha}L\hat{\rho}L^{\dagger}{Q},\\
            &\lambda(P_{\alpha}\rho_3P_{\alpha}^{\dagger}+P_{-\alpha}\rho_3P_{-\alpha}^{\dagger})+(\lambda+2\epsilon\alpha^2)(P_{\alpha}\rho_3 P_{-\alpha}^{\dagger}+P_{-\alpha}\rho_3P_{\alpha}^{\dagger})={P}L\hat{\rho}L^{\dagger}{P}.
        \end{aligned}
    \end{equation*}
    Since $\lambda\hat{\mathbf{I}}-\hat{\mathcal{L}}_{\epsilon}$ is invertible on $\hat{\mathcal{K}}_{L}$, $\lambda+\hat{V}_{\alpha}$ and $\lambda+\hat{V}_{-\alpha}$ are invertible on $\hat{\mathcal{H}}$, $\lambda\neq 0$ and $\lambda+2\epsilon\alpha^2\neq 0$, it can be easily solved that $\rho=0$. Therefore $\lambda\mathbf{I}-\tilde{\mathcal{L}}_{\epsilon}$ is injective. Hence $\lambda\mathbf{I}-\tilde{\mathcal{L}}_{\epsilon}$ is invertible and its right inverse $\tilde{R}_{\lambda}^{\epsilon}$ coincides with its inverse.
    \end{proof}
\subsection{Perturbed Lindbladian $\mathcal{L}_{\epsilon}$}
Let us introduce the following perturbations
\begin{equation}
    \begin{aligned}
\delta_1(\rho)&=\frac{1}{2}\rho {P}N{Q}+\frac{1}{2}{Q}N{P}\rho,\\
\delta_2(\rho)&=\frac{1}{2}{P}N{Q}\rho+\frac{1}{2}\rho {Q}N{P},\\
\delta_{3}(\rho)&=-{P}a{Q}\rho a^{\dagger}-a\rho {Q}a^{\dagger}{P}+{P}a{Q}\rho {Q}a^{\dagger}{P},\\
\delta_4(\rho)&=\frac{1}{2}{P}(N-\alpha^2){P}\rho+\frac{1}{2}\rho {P}(N-\alpha^2){P}.
\end{aligned}
\end{equation}
Then we have 
\[\tilde{\mathcal{L}}_{\epsilon}=\mathcal{L}_{\epsilon}+\epsilon(\delta_1+\delta_2+\delta_3+\delta_4),\]
and
\[\lambda\mathbf{I}-\mathcal{L}_{\epsilon}=\lambda\mathbf{I}-\tilde{\mathcal{L}}_{\epsilon}+\epsilon(\delta_1+\delta_2+\delta_3+\delta_4).
\]
We need to prove that these perturbations are \textbf{small} with respect to $(\lambda\mathbf{I}-\tilde{\mathcal{L}}_{\epsilon})$ under the operator norm $\|\cdot\|_{\mathrm{op},\alpha}$ for $\lambda$ in a certain sub-domain of $\mathbb{C}$. First, let us introduce some auxiliary lemmas.
\begin{lemma}
\label{lemma:bound_2}
    There exist $\alpha_0>0$ and $C>0$ such that for all $\alpha\geq \alpha_0$, we have
    \[\|P_{\alpha}(N+1)^{-\frac{1}{2}}\|_{\infty}\leq \frac{C}{\alpha},\quad \|P_{-\alpha}(N+1)^{-\frac{1}{2}}\|_{\infty}\leq \frac{C}{\alpha}.\]
\end{lemma}
\begin{proof}
For $\alpha>0$, by the definition of the coherent states, we have
\[\ket{\pm\alpha}=e^{-\frac{\alpha^2}{2}}\sum_{n=0}^{\infty}\frac{(\pm\alpha)^n}{\sqrt{n!}}\ket{n}.\]
Applying $(N+1)^{-\frac{1}{2}}$ to the coherent states, we obtain
\[(N+1)^{-\frac{1}{2}}\ket{\pm\alpha}=e^{-\frac{\alpha^2}{2}}\sum_{n=0}^{\infty}\frac{(\pm\alpha)^n}{\sqrt{(n+1)!}}\ket{n}.\]
We can deduce that
\begin{equation*}
    \begin{aligned}
\|(N+1)^{-1/2}|\pm\alpha\rangle\|^2
=e^{-\alpha^2}\sum_{n=0}^{\infty}
\frac{\alpha^{2n}}{(n+1)!}
=\frac{1-e^{-\alpha^2}}{\alpha^2}
<\frac1{\alpha^2}.
    \end{aligned}
\end{equation*}
Therefore, $\|(N+1)^{-1/2}\ket{\pm\alpha}\|<\frac
{1}{\alpha}$. By the definition of $P_{\alpha}$ and $P_{-\alpha}$ given in Remark~\ref{remark:decomposition}, we conclude that there exist $\alpha_0>0$ and $C>0$ such that for all $\alpha\geq \alpha_0$,
\[\|P_{\alpha}(N+1)^{-\frac{1}{2}}\|_{\infty}\leq \frac{C}{\alpha},\quad \|P_{-\alpha}(N+1)^{-\frac{1}{2}}\|_{\infty}\leq \frac{C}{\alpha}.\]
\end{proof}

\begin{lemma}
\label{lemma:bound_3}
    There exist $\alpha_0> 0$ and $C_1,C_2>0$ such that for all $\alpha\geq \alpha_0$
    \begin{equation*}
        \|{Q}a^{\dagger}{P}\|_{\infty}\leq C_1,\quad \|{Q}N{P}\|_{\infty}\leq C_2\alpha,\quad\|\hat
        S{Q}N{P}\|\leq 2\alpha^2.
    \end{equation*}
\end{lemma}
\begin{proof}
For $\alpha>0$, we have
\[{Q}=L^{\dagger}(LL^{\dagger})^{-1}L,\quad \hat{S}=L^{\dagger}(LL^{\dagger})^{-\frac{1}{2}}L.\]
Moreover,
\begin{equation*}
La^{\dagger}=a^{\dagger}L+2a,
\qquad
LN=(N+2)L+2\alpha^2,
\qquad
L{P}=0.
\end{equation*}
It follows that
    \begin{equation*}
        \begin{aligned}
            {Q}a^{\dagger}{P}&=L^{\dagger}(LL^{\dagger})^{-1}La^{\dagger}{P}=L^{\dagger}(LL^{\dagger})^{-1}(a^{\dagger}L+2a){P}=2L^{\dagger}(LL^{\dagger})^{-1}a{P}\\
            {Q}N{P}&=L^{\dagger}(LL^{\dagger})^{-1}LN{P}=L^{\dagger}(LL^{\dagger})^{-1}((N+2)L+2\alpha^2){P}\\&=2\alpha^2L^{\dagger}(LL^{\dagger})^{-1}{P},\\
            \hat{S}{Q}N{P}&=L^{\dagger}(LL^{\dagger})^{-\frac{1}{2}}LN{P}=L^{\dagger}(LL^{\dagger})^{-\frac{1}{2}}((N+2)L+2\alpha^2){P}
            \\&=2\alpha^2L^{\dagger}(LL^{\dagger})^{-\frac{1}{2}}
{P}.        \end{aligned}
    \end{equation*}
{Since $LL^{\dagger}=L^{\dagger}L+4N+2\geq 2(N+1)$, we have
\[a^{\dagger}(LL^{\dagger})^{-1}a\leq \frac{1}{2}a^{\dagger}(N+1)^{-1}a\leq \frac{1}{2}.\]}
There exist $\alpha_0>0$ and $C_1,C_2>0$ such that, for all $\alpha\geq \alpha_0$,
\begin{equation*}
    \|L^{\dagger}(LL^{\dagger})^{-\frac{1}{2}}\|_\infty\leq 1,\quad \|(LL^{\dagger})^{-\frac{1}{2}}a\|_{\infty}\leq \frac{C_1}{2},\quad \|L^{\dagger}(LL^{\dagger})^{-1}\|_\infty\leq \frac{C_2}{2\alpha}.
\end{equation*}
Consequently,
\begin{equation*}
\begin{aligned}    \|{Q}a^{\dagger}{P}\|_{\infty}&\leq 2\|L^{\dagger}(LL^{\dagger})^{-\frac{1}{2}}\|_{\infty}\cdot\|(LL^{\dagger})^{-\frac{1}{2}}a\|_{\infty}\leq C_1,\\
\|{Q}N{P}\|_\infty&\leq 2\alpha^2\|L^{\dagger}(LL^{\dagger})^{-1}\|_{\infty}\leq C_2 \alpha,\\
\|\hat{S}{Q}N{P}\|_{\infty}&\leq 2\alpha^2\|L^{\dagger}(LL^{\dagger})^{-\frac{1}{2}}\|_\infty\leq2\alpha^2.
\end{aligned}
\end{equation*}
\end{proof}

\begin{lemma}
\label{lemma:bound_4}
    For all $\alpha>0$,
    \begin{equation*}
        \left\|{P}(N-\alpha^2){P}\right\|_{\infty}=\frac{2s}{1-s}\alpha^2,
    \end{equation*}
    where $s=e^{-2\alpha^2}.$
\end{lemma}
\begin{proof}
    Take $\alpha>0$ and set $s=e^{-2\alpha^2}$. Define
    \begin{equation*}
        \ket{+}=\frac{\ket{\alpha}+\ket{-\alpha}}{\sqrt{2+2s}},\quad \ket{-}=\frac{\ket{\alpha}-\ket{-\alpha}}{\sqrt{2-2s}}.
    \end{equation*}
Since $\braket{\alpha}{-\alpha}=s$, the vectors $\ket{+}$ and $\ket{-}$ form an orthonormal basis of ${P}\mathcal{H}$. A direct computation gives
\begin{equation*}
    {P}N {P}=\frac{1-s}{1+s}\alpha^2\ket{+}\bra{+}+\frac{1+s}{1-s}\alpha^2\ket{-}\bra{-}.
\end{equation*}
Hence
\[{P}(N-\alpha^2){P}=-\frac{2s}{1+s}\alpha^2\ket{+}\bra{+}+\frac{2s}{1-s}\alpha^2\ket{-}\bra{-}.\]
Since $\ket{+}$ and $\ket{-}$ are orthonormal and
\begin{equation*}
\frac{2s}{1-s}
\geq
\frac{2s}{1+s},
\end{equation*}
we conclude that
\begin{equation*}
\left\|
{P}(N-\alpha^2){P}
\right\|_{\infty}
=
\frac{2s}{1-s}\alpha^2.
\end{equation*}
\end{proof}

\begin{remark}
\label{remark:bounds}
    By Proposition~\ref{prop:resolvent_tilde}, Lemma~\ref{lemma:hat_L}, Lemma~\ref{lemma:gap_1}, Lemma~\ref{lemma:bound_1}, Lemma~\ref{lemma:bound_2}, Lemma~\ref{lemma:bound_3}, Lemma~\ref{lemma:bound_4} and Remark~\ref{remark:decomposition}, there exist $\alpha_0>0$, $\epsilon_0>0$ and $c_1,c_2,c_3,c_4$, $c_5,c_6,c_7,c_7$ such that for all $\alpha\geq \alpha_0$, $\epsilon\in[0,\epsilon_0]$ and $\lambda\in D_{\frac{1}{2}}\setminus\{-2\epsilon\alpha^2,0\}$, $\lambda\hat{\mathbf{I}}-\hat{\mathcal{L}}_{\epsilon}$ is invertible on $\hat{\mathcal{K}}_{L}$, $\lambda+\hat{V}_{\alpha}$ and $\lambda+\hat{V}_{-\alpha}$ are invertible on $\hat{H}$ and $\lambda\mathbf{I}-\tilde{\mathcal{L}}_{\epsilon}$ is invertible on $\mathcal{K}_{L}$. Furthermore, 
    \begin{equation*}
\begin{aligned}
&\|(N+1)^{\frac{1}{2}}L(\lambda\hat{\mathbf{I}}-\hat{\mathcal{L}}_{\epsilon})^{-1}(\hat{f})L^{\dagger}(N+1)^{\frac{1}{2}}\|_1\leq \|\hat{f}\|_{L},\\
&\|(L^{\dagger}L+1)^{\frac{1}{2}}a(\lambda\hat{\mathbf{I}}-\hat{\mathcal{L}}_{\epsilon})^{-1}(\hat{f})a^{\dagger}(L^{\dagger}L+1)^{\frac{1}{2}}\|_1\leq3\|\hat{f}\|_{L},\\
&\|\hat{S}(\lambda\hat{\mathbf{I}}-\hat{\mathcal{L}}_{
\epsilon
})^{-1}(\hat{f})\hat{S}\|_1=\|(\lambda\hat{\mathbf{I}}-\hat{\mathcal{L}}_{\epsilon})^{-1}(\hat{f})\|_L\leq\|\hat{f}\|_{L},
\end{aligned}
\end{equation*}
for all $\hat{f}\in\hat{\mathcal{K}}_{L}$ and 
\begin{equation*}
        \begin{aligned}
            &\|\hat{S}^{-1}\|_{\infty}\leq \frac{c_1}{\alpha},\quad\|(\lambda+\hat{V}_{\alpha}
            )^{-1}\|_{\infty}\leq\frac{c_2}{\alpha^2},\quad\|(\lambda+\hat{V}_{-\alpha})^{-1}\|_{\infty}\leq \frac{c_2}{\alpha^2},\\&\|\hat{S}(\lambda+\hat{V}_{\alpha})^{-1}\|_{\infty}\leq \frac{c_3}{\alpha},\quad \|\hat{S}(\lambda+\hat{V}_{-\alpha})^{-1}\|_{\infty}\leq \frac{c_3}{\alpha},\\&\|P_{\alpha}(N+1)^{-\frac{1}{2}}\|_{\infty}\leq \frac{c_4}{\alpha},\quad \|P_{-\alpha}(N+1)^{-\frac{1}{2}}\|_{\infty}\leq \frac{c_4}{\alpha},\quad\|{Q}a^{\dagger}{P}\|_{\infty}\leq c_5,\\&\|{Q}N{P}\|_{\infty}\leq c_6\alpha,\quad \|\hat{S}{Q}N{P}\|_{\infty}\le 2\alpha^2,\quad \|P_{\alpha}\|_{\infty}\leq c_7,\quad\|P_{-\alpha}\|_{\infty}\leq c_7. 
        \end{aligned}
    \end{equation*}
\end{remark}
\paragraph{$\alpha$-Norm}
For $\rho\in\mathcal{K}_{L}$, let us define the its $\alpha$-norm $\|\rho\|_{\alpha}$
\begin{equation}
\|\rho\|_{\alpha}=\|\rho\|_{L}+\alpha^2\|{P}\rho {P}\|_1,
\end{equation}
which is equivalent to the norm $\|\cdot\|_{L}$ for each fixed $\alpha$. Let $\|\cdot\|_{\mathrm{op},\alpha}$ denote the operator norm associated with the norm $\|\cdot\|_{\alpha}$.

\begin{remark}
    Under the norm $\|\cdot\|_{\alpha}$, for all $f\in\mathcal{K}_{L}$
    \begin{equation*}
\begin{aligned}\|{Q}f{Q}\|_{\alpha}&=\|{Q}f{Q}\|_{L}=\|\hat{S}{Q}f {Q}\hat{S}\|_1=\|{Q}SfS{Q}\|_1\leq\|f\|_L\leq \|f\|_{\alpha},\\
\|{Q}f{P}\|_{\alpha}&=\|{Q}f{P}\|_{L}=\|\hat{S}{Q}f{P}\|_1=\|{Q}SfS{P}\|_1\leq\|f\|_L\leq \|f\|_{\alpha},\\
\|{P}f{P}\|_{\alpha}&=(1+\alpha^2)\|{P}f{P}\|_{1}= \alpha^2\|{P}f{P}\|_1+\|{P}SfS{P}\|_1\leq\|f\|_{\alpha}.
    \end{aligned}
\end{equation*}
\end{remark}

\begin{lemma}
\label{lemma:boundedness}
    There exist $\alpha_0>0$, $\epsilon_0>0$ and $C_1,C_2>0$ such that for all $\alpha\geq \alpha_0$, $\epsilon\in[0,\epsilon_0]$ and $\lambda\in D_{\frac{1}{2}}\setminus\{-2\epsilon\alpha^2,0\}$, $\mathbf{\lambda I-\tilde{\mathcal{L}}_{\epsilon}}$ is invertible on $\mathcal{K}_{L}$ and
    \[\|(\lambda\mathbf{I}-\tilde{\mathcal{L}}_{\epsilon})^{-1}\|_{\mathrm{op},\alpha}\leq C_1+\frac{C_2}{|\lambda|}+\frac{C_2}{|\lambda+2\epsilon\alpha^2|}.\]
\end{lemma}
\begin{proof}
Take $\alpha_0$, $\epsilon_0$ and $c_1,c_2,c_3,c_4,c_7$ as given in Remark~\ref{remark:bounds}. Then for all $\alpha\geq\alpha_0$, $\epsilon\in[0,\epsilon_0]$ and $\lambda\in D_{\frac{1}{2}}\setminus\{-2\epsilon\alpha^2,0\}$, $\lambda\mathbf{I}-\tilde{\mathcal{L}}_{\epsilon}$ is invertible on $\mathcal{K}_{L}$. Take $f\in\mathcal{K}_{L}$ and define $\hat{\rho}(\lambda)=(\lambda\hat{\mathbf{I}}-\hat{\mathcal{L}}_{\epsilon})^{-1}({Q}f{Q})$. Applying the bounds given in Remark~\ref{remark:bounds}, we obtain
\begin{equation*}
\begin{aligned}
    \|\hat{\rho}(\lambda)\|_{\alpha}=\|\hat{\rho}(\lambda)\|_{L}=\|(\lambda\hat{\mathbf{I}}-\hat{\mathcal{L}}_{\epsilon})^{-1}({Q}f{Q})\|_{L}\leq\|{Q}f{Q}\|_{L}\leq \|f\|_{\alpha},
\end{aligned}
\end{equation*}
and
\begin{equation*}
    \begin{aligned}
        &\|(\lambda+\hat{V}_{\alpha})^{-1}{Q}L\hat{\rho}(\lambda)L^{\dagger}P_{\alpha}^{\dagger}\|_{\alpha}=\|\hat{S}(\lambda+\hat{V}_{\alpha})^{-1}{Q}L\hat{\rho}(\lambda)L^{\dagger}P_{\alpha}^{\dagger}\|_{1}\\
        &=\|\hat{S}(\lambda+\hat{V}_{\alpha})^{-1}{Q}(N+1)^{-\frac{1}{2}}(N+1)^{\frac{1}{2}}L\hat{\rho}(\lambda)L^{\dagger}(N+1)^{\frac{1}{2}}(N+1)^{-\frac{1}{2}}P_{\alpha}^{\dagger}\|_1\\
        &\leq\|\hat{S}(\lambda+\hat{V}_{\alpha})^{-1}\|_{\infty}\cdot\|{Q}(N+1)^{-\frac{1}{2}}\|_{\infty}\cdot\|(N+1)^{-\frac{1}{2}}P_{\alpha}^{\dagger}\|_{\infty}\cdot\|(N+1)^{\frac{1}{2}}L\hat{\rho}(\lambda)L^{\dagger}(N+1)^{\frac{1}{2}}\|_{1}\\
        &\leq\frac{c_3}{\alpha}\cdot 1\cdot\frac{c_4}{\alpha}\cdot\|{Q}f{Q}\|_{L}\leq \frac{c_3c_4}{\alpha^2}\|f\|_{\alpha},
    \end{aligned}
\end{equation*}
and
\begin{equation*}
    \begin{aligned}
        &\|(\lambda+\hat{V}_{\alpha})^{-1}{Q}fP_{\alpha}^{\dagger}\|_{\alpha}=\|\hat{S}(\lambda+\hat{V}_{\alpha})^{-1}{Q}fP_{\alpha}^{\dagger}\|_1\\&=\|\hat{S}(\lambda+\hat{V}_{\alpha})^{-1}\hat{S}^{-1}\hat{S}{Q}f{P}P_{\alpha}^{\dagger}\|_1\\
        &\leq\|\hat{S}(\lambda+\hat{V}_{\alpha})^{-1}\|_{\infty}\cdot\|\hat{S}^{-1}\|_{\infty}\cdot\|P_{\alpha}^{\dagger}\|_{\infty}\cdot\|\hat{S}{Q}f{P}\|_{1}\\
        &\leq \frac{c_3}{\alpha}\cdot\frac{c_1}{\alpha}\cdot c_7\cdot\|{Q}f{P}\|_{\alpha}\leq\frac{c_1c_3c_7}{\alpha^2}\|f\|_{\alpha},
    \end{aligned}
\end{equation*}
and
\begin{equation*}
    \begin{aligned}
        &\|P_{\alpha}L\hat{\rho}(\lambda)L^{\dagger}P_{\alpha}^{\dagger}\|_{\alpha}=(1+\alpha^2)\|P_{\alpha}L\hat{\rho}(\lambda)L^{\dagger}P_{\alpha}^{\dagger}\|_1\\&=(1+\alpha^2)\|P_{\alpha}(N+1)^{-\frac{1}{2}}(N+1)^{\frac{1}{2}}L\hat{\rho}(\lambda)L^{\dagger}(N+1)^{\frac{1}{2}}(N+1)^{-\frac{1}{2}}P_{\alpha}^{\dagger}\|_{1}\\
        &\leq(1+\alpha^2)\|P_{\alpha}(N+1)^{-\frac{1}{2}}\|_{\infty}\cdot\|(N+1)^{-\frac{1}{2}}P_{\alpha}^{\dagger}\|_{\infty}\cdot\|(N+1)^{\frac{1}{2}}L\hat{\rho}(\lambda)L^{\dagger}(N+1)^{\frac{1}{2}}\|_1\\
        &\leq (1+\alpha^2)\cdot\left(\frac{c_4}{\alpha}\right)^2\cdot\|{Q}f{Q}\|_{L}\leq\frac{1+\alpha^2}{\alpha^2}c_4^2\|f\|_{\alpha},
    \end{aligned}
\end{equation*}
and
\begin{equation*}
\begin{aligned}
    &\|P_{\alpha}fP_{\alpha}^{\dagger}\|_{\alpha}=(1+\alpha^2)\|P_{\alpha}fP_{\alpha}^{\dagger}\|_1=(1+\alpha^2)\|P_{\alpha}{P}f{P}P_{\alpha}^{\dagger}\|_1\\&\leq(1+\alpha^2)\|P_{\alpha}\|_{\infty}\cdot\|P_{\alpha}^{\dagger}\|_{\infty}\cdot\|{P}f{P}\|_{1}\leq c_7^2\|{P}f{P}\|_{\alpha}\le c_7^2\|f\|_{\alpha}.
\end{aligned}
\end{equation*}
Similarly, we can obtain for all $f\in\mathcal{K}_{L}$
\begin{equation*}
    \begin{aligned}
       &\|(\lambda+\hat{V}_{-\alpha})^{-1}{Q}(L\hat{\rho}(\lambda)L^{\dagger}+f)P_{-\alpha}^{\dagger}\|_{\alpha}\leq\left(\frac{c_3c_4}{\alpha^2}+\frac{c_1c_3c_7}{\alpha^2}\right)\|f\|_{\alpha},\\
       &\|P_{\alpha}(L\hat{\rho}(\lambda)L^{\dagger}+f){Q}(\lambda+\hat{V}_{\alpha}^{\dagger})^{-1}\|_{\alpha}\leq\left(\frac{c_3c_4}{\alpha^2}+\frac{c_1c_3c_7}{\alpha^2}\right)\|f\|_{\alpha},\\
       &\|P_{-\alpha}(L\hat{\rho}(\lambda)L^{\dagger}+f){Q}(\lambda+\hat{V}_{-\alpha}^{\dagger})^{-1}\|_{\alpha}\leq\left(\frac{c_3c_4}{\alpha^2}+\frac{c_1c_3c_7}{\alpha^2}\right)\|f\|_{\alpha},
    \end{aligned}
\end{equation*}
and
\begin{equation*}
    \begin{aligned}
        &\|P_{-\alpha}(L\hat{\rho}(\lambda)L^{\dagger}+f)P_{-\alpha}^{\dagger}\|_{\alpha}{\leq}\left(\frac{1+\alpha^2}{\alpha^2}c_4^2+c_7^2\right)\|f\|_{\alpha},\\
        &\|P_{\alpha}(L\hat{\rho}(\lambda)L^{\dagger}+f)P_{-\alpha}^{\dagger}\|_{\alpha}{\leq}\left(\frac{1+\alpha^2}{\alpha^2}c_4^2+c_7^2\right)\|f\|_{\alpha},\\
        &\|P_{-\alpha}(L\hat{\rho}(\lambda)L^{\dagger}+f)P_{\alpha}^{\dagger}\|_{\alpha}{\leq}\left(\frac{1+\alpha^2}{\alpha^2}c_4^2+c_7^2\right)\|f\|_{\alpha}.
    \end{aligned}
\end{equation*}
Take $C_1,C_2>0$ such that for all $\alpha\geq\alpha_0$
\begin{equation*}
    \begin{aligned}
        1+4\left(\frac{c_3c_4}{\alpha^2}+\frac{c_1c_3c_7}{\alpha^2}\right)\leq C_1,\quad 2\left(\frac{1+\alpha^2}{\alpha^2}c_4^2+c_7^2\right)\leq C_2.
    \end{aligned}
\end{equation*}
Using equation~\eqref{eq:resolvent_tilde} given in Proposition~\ref{prop:resolvent_tilde}, we conclude that for all $\alpha\geq \alpha_0$, $\epsilon\in[0,\epsilon_0]$ and $\lambda\in D_{\frac{1}{2}}\setminus\{-2\epsilon\alpha^2,0\}$, 
\begin{equation*}
    \|(\lambda\mathbf{I}-\tilde{\mathcal{L}}_{\epsilon})^{-1}\|_{\mathrm{op},\alpha}\leq C_1+\frac{C_2}{|\lambda|}+\frac{C_2}{|\lambda+2\epsilon\alpha^2|}.
\end{equation*}
\end{proof}

\begin{lemma}[relative boundedness of $\delta_1$]
\label{lemma:boundedness1}
    There exist $\alpha_0> 0$, $\epsilon_0>0$ and $C_1,C_2>0$ such that for all $\alpha\geq\alpha_0$, $\epsilon\in[0,\epsilon_0]$ and $\lambda\in D_{\frac{1}{2}}\setminus\{-2\epsilon\alpha^2,0\}$, $\lambda\mathbf{I}-\tilde{\mathcal{L}}_{\epsilon}$ is invertible and
    \[\|\delta_1\circ(\lambda\mathbf{I}-\tilde{\mathcal{L}}_{\epsilon})^{-1}\|_{\mathrm{op},\alpha}\leq \left(C_1+\frac{C_2}{|\lambda|}+\frac{C_2}{|\lambda+2\epsilon\alpha^2|}\right).\]
\end{lemma}
\begin{proof}
Take $\alpha_0$, $\epsilon_0$ and $c_1,c_3,c_4,c_7>0$ as given in Remark~\ref{remark:bounds}. Then, for all $\lambda\in D_{\frac{1}{2}}\setminus\{-2\epsilon\alpha^2,0\}$ and $\epsilon\in[0,\epsilon_0]$, $\alpha\geq \alpha_0$, $\lambda\mathbf{I}-\tilde{\mathcal{L}}_{\epsilon}$ is invertible on $\mathcal{K}_{L}$. Moreover, by equation~\eqref{eq:resolvent_tilde} in Proposition~\ref{prop:resolvent_tilde}, for all $f\in\mathcal{K}_{L}$,
   \begin{equation*}
    \begin{aligned}
        &(\lambda \mathbf{I}-\tilde{\mathcal{L}}_{\epsilon})^{-1}(f){P}N{Q}=(\lambda+\hat{V}_{\alpha})^{-1}{Q}(L\hat{\rho}(\lambda)L^{\dagger}+f)P_{\alpha}^{\dagger}N{Q}\\&+(\lambda+\hat{V}_{-\alpha})^{-1}{Q}(L\hat{\rho}(\lambda)L^{\dagger}+f)P_{-\alpha}^{\dagger}N {Q}+\frac{1}{\lambda}P_{\alpha}(L\hat{\rho}(\lambda)L^{\dagger}+f)P_{\alpha}^{\dagger}N{Q}\\&+\frac{1}{\lambda}P_{-\alpha}(L\hat{\rho}(\lambda)L^{\dagger}+f)P_{-\alpha}^{\dagger}N{Q}+\frac{1}{\lambda+2\epsilon\alpha^2}P_{\alpha}(L\hat{\rho}(\lambda)L^{\dagger}+f)P_{-\alpha}^{\dagger}N{Q}\\&+\frac{1}{\lambda+2\epsilon\alpha^2}P_{-\alpha}(L\hat{\rho}(\lambda)L^{\dagger}+f)P_{\alpha}^{\dagger}N{Q},
    \end{aligned}       
   \end{equation*}
   where $\hat{\rho}(\lambda)=(\lambda\hat{\mathbf{I}}-\hat{\mathcal{L}}_{\epsilon})^{-1}({Q}f {Q})$.
Applying the bounds given in Remark~\ref{remark:bounds}, we obtain
\begin{equation*}
    \begin{aligned}
        \|&(\lambda+\hat{V}_{\alpha})^{-1}{Q}L\hat{\rho}(\lambda)L^{\dagger}P_{\alpha}^{\dagger}N{Q}\|_{\alpha}=\|\hat{S}(\lambda+\hat{V}_{\alpha})^{-1}{Q}L\hat{\rho}(\lambda)L^{\dagger}P_{\alpha}^{\dagger}N{Q}\hat{S}\|_1\\=&\|\hat{S}(\lambda+\hat{V}_{\alpha})^{-1}{Q}(N+1)^{-\frac{1}{2}}(N+1)^{\frac{1}{2}}L\hat{\rho}(\lambda)L^{\dagger}(N+1)^{\frac{1}{2}}(N+1)^{-\frac{1}{2}}P_{\alpha}^{\dagger}{P}N{Q}\hat{S}\|_1 \\
        \leq& \|\hat{S}(\lambda+\hat{V}_{\alpha})^{-1}\|_{\infty}\cdot\|{Q}(N+1)^{-\frac{1}{2}}\|_{\infty}\cdot\|(N+1)^{-\frac{1}{2}}P_{\alpha}^{\dagger}\|_{\infty}\\&\cdot\|{P}N{Q}\hat{S}\|_{\infty}\cdot\|(N+1)^{\frac{1}{2}}L\hat{\rho}(\lambda)L^{\dagger}(N+1)^{\frac{1}{2}}\|_{1}\\
        \leq& \frac{c_3}{\alpha}\cdot    1 \cdot\frac{c_4}{\alpha}\cdot (2\alpha^2)\cdot\|{Q}f{Q}\|_{L}= 2c_3c_4\|{Q}f{Q}\|_{\alpha}\leq 2c_3c_4\|f\|_\alpha,
    \end{aligned}
\end{equation*}
and 
\begin{equation*}
    \begin{aligned}
        &\|(\lambda+\hat{V}_{\alpha})^{-1}{Q}fP_{\alpha}^{\dagger}N{Q}\|_{\alpha}=\|\hat{S}(\lambda+\hat{V}_{\alpha})^{-1}{Q}fP_{\alpha}^{\dagger}N{Q}\hat{S}\|_1\\
        &=\|\hat{S}(\lambda+\hat{V}_{\alpha})^{-1}\hat{S}^{-1}\hat{S}{Q}f{P}P_{\alpha}^{\dagger}{P}N{Q}\hat{S}\|_1\\
        &\leq\|\hat{S}(\lambda+\hat{V}_{\alpha})^{-1}\|_{\infty}\cdot\|\hat{S}^{-1}\|_{\infty}\cdot\|P^{\dagger}_{\alpha}\|_{\infty}\cdot\|{P}N{Q}\hat{S}\|_{\infty}\cdot\|\hat{S}{Q}f{P}\|_1,\\
        &\leq \frac{c_3}{\alpha}\cdot\frac{c_1}{\alpha}\cdot c_7\cdot (2\alpha^2)\cdot\|{Q}f {P}\|_{L}=2c_1c_3c_7\|{Q}f{P}\|_{\alpha}\leq2c_1c_3c_7\|f\|_\alpha,
        \end{aligned}
\end{equation*}
and
{
\begin{equation*}
\begin{aligned}
&\|P_{\alpha}L\hat{\rho}(\lambda)L^{\dagger}P_{\alpha}^{\dagger}NQ\|_\alpha
=\|P_{\alpha}L\hat{\rho}(\lambda)L^{\dagger}P_{\alpha}^{\dagger}NQ\hat S\|_1\\
&=\|P_{\alpha}(N+1)^{-\frac12}(N+1)^{\frac12}L\hat\rho(\lambda)L^\dagger
(N+1)^{\frac12}(N+1)^{-\frac12}P_\alpha^\dagger PNQ\hat S\|_1\\
&\leq\|P_\alpha(N+1)^{-\frac12}\|_\infty
\|(N+1)^{-\frac12}P_\alpha^\dagger\|_\infty\|PNQ\hat S\|_\infty\\
&\phantom{\leq}\cdot\|(N+1)^{\frac12}L\hat\rho(\lambda)L^\dagger(N+1)^{\frac12}\|_1\\
&\leq\left(\frac{c_4}{\alpha}\right)^2(2\alpha^2)\|QfQ\|_L
=2c_4^2\|QfQ\|_\alpha\leq2c_4^2\|f\|_\alpha.
\end{aligned}
\end{equation*}
}
and
\begin{equation*}
    \begin{aligned}
        &\|P_{\alpha}fP_{\alpha}^{\dagger}N{Q}\|_{\alpha}=\|P_{\alpha}fP_{\alpha}^{\dagger}N{Q}\hat{S}\|_1=\|P_{\alpha}{P}f{P}P_{\alpha}^{\dagger}{P}N{Q}\hat{S}\|_1\\&\leq\|P_{\alpha}\|_{\infty}\cdot\|P_{\alpha}^{\dagger}\|_{\infty}\cdot\|{P}N{Q}\hat{S}\|_{\infty}\cdot\|{P}f{P}\|_1\\&\leq c_7^2\cdot(2\alpha^2)\cdot\left(\frac{1}{1+\alpha^2}\cdot\|{P}f{P}\|_{\alpha}\right)\leq\frac{2\alpha^2}{1+\alpha^{{2}}}c_7^2\|f\|_{\alpha}.    \end{aligned}
\end{equation*}
Similarly, we can prove that
\begin{equation*}
    \begin{aligned}
        &\|(\lambda+\hat{V}_{-\alpha})^{-1}{Q}(L\hat{\rho}(\lambda)L^{\dagger}+f)P_{-\alpha}^{\dagger}N {Q}\|_{\alpha}\leq2(c_3c_4+c_1c_3c_7)\|f\|_{\alpha},\\
        &\|P_{-\alpha}(L\hat{\rho}(\lambda)L^{\dagger}+f)P_{-\alpha}^{\dagger}N{Q}\|_{\alpha}\leq \left({2c_4^2}+\frac{2\alpha^2}{1+\alpha^2}c_7^2\right)\|f\|_{\alpha},\\
        &\|P_{\alpha}(L\hat{\rho}(\lambda)L^{\dagger}+f)P_{-\alpha}^{\dagger}N{Q}\|_{\alpha}\leq \left({2c_4^2}+\frac{2\alpha^2}{1+\alpha^2}c_7^2\right)\|f\|_{\alpha},\\
        &\|P_{-\alpha}(L\hat{\rho}(\lambda)L^{\dagger}+f)P_{\alpha}^{\dagger}N{Q}\|_{\alpha}\leq \left({2c_4^2}+\frac{2\alpha^2}{1+\alpha^2}c_7^2\right)\|f\|_{\alpha}.
    \end{aligned}
\end{equation*}
Let us take $C_1=4(c_3c_4+c_1c_3c_7)$. There exist $C_2$ such that
\[2\left({2c_4^2}+\frac{2\alpha^2}{1+\alpha^2}c_7^2\right)\leq C_2,\quad\forall \alpha\geq\alpha_0.\]
We can deduce that for all $\alpha\geq\alpha_0$, $\epsilon\in[0,\epsilon_0]$ and $\lambda\in D_{\frac{1}{2}}\setminus\{-2\epsilon\alpha^2,0\},$
\begin{equation*}
    \|(\lambda\mathbf{I}-\tilde{\mathcal{L}}_{{\epsilon}})^{-1}(f){P}N{Q}\|_{\alpha}\leq \left(C_1+\frac{C_2}{{|\lambda|}}+\frac{C_2}{{|\lambda+2\epsilon\alpha^2|}}\right)\|f\|_{\alpha},
\end{equation*}
{Using the block formula~\eqref{eq:resolvent_tilde} and the same trace-norm product estimates with the right-sided resolvent factors, we also obtain}
\begin{equation*}
    \|{Q}N{P}(\lambda\mathbf{I}-\tilde{\mathcal{L}}_{{\epsilon}})^{-1}(f)\|_{\alpha}\leq \left(C_1+\frac{C_2}{{|\lambda|}}+\frac{C_2}{{|\lambda+2\epsilon\alpha^2|}}\right)\|f\|_{\alpha}.
\end{equation*}
We conclude the proof by observing that for all $f\in\mathcal{K}_L$,
\[\delta_1\circ(\lambda\mathbf{I}-\tilde{\mathcal{L}}_{\epsilon})^{-1}(f)=\frac{1}{2}(\lambda\mathbf{I}-\tilde{\mathcal{L}}_{{\epsilon}})^{-1}(f){P}N{Q}+\frac{1}{2}{Q}N{P}(\lambda\mathbf{I}-\tilde{\mathcal{L}}_{{\epsilon}})^{-1}(f).\]
\end{proof}

\begin{lemma}[relative boundedness of $\delta_2$]
\label{lemma:boundedness2}
    There exist $\alpha_0>0$, $\epsilon_0>0$ and $C>0$ such that for all $\alpha\geq \alpha_0$, $\epsilon\in[0,\epsilon_0]$ and $\lambda\in D_{\frac{1}{2}}\setminus\{-2\epsilon\alpha^2,0\}$, $\lambda\mathbf{I}-\tilde{\mathcal{L}}_{\epsilon}$ is invertible on $\mathcal{K}_{L}$ and
    \[\|\delta_2\circ(\lambda\mathbf{I}-\tilde{\mathcal{L}}_{\epsilon})^{-1}\|_{\mathrm{op},\alpha}\leq C.\]
\end{lemma}
\begin{proof}
Take $\alpha_0$, $\epsilon_0$ and $c_1,c_2,c_4,c_6,c_7$ as given in Remark~\ref{remark:bounds}.  Then for all $\alpha\geq \alpha_0$, $\epsilon\in[0,\epsilon_0]$ and $\lambda\in D_{\frac{1}{2}}{\setminus\{-2\epsilon\alpha^2,0\}}$, $\lambda\mathbf{I}-\tilde{\mathcal{L}}_{\epsilon}$ is invertible on $\mathcal{K}_{L}$. Moreover, by equation~\eqref{eq:resolvent_tilde} in Proposition~\ref{prop:resolvent_tilde}, for all $f\in\mathcal{K}_{L}$,
    \begin{equation*}
        \begin{aligned}
            &{P}N{Q}(\lambda\mathbf{I}-\tilde{\mathcal{L}}_{\epsilon})^{-1}(f)={P}N{Q}\hat{\rho}(\lambda)+{P}N{Q}(\lambda+\hat{V}_{\alpha})^{-1}{Q}(L\hat{\rho}(\lambda)L^{\dagger}+f)P_{\alpha}^{\dagger}\\
            &+{P}N{Q}(\lambda+\hat{V}_{-\alpha})^{-1}{Q}(L\hat{\rho}(\lambda)L^{\dagger}+f)P_{-\alpha}^{\dagger},
        \end{aligned}
    \end{equation*}
 where $\hat{\rho}(\lambda)=(\lambda\hat{\mathbf{I}}-\hat{\mathcal{L}}_{\epsilon})^{-1}({Q}f {Q})$. Applying the bounds given in Remark~\ref{remark:bounds}, we obtain
\begin{equation*}
    \begin{aligned}
&\|{P}N{Q}\hat{\rho}(\lambda)\|_{\alpha}=\|{P}N{Q}\hat{\rho}(\lambda)\hat{S}\|_1=\|{P}N{Q}\hat{S}^{-1}\hat{S}\hat{\rho}(\lambda)\hat{S}\|_1\\
&\leq\|{P}N{Q}\|_{\infty}\cdot\|\hat{S}^{-1}\|_{\infty}\cdot\|\hat{S}\hat{\rho}(\lambda)\hat{S}\|_{1}\leq c_6\alpha\cdot\frac{c_1}{\alpha}\cdot\|{Q}f{Q}\|_{L}\\&=c_1c_6\|{Q}f{Q}\|_{\alpha}\leq c_1c_6\|f\|_{\alpha},
    \end{aligned}
\end{equation*}
and
\begin{equation*}
\begin{aligned}
&\|{P}N{Q}(\lambda+\hat{V}_{\alpha})^{-1}{Q}L\hat{\rho}(\lambda)L^{\dagger}P_{\alpha}^{\dagger}\|_{\alpha}=(1+\alpha^2)\|{P}N{Q}(\lambda+\hat{V}_{\alpha})^{-1}{Q}L\hat{\rho}(\lambda)L^{\dagger}P_{\alpha}^{\dagger}\|_{1}\\
&=(1+\alpha^2)\|{P}N{Q}(\lambda+\hat{V}_{\alpha})^{-1}{Q}(N+1)^{-\frac{1}{2}}(N+1)^{\frac{1}{2}}L\hat{\rho}(\lambda)L^{\dagger}(N+1)^{\frac{1}{2}}(N+1)^{-\frac{1}{2}}P_{\alpha}^{\dagger}\|_1\\
&\leq(1+\alpha^2)\|{P}N{Q}\|_{\infty}\cdot\|(\lambda+\hat{V}_{\alpha})^{-1}\|_{\infty}\cdot\|{Q}(N+1)^{-\frac{1}{2}}\|_{\infty}\\
&\phantom{\leq}\cdot\|(N+1)^{-\frac{1}{2}}P_{\alpha}^{\dagger}\|_{\infty}\cdot\|(N+1)^{\frac{1}{2}}L\hat{\rho}(\lambda)L^{\dagger}(N+1)^{\frac{1}{2}}\|_1\\
&\leq(1+\alpha^2)\cdot c_6\alpha\cdot\frac{c_2}{\alpha^2}\cdot 1\cdot\frac{c_4}{\alpha}\cdot\|{Q}f{Q}\|_{L}=\frac{1+\alpha^2}{\alpha^2}c_2c_4c_6\|{Q}f{Q}\|_{\alpha}\\&\leq \frac{1+\alpha^2}{\alpha^2}c_2c_4c_6\|f\|_{\alpha},
\end{aligned}
\end{equation*}
and
\begin{equation*}
    \begin{aligned}
        &\|{P}N{Q}(\lambda+\hat{V}_{\alpha})^{-1}{Q}fP_{\alpha}^{\dagger}\|_{\alpha}=(1+\alpha^2)\|{P}N{Q}(\lambda+\hat{V}_{\alpha})^{-1}{Q}fP_{\alpha}^{\dagger}\|_{1}\\
        &=(1+\alpha^2)\|{P}N{Q}(\lambda+\hat{V}_{\alpha})^{-1}\hat{S}^{-1}\hat{S}{Q}f{P}P_{\alpha}^{\dagger}\|_1\\
        &\leq (1+\alpha^2)\|{P}N{Q}\|_{\infty}\cdot\|(\lambda+\hat{V}_{\alpha})^{-1}\|_{\infty}\cdot\|\hat{S}^{-1}\|_{\infty}\cdot\|P_{\alpha}^{\dagger}\|_{\infty}\cdot\|\hat{S}{Q}f{P}\|_{1}\\
        &\leq (1+\alpha^2)\cdot c_6\alpha\cdot\frac{c_2}{\alpha^2}\cdot\frac{c_1}{\alpha}\cdot c_7\cdot\|{Q}f{P}\|_\alpha\leq \frac{1+\alpha^2}{\alpha^2}c_1c_2c_6c_7\|f\|_{\alpha}.
    \end{aligned}
\end{equation*}
Similarly, we can prove
\[\|{P}N{Q}(\lambda+\hat{V}_{-\alpha})^{-1}{Q}(L\hat{\rho}(\lambda)L^{\dagger}+f)P_{-\alpha}^{\dagger}\|_{\alpha}\le\left(\frac{1+\alpha^2}{\alpha^2}c_2c_4c_6+\frac{1+\alpha^2}{\alpha^2}c_1c_2c_6c_7\right)\|f\|_{\alpha}.\]
There exist $C$ such that
\[c_1c_6+2\left(\frac{1+\alpha^2}{\alpha^2}c_2c_4c_6+\frac{1+\alpha^2}{\alpha^2}c_1c_2c_6c_7\right)\leq C,\quad\forall \alpha\geq \alpha_0.\]
Therefore, for all $\alpha\geq \alpha_0$, $\epsilon\in[0,\epsilon_0]$, $\lambda\in D_{\frac{1}{2}}{\setminus\{-2\epsilon\alpha^2,0\}}$ and $f\in\mathcal{K}_{L}$,
\[\|{P}N{Q}(\lambda\mathbf{I}-\tilde{\mathcal{L}}_{\epsilon})^{-1}(f)\|_{\alpha}\leq C\|f\|_{\alpha},\]
{Similarly, direct estimation of the block formula with $QNP$ on the right gives}
\[\|(\lambda\mathbf{I}-\tilde{\mathcal{L}}_{\epsilon})^{-1}(f){Q}N{P}\|_{\alpha}\leq C\|f\|_{\alpha}.\]
We conclude the proof by observing that for all $f\in\mathcal{K}_{L}$
\[\delta_2\circ(\lambda\mathbf{I}-\tilde{\mathcal{L}}_{\epsilon})^{-1}(f)=\frac{1}{2}{P}N{Q}(\lambda\mathbf{I}-\tilde{\mathcal{L}}_{\epsilon})^{-1}(f)+\frac{1}{2}(\lambda\mathbf{I}-\tilde{\mathcal{L}}_{\epsilon})^{-1}(f){Q}N{P}.\]
\end{proof}

\begin{lemma}[relative boundedness of $\delta_3$]
\label{lemma:boundedness3}
    There exist $\alpha_0>0$, $\epsilon_0>0$ and $C>0$ such that for all $\alpha\geq \alpha_0$, $\epsilon\in[0,\epsilon_0]$ and $\lambda\in D_{\frac{1}{2}}\setminus\{-2\epsilon\alpha^2,0\}$, $\lambda\mathbf{I}-\tilde{\mathcal{L}}_{\epsilon}$ is invertible on $\mathcal{K}_L$ and
    \[\|\delta_3\circ(\lambda\mathbf{I}-\tilde{\mathcal{L}}_{\epsilon})^{-1}\|_{\mathrm{op},\alpha}\leq C.\]
\end{lemma}
\begin{proof}
Take $\alpha_0$, $\epsilon_0$ and $c_1,c_2,c_4,c_5,c_7$ as given in Remark~\ref{remark:bounds}. Then for all $\alpha\geq \alpha_0$, $\epsilon\in[0,\epsilon_0]$ and $\lambda\in D_{\frac{1}{2}}\setminus\{-2\epsilon\alpha^2,0\}$, $\lambda\mathbf{I}-\tilde{\mathcal{L}_{\epsilon}}$ is invertible. Moreover, by equation~\eqref{eq:resolvent_tilde} in Proposition~\ref{prop:resolvent_tilde}, for all $f\in\mathcal{K}_{L}$,
\begin{equation*}
\begin{aligned}
    &-{P}a{Q}(\lambda\mathbf{I}-\tilde{\mathcal{L}}_{\epsilon})^{-1}(f)a^{\dagger}+\frac{1}{2}{P}a{Q}(\lambda\mathbf{I}-\tilde{\mathcal{L}}_{\epsilon})^{-1}(f){Q}a^{\dagger}{P}\\&=-{P}a{Q}\hat{\rho}(\lambda)a^{\dagger}-{P}a{Q}(\lambda+\hat{V}_{\alpha})^{-1}{Q}(L\hat{\rho}(\lambda)L^{\dagger}+f)P_{\alpha}^{\dagger}a^{\dagger}\\&\phantom{=}-{P}a{Q}(\lambda+\hat{V}_{-\alpha})^{-1}{Q}(L\hat{\rho}(\lambda)L^{\dagger}+f)P_{-\alpha}^{\dagger}a^{\dagger}+\frac{1}{2}{P}a{Q}\hat{\rho}(\lambda){Q}a^{\dagger}{P}\\
    &=-{P}a\hat{\rho}(\lambda)a^{\dagger}{Q}-\frac{1}{2}{P}a{Q}\hat{\rho}(\lambda){Q}a^{\dagger}{P}-\alpha {P}a{Q}(\lambda+\hat{V}_{\alpha})^{-1}{Q}(L\hat{\rho}(\lambda)L^{\dagger}+f)P_{\alpha}^{\dagger}\\
    &\phantom{=}+\alpha {P}a{Q}(\lambda+\hat{V}_{-\alpha})^{-1}{Q}(L\hat{\rho}(\lambda)L^{\dagger}+f)P_{-\alpha}^{\dagger},
\end{aligned}
\end{equation*}
where $\hat{\rho}(\lambda)=(\lambda\hat{\mathbf{I}}-\hat{\mathcal{L}}_{\epsilon})^{-1}({Q}f {Q})$. Applying the bounds given in Remark~\ref{remark:bounds}, we obtain
\begin{equation*}
\begin{aligned}
    &\|{P}a\hat{\rho}(\lambda)a^{\dagger}{Q}\|_{\alpha}=\|{P}a\hat{\rho}(\lambda)a^{\dagger}{Q}\hat{S}\|_{1}\\&=\|{P}(L^{\dagger}L+1)^{\frac{1}{2}}a\hat{\rho}(\lambda)a^{\dagger}(L^{\dagger}L+1)^{\frac{1}{2}}(L^{\dagger}L+1)^{-\frac{1}{2}}(L^{\dagger}L)^{\frac{1}{2}}\|_{1}\\
    &\leq\|{P}\|_{\infty}\cdot\|(L^{\dagger}L+1)^{-\frac{1}{2}}(L^{\dagger}L)^{\frac{1}{2}}\|_{\infty}\cdot\|(L^{\dagger}L+1)^{\frac{1}{2}}a\hat{\rho}(\lambda)a^{\dagger}(L^{\dagger}L+1)^{\frac{1}{2}}\|_1\\
    &\leq 1\cdot 1\cdot{3}\|{Q}f{Q}\|_{L}={3}\|{Q}f{Q}\|_{\alpha}\leq {3}\|f\|_{\alpha},
\end{aligned}
\end{equation*}
and
\begin{equation*}
    \begin{aligned}
        &\|{P}a{Q}\hat{\rho}(\lambda){Q}a^{\dagger}{P}\|_{\alpha}=(1+\alpha^2)\|{P}a {Q}\hat{\rho}(\lambda){Q}a^{\dagger}{P}\|_{1}\\
        &=(1+\alpha^2)\|{P}a{Q}\hat{S}^{-1}\hat{S}\hat{\rho}(\lambda)\hat{S}\hat{S}^{-1}{Q}a^{\dagger}{P}\|_1\\
        &\leq (1+\alpha^2)\|{P}a{Q}\|_{\infty}\cdot\|\hat{S}^{-1}\|^2\cdot\|{Q}a^{\dagger}{P}\|_\infty\cdot\|\hat{S}\hat{\rho}(\lambda)\hat{S}\|_{1}\\
        &\leq (1+\alpha^2)\cdot c_5\cdot\left(\frac{c_1}{\alpha}\right)^2\cdot c_5\cdot\|{Q}f{Q}\|_{L}=\frac{1+\alpha^2}{\alpha^2}c_1^2c_5^2\|{Q}f{Q}\|_{\alpha}\leq \frac{1+\alpha^2}{\alpha^2}c_1^2c_5^2\|f\|_{\alpha},
    \end{aligned}
\end{equation*}
and
\begin{equation*}
    \begin{aligned}
        &\|{P}a {Q}(\lambda+\hat{V}_{\alpha})^{-1}{Q}L\hat{\rho}(\lambda)L^{\dagger}P_{\alpha}^{\dagger}\|_{\alpha}=(1+\alpha^2)\|{P}a{Q}(\lambda+\hat{V}_{\alpha})^{-1}{Q}L\hat{\rho}(\lambda)L^{\dagger}P_{\alpha}^{\dagger}\|_{1}\\
        &=(1+\alpha^2)\|{P}a {Q}(\lambda+\hat{V}_{\alpha})^{-1}{Q}(N+1)^{-\frac{1}{2}}(N+1)^{\frac{1}{2}}L\hat{\rho}(\lambda)L^{\dagger}(N+1)^{\frac{1}{2}}(N+1)^{-\frac{1}{2}}P_{\alpha}^{\dagger}\|_1\\
        &\leq(1+\alpha^2)\|{P}a{Q}\|_{\infty}\cdot\|(\lambda+\hat{V}_{\alpha})^{-1}\|_{\infty}\cdot\|{Q}(N+1)^{-\frac{1}{2}}\|_{\infty}\\
        &\phantom{\leq}\cdot\|(N+1)^{-\frac{1}{2}}P_{\alpha}^{\dagger}\|_{\infty}\cdot\|(N+1)^{\frac{1}{2}}L\hat{\rho}(\lambda)L^{\dagger}(N+1)^{\frac{1}{2}}\|_{1}\\
        &\leq (1+\alpha^2)\cdot c_5\cdot\frac{c_2}{\alpha^2}\cdot 1\cdot \frac{c_4}{\alpha}\|{Q}f{Q}\|_{\alpha}\\&=\frac{1+\alpha^2}{\alpha^3}c_2c_4c_5\|{Q}f{Q}\|_{\alpha}\leq\frac{1+\alpha^2}{\alpha^3}c_2c_4c_5\|f\|_{\alpha},
    \end{aligned}
\end{equation*}
and
\begin{equation*}
\begin{aligned}
    &\|{P}a{Q}(\lambda+\hat{V}_{\alpha})^{-1}{Q}fP_{\alpha}^{\dagger}\|_{\alpha}=(1+\alpha^2)\|{P}a{Q}(\lambda+\hat{V}_{\alpha})^{-1}{Q}fP_{\alpha}^{\dagger}\|_{1}\\
    &=(1+\alpha^2)\|{P}a {Q}(\lambda+\hat{V}_{\alpha})^{-1}\hat{S}^{-1}\hat{S}{Q}f{P}P_{\alpha}^{\dagger}\|_1\\
    &\leq (1+\alpha^2)\|{P}a{Q}\|_{\infty}\cdot\|(\lambda+\hat{V}_{\alpha})^{-1}\|_{\infty}\cdot\|\hat{S}^{-1}\|\cdot\|P_{\alpha}^{\dagger}\|\cdot\|\hat{S}{Q}f{P}\|_1\\
    &\leq (1+\alpha^2)\cdot c_5\cdot\frac{c_2}{\alpha^2}\cdot\frac{c_1}{\alpha}\cdot c_7\cdot\|{Q}f{P}\|_{\alpha}\\&=\frac{1+\alpha^2}{\alpha^3}c_1c_2c_5c_7\|{Q}f{P}\|_{\alpha}\leq \frac{1+\alpha^2}{\alpha^3}c_1c_2c_5c_7\|f\|_{\alpha}
\end{aligned}
\end{equation*}
{Using $P_{\pm\alpha}^{\dagger}a^{\dagger}=\pm\alpha P_{\pm\alpha}^{\dagger}$,} similarly, we can prove
\[\|{P}a{Q}(\lambda+\hat{V}_{-\alpha})^{-1}{Q}L\hat{\rho}(\lambda)L^{\dagger}P_{-\alpha}^{\dagger}a^{\dagger}\|_{\alpha}\le\frac{1+\alpha^2}{\alpha^2}c_2c_4c_5\|f\|_{\alpha},\]
and \[\|{P}a{Q}(\lambda+\hat{V}_{-\alpha})^{-1}{Q}fP_{-\alpha}^{\dagger}a^{\dagger}\|_{\alpha}\leq\frac{1+\alpha^2}{\alpha^2}c_1c_2c_5c_7\|f\|_{\alpha}.\]
Take $C>0$ such that 
\[{6}+\frac{1+\alpha^2}{\alpha^2}c_1^2c_5^2+{4}\left(\frac{1+\alpha^2}{\alpha^2}c_2c_4c_5+\frac{1+\alpha^2}{\alpha^2}c_1c_2c_5c_7\right)\leq C,\quad \forall\alpha\geq \alpha_0.\]
Then for all $\alpha\geq \alpha_0$, $\epsilon\in[0,\epsilon_0]$, $\lambda\in D_{\frac{1}{2}}\setminus\{-2\epsilon\alpha^2,0\}$ and $f\in\mathcal{K}_{L}$, 
\[\|-{P}a{Q}(\lambda\mathbf{I}-\tilde{\mathcal{L}}_{\epsilon})^{-1}(f)a^{\dagger}+\frac{1}{2}{P}a{Q}(\lambda\mathbf{I}-\tilde{\mathcal{L}}_{\epsilon})^{-1}(f){Q}a^{\dagger}{P}\|_{\alpha}\leq\frac{C}{2}\|f\|_{\alpha},\]
{Estimating the second contribution directly with the same trace-norm product bounds and the right-sided resolvent estimates gives}
\[\|-a(\lambda\mathbf{I}-\tilde{\mathcal{L}}_{\epsilon})^{-1}(f){Q}a^{\dagger}{P}+\frac{1}{2}{P}a{Q}(\lambda\mathbf{I}-\tilde{\mathcal{L}}_{\epsilon})^{-1}(f){Q}a^{\dagger}{P}\|_{\alpha}\leq \frac{C}{2}\|f\|_{\alpha}.\]
We can conclude the proof by observing that for all $f\in\mathcal{K}_{L}$
\begin{equation*}
    \begin{aligned}
    &\delta_3\circ(\lambda\mathbf{I}-\tilde{\mathcal{L}}_{\epsilon})^{-1}(f)=-{P}a{Q}(\lambda\mathbf{I}-\tilde{\mathcal{L}}_{\epsilon})^{-1}(f)a^{\dagger}-a(\lambda\mathbf{I}-\tilde{\mathcal{L}}_{\epsilon})^{-1}(f){Q}a^{\dagger}{P}\\
    &\phantom{=}+{P}a{Q}(\lambda\mathbf{I}-\tilde{\mathcal{L}}_{\epsilon})^{-1}(f){Q}a^{\dagger}{P}.
    \end{aligned}
\end{equation*}
\end{proof}

\begin{lemma}[relative boundedness of $\delta_4$]
\label{lemma:boundedness4}
    There exist $\alpha_0> 0$, $\epsilon_0>0$ and $C_1,C_2>0$ such that for all $\alpha\geq\alpha_0$, $\epsilon\in[0,\epsilon_0]$ and $\lambda\in D_{\frac{1}{2}}\setminus\{-2\epsilon\alpha^2,0\}$, $\lambda\mathbf{I}-\tilde{\mathcal{L}}_{\epsilon}$ is invertible on $\mathcal{K}_{L}$ and
    \[\|\delta_4\circ(\lambda\mathbf{I}-\tilde{\mathcal{L}}_{\epsilon})^{-1}\|_{\mathrm{op},\alpha}\leq \left(C_1+\frac{C_2}{|\lambda|}+\frac{C_2}{|\lambda+2\epsilon\alpha^2|}\right).\]
\end{lemma}
\begin{proof}
    By Lemma~\ref{lemma:boundedness}, there exist $\alpha_0>0$, $\epsilon_0{>0}$ and $c_1,c_2>0$ such that for all $\alpha\geq \alpha_0$, $\epsilon\in[0,\epsilon_0]$ and $\lambda\in D_{\frac{1}{2}}\setminus\{-2\epsilon\alpha^2,0\}$, $\lambda\mathbf{I}-\tilde{\mathcal{L}}_{\epsilon}$ is invertible on $\mathcal{K}_{L}$ and for all $f\in\mathcal{K}_{L}$
    \[\|(\lambda\mathbf{I}-\tilde{\mathcal{L}}_{\epsilon})^{-1}(f)\|_{\alpha}\leq \left(c_1+\frac{c_2}{|\lambda|}+\frac{c_2}{|\lambda+2\epsilon\alpha^2|}\right)\|f\|_{\alpha}.\]
    By Lemma~\ref{lemma:bound_4}, there exists $c_0>0$ such that
    \[\|{P}(N-\alpha^2){P}\|_{\infty}\leq c_0,\quad\forall \alpha\geq 
    \alpha_0.\]
    Then for all $f\in\mathcal{K}_{L}$, we have
    \begin{equation*}
    \begin{aligned}
        &\|{P}(\alpha^2-N){P}(\lambda\mathbf{I}-\tilde{\mathcal{L}}_{\epsilon})^{-1}(f)\|_{\alpha}\\&=\|{P}(\alpha^2-N){P}(\lambda\mathbf{I}-\tilde{\mathcal{L}}_{\epsilon})^{-1}(f)S\|_1+\alpha^2\|{P}(\alpha^2-N){P}(\lambda\mathbf{I}-\tilde{\mathcal{L}}_{\epsilon})^{-1}(f){P}\|_{1}\\
        &\leq \|{P}(\alpha^2-N){P}\|_{\infty}\cdot(\|{P}(\lambda\mathbf{I}-\tilde{\mathcal{L}}_{\epsilon})^{-1}(f)S\|_{1}+\alpha^2\|{P}(\lambda\mathbf{I}-\tilde{\mathcal{L}}_{\epsilon})^{-1}(f){P}\|_1)\\
        &\leq\|{P}(\alpha^2-N){P}\|_{\infty}\cdot\|(\lambda\mathbf{I}-\tilde{\mathcal{L}}_{\epsilon})^{-1}(f)\|_\alpha\leq c_0\left(c_1+\frac{c_2}{|\lambda|}+\frac{c_2}{|\lambda+2\epsilon\alpha^2|}\right)\|f\|_{\alpha}{.}
    \end{aligned}
    \end{equation*}
    {
For right multiplication, $SP=PS=P$ gives, for all $\rho\in\mathcal K_L$,
\[
\|\rho P(\alpha^2-N)P\|_\alpha
\leq\|P(\alpha^2-N)P\|_\infty\|\rho\|_\alpha
\leq c_0\|\rho\|_\alpha.
\]
Applying this to $\rho=(\lambda\mathbf I-\tilde{\mathcal L}_\epsilon)^{-1}(f)$ and averaging the two estimates gives the bound for $\delta_4$.
\par}
    We can conclude the proof by taking $C_1=c_0c_1$ and $C_2=c_0c_2$.
\end{proof}

\begin{proof}[Proof of Theorem~\ref{theorem:main_1}]
Set $\delta=\delta_1+\delta_2+\delta_3+\delta_4$. By Lemma~\ref{lemma:boundedness1}, Lemma~\ref{lemma:boundedness2}, Lemma~\ref{lemma:boundedness3} and Lemma~\ref{lemma:boundedness4}, there exist $\alpha_0>0$, $\epsilon'_0>0$ and $C'_1,C'_2>0$ such that for all $\alpha\geq \alpha_0$, $\epsilon\in[0,\epsilon'_0]$ and $\lambda\in D_{\frac{1}{2}}\setminus\{-2\epsilon\alpha^2,0\}$, $\lambda\mathbf{I}-\tilde{\mathcal{L}}_{\epsilon}$ is invertible on $\mathcal{K}_{L}$ and 
\begin{equation*}
    \|\delta\circ(\lambda\mathbf{I}-\tilde{\mathcal{L}}_{\epsilon})^{-1}\|_{\mathrm{op},\alpha}\leq C_1'+\frac{C'_2}{|\lambda|}+\frac{C_2'}{|\lambda+2\epsilon\alpha^2|}.
\end{equation*}
Take 
\[\epsilon_0=\min\left(\epsilon_0',\frac{1}{4C'_1}\right),\quad r_1=4C'_2,\quad r_2=4C'_2.\]
Then for all $\alpha\ge\alpha_0$, $\epsilon\in(0,\epsilon_0]$ and $\lambda\in D_{\frac{1}{2}}\setminus\left(\bar{B}(0,\epsilon r_1)\cup \bar{B}(-2\epsilon\alpha^2,\epsilon r_2)\right)$, we have
\begin{equation*}
    \begin{aligned}
        &\|\epsilon\delta\circ(\lambda\mathbf{I}-\tilde{\mathcal{L}}_{\epsilon})^{-1}\|_{\mathrm{op},\alpha}\leq \epsilon C_1'+\frac{\epsilon C'_2}{|\lambda|}+\frac{\epsilon C'_2}{|\lambda+2\epsilon\alpha^2|}< \epsilon C_1'
+\frac{\epsilon C_2'}{\epsilon r_1}+\frac{\epsilon C_2'}{\epsilon r_2}\leq \frac{3}{4}.
\end{aligned}
\end{equation*}
{The maximal weak realizations of \(\mathcal L_\epsilon\) and \(\tilde{\mathcal L}_\epsilon\) have the same domain. Indeed, the relative resolvent estimates imply
\[
\delta\bigl(D(\tilde{\mathcal L}_\epsilon)\bigr)\subset\mathcal K_L.
\]
Hence\(D(\tilde{\mathcal L}_\epsilon)\subset D(\mathcal L_\epsilon)\). Conversely, take \(\rho\in D(\mathcal L_\epsilon)\), and set \(f=\mathcal L_\epsilon(\rho)\) and \(\hat\rho=Q\rho Q\). Since \(\hat S QNP\) and its adjoint are bounded,
\[
QNP\rho Q+Q\rho PNQ\in\hat{\mathcal K}_L.
\]
Testing the equation against vectors in \(Q\mathcal D_\infty\), we obtain \(\hat\rho\in D(\hat{\mathcal L}_\epsilon)\) and
\[
\hat{\mathcal L}_\epsilon(\hat\rho)
=
QfQ+\frac{\epsilon}{2}
\bigl(QNP\rho Q+Q\rho PNQ\bigr)=h
\in\hat{\mathcal K}_L.
\]
Thus \(\hat\rho=\hat{\mathcal L}_\epsilon^{-1}(h)\), and Lemma~\ref{lemma:hat_L}, together with \(S^2\le L^\dagger L+I_{\mathcal H}\), gives
\[
a\hat\rho a^\dagger\in\mathcal K_L.
\]
The maps \(\delta_1,\delta_2,\delta_4\) are bounded on \(\mathcal K_L\) for each fixed \(\alpha\). Moreover,
\[
P\delta_3(\rho)Q=-Pa\hat\rho a^\dagger Q,
\qquad
Q\delta_3(\rho)P=-Qa\hat\rho a^\dagger P,
\]
while \(Q\delta_3(\rho)Q=0\) and \(P\delta_3(\rho)P\) acts on the finite-dimensional space \(P\mathcal H\). Hence \(\delta_3(\rho)\in\mathcal K_L\), and consequently \(\delta(\rho)\in\mathcal K_L\). Using our definition of maximal realization, we obtain \(\rho\in D(\tilde{\mathcal L}_\epsilon)\), with
\[
\tilde{\mathcal L}_\epsilon(\rho)=f+\epsilon\delta(\rho).
\]
This proves the reverse inclusion and therefore the equality of the domains.}
Moreover
$\mathbf{I}+\epsilon\delta(\lambda-\tilde{\mathcal{L}}_{\epsilon})^{-1}$ is invertible on $\mathcal{K}_{L}$. Since 
\begin{equation*}
    \begin{aligned}   
\lambda\mathbf{I}-\mathcal{L}_{\epsilon}&=\lambda\mathbf{I}-\tilde{\mathcal{L}}_{\epsilon}+\epsilon\delta=\left(\mathbf{I}+\epsilon\delta\circ(\lambda\mathbf{I}-\tilde{\mathcal{L}}_{\epsilon})^{-1}\right)\circ(\lambda\mathbf{I}-\tilde{\mathcal{L}}_{\epsilon}),
    \end{aligned}
\end{equation*}
$\lambda\mathbb{\mathbf{I}-\mathcal{L}}_{\epsilon}$ is invertible on $\mathcal{K}_{L}$ and
\[(\lambda\mathbf{I}-\mathcal{L}_{\epsilon})^{-1}=(\lambda\mathbf{I}-\tilde{\mathcal{L}}_{\epsilon})^{-1}\circ\left(\mathbf{I}+\epsilon\delta\circ(\lambda\mathbf{I}-\tilde{\mathcal{L}}_{\epsilon})^{-1}\right)^{-1}.\]
 We can conclude that for all $\alpha\geq \alpha_0$ and $\epsilon\in(0,\epsilon_0]$, $D_{\frac{1}{2}}\setminus\left(\bar{B}(0,\epsilon r_1)\cup\bar{B}(-2\epsilon\alpha^2,\epsilon r_2)\right)$ is contained in the resolvent set of $\mathcal{L}_{\epsilon}$.

\end{proof}
\begin{comment}
% REVIEW-BEGIN 34
\begin{reviewnote}{34}{Justification: common domains in the perturbation step}
The numerical Neumann estimate is consistent once the earlier constants have been enlarged. The operator factorization also requires that $\mathcal L_\epsilon$ and $\tilde{\mathcal L}_\epsilon$ have the compatible common domain used in the perturbation argument, and that $\delta$ maps that domain into $\mathcal K_L$. A bound for $\delta(\lambda\mathbf I-\tilde{\mathcal L}_\epsilon)^{-1}$ supplies a route to a closed perturbed realization; identify this realization with the original generator. This closes the domain issue raised at the start of Section 2 rather than treating the formal identity alone as a proof of invertibility.
\end{reviewnote}
% REVIEW-END 34
\end{comment}
\begin{prop}
\label{prop:projection}
There exist $\alpha_0>0$, $\epsilon_0>0$ and $r>0$ such that for all $\alpha\geq \alpha_0$ and $\epsilon\in(0,\epsilon_0]$, $\Gamma_{\epsilon r}$ is contained in the resolvent sets of both $\mathcal{L}_{\epsilon}$ and $\tilde{\mathcal{L}}_{\epsilon}$ as operators on $\mathcal{K}_{L}$ and the contour $\Gamma_{\epsilon r}$ separates the two excluded disks in \eqref{eq:theorem_1}. Then the spectral projections
    \begin{equation*}
    \mathcal{P}_{\epsilon,\alpha}=\frac{1}{2\pi i}\oint_{\Gamma_{\epsilon r}}(\xi\mathbf{I}-\mathcal{L}_{\epsilon})^{-1}\mathrm{d}\xi,\quad \tilde{\mathcal{P}}_{\epsilon,\alpha}=\frac{1}{2\pi i}\oint_{\Gamma_{\epsilon r}}(\xi\mathbf{I}-\tilde{\mathcal{L}}_{\epsilon})^{-1}\mathrm{d}\xi
\end{equation*}
    are well defined and
    \[\left\|\mathcal{P}_{\epsilon,\alpha}-\tilde{\mathcal{P}}_{\epsilon,\alpha}\right\|_{\mathrm{op},\alpha}\leq \frac{1}{2}.\]
\end{prop}
\begin{proof}
Set $\delta=\delta_1+\delta_2+\delta_3+\delta_4$. By Lemma~\ref{lemma:boundedness}, Lemma~\ref{lemma:boundedness1}, Lemma~\ref{lemma:boundedness2}, Lemma~\ref{lemma:boundedness3} and Lemma~\ref{lemma:boundedness4}, there exist $\alpha'_0>0$, $\epsilon'_0>0$ and $C'_1,C'_2,C_1'',C_2''>0$ such that for all $\alpha\geq \alpha'_0$, $\epsilon\in[0,\epsilon'_0]$ and $\lambda\in D_{\frac{1}{2}}\setminus\{-2\epsilon\alpha^2,0\}$, $\lambda\mathbf{I}-\tilde{\mathcal{L}}_{\epsilon}$ is invertible on $\mathcal{K}_{L}$ and 
\begin{equation*}
\begin{aligned}
&\|(\lambda\mathbf{I}-\tilde{\mathcal{L}}_{\epsilon})^{-1}\|_{\mathrm{op},\alpha}\le C_1'+\frac{C_2'}{|\lambda|}+\frac{C_2'}{|\lambda+2\epsilon\alpha^2|},\\
    &\|\delta\circ(\lambda\mathbf{I}-\tilde{\mathcal{L}}_{\epsilon})^{-1}\|_{\mathrm{op},\alpha}\leq C_1''+\frac{C''_2}{|\lambda|}+\frac{C_2''}{|\lambda+2\epsilon\alpha^2|}.
\end{aligned}
\end{equation*}
By Theorem~\ref{theorem:main_1}, there exist $\alpha_0''>0$, $\epsilon_0''>0$ and $r_1,r_2>0$ such that for all $\alpha\geq \alpha_0''$ and $\epsilon\in(0,\epsilon_0'']$, \begin{equation}
    \label{eq:resolvent_region_recalled}
        D_{\frac{1}{2}}\setminus\left(\bar{B}(0,\epsilon r_1)\cup \bar{B}(-2\epsilon\alpha^2,\epsilon r_2)\right)
    \end{equation}
    is contained in the resolvent set of $\mathcal{L}_{\epsilon}$ as an operator on $\mathcal{K}_{L}$.
Now let us take
\begin{equation}
\begin{aligned}
\label{eq:parameters}
    r&=\max(8C_2'',64C_2'C_2'',2r_1,2r_2),\quad\alpha_0=\max(\sqrt{r},\alpha'_0,\alpha_0''),\\
    \epsilon_0&=\min\left(\frac{1}{4C_1''},\frac{1}{2r},\frac{1}{8C_1'C_1''},\frac{1}{32C'_1C_2''},\frac{1}{32C_2'C_1''},\epsilon_0',\epsilon_0''\right).
\end{aligned}
\end{equation}
Take $\alpha\geq \alpha_0$, $\epsilon\in(0,\epsilon_0]$ and $\lambda\in\Gamma_{\epsilon r}$. {By the choices in~\eqref{eq:parameters},
\[
|\lambda|=\epsilon r\leq\frac12,\qquad |\lambda|=\epsilon r>\epsilon r_1,
\]
and
\[
|\lambda+2\epsilon\alpha^2|
\geq2\epsilon\alpha^2-|\lambda|
\geq2\epsilon\alpha_0^2-\epsilon r
\geq\epsilon r>\epsilon r_2.
\]
Thus $\Gamma_{\epsilon r}\subset D_{1/2}$ encloses the disk centered at $0$
and leaves the disk centered at $-2\epsilon\alpha^2$ outside.
It therefore separates the two excluded disks in~\eqref{eq:theorem_1}.
\par} 
Notice that
\[\lambda\mathbf{I}-\mathcal{L}_{\epsilon}=\lambda\mathbf{I}-\tilde{\mathcal{L}}_{\epsilon}+\epsilon\delta=\left(\mathbf{I}+\epsilon\delta\circ(\lambda\mathbf{I}-\tilde{\mathcal{L}}_{\epsilon})^{-1}\right)\circ(\lambda\mathbf{
I
}-\tilde{\mathcal{L}}_{\epsilon}).\]
Using equation~\eqref{eq:parameters}, we obtain
\begin{equation}
\label{eq:projection_1}
    \begin{aligned}
        \|(\lambda\mathbf{I}-\tilde{\mathcal{L}}_{\epsilon})^{-1}\|_{\mathrm{op},\alpha}&\leq C_1'+\frac{C_2'}{|\lambda|}+\frac{C_2'}{|2\epsilon\alpha^2+\lambda|}\leq C_1'+\frac{2C_2'}{\epsilon r},\\\|\epsilon\delta\circ(\lambda\mathbf{I}-\tilde{\mathcal{L}}_{\epsilon})^{-1}\|_{\mathrm{op},\alpha}&\leq \epsilon C_1''+\frac{\epsilon C_2''}{|\lambda|}+\frac{\epsilon C_2''}{|2\epsilon\alpha^2+\lambda|}\leq \epsilon C_1''+2\frac{C_2''}{r}\leq \frac{1}{2}.
    \end{aligned}
\end{equation}
We can then deduce that 
$\mathbf{I}+\epsilon\delta\circ(\lambda-\tilde{\mathcal{L}}_{\epsilon})^{-1}$ is invertible on $\mathcal{K}_{L}$. Hence $\lambda\mathbf{I}-\mathcal{L}_{\epsilon}$ is invertible on $\mathcal{K}_{L}$ and
\[(\lambda\mathbf{I}-\mathcal{L}_{\epsilon})^{-1}=(\lambda\mathbf{I}-\tilde{\mathcal{L}}_{\epsilon})^{-1}\circ\left(\mathbf{I}+\epsilon\delta\circ(\lambda\mathbf{I}-\tilde{\mathcal{L}}_{\epsilon})^{-1}\right)^{-1}.\]
Consequently
\begin{equation*}
    (\lambda\mathbf{I}-\mathcal{L}_{\epsilon})^{-1}-(\lambda\mathbf{I}-\tilde{\mathcal{L}}_{\epsilon})^{-1}=(\lambda\mathbf{I}-\tilde{\mathcal{L}}_{\epsilon})^{-1}\left(\left(\mathbf{I}+\epsilon\delta\circ(\lambda\mathbf{I}-\tilde{\mathcal{L}_{\epsilon}})^{-1}\right)^{-1}-\mathbf{I}\right).
\end{equation*}
Using equation~\eqref{eq:projection_1} and then equation~\ref{eq:parameters}, we obtain
\begin{equation*}
\begin{aligned}
    \left\|\left(\mathbf{I}+\epsilon\delta\circ(\lambda\mathbf{I}-\tilde{\mathcal{L}}_{\epsilon})^{-1}\right)^{-1}-\mathbf{I}\right\|_{\mathrm{op},\alpha}&\leq \frac{\left\|\epsilon\delta\circ(\lambda\mathbf{I}-\tilde{\mathcal{L}}_{\epsilon})^{-1}\right\|_{\mathrm{op},\alpha}}{1-\left\|\epsilon\delta\circ(\lambda\mathbf{I}-\tilde{\mathcal{L}}_{\epsilon})^{-1}\right\|_{\mathrm{op},\alpha}}\\&\leq 2\left\|\epsilon\delta\circ(\lambda\mathbf{I}-\tilde{\mathcal{L}}_{\epsilon})^{-1}\right\|_{\mathrm{op},\alpha}\leq 2\epsilon C_1''+4\frac{C_2''}{r},
\end{aligned}
\end{equation*}
and
\begin{equation*}
\begin{aligned}
    &\epsilon r\left\|(\lambda\mathbf{I}-\mathcal{L}_{\epsilon})~^{-1}-(\lambda\mathbf{I}-\tilde{\mathcal{L}}_{\epsilon})^{-1}\right\|_{\mathrm{op},\alpha}\\&\leq \epsilon r\left\|(\lambda\mathbf{I}-\tilde{\mathcal{L}}_{\epsilon})^{-1}\right\|_{\mathrm{op},\alpha}\cdot \left\|\left(\mathbf{I}+\epsilon\delta\circ(\lambda-\tilde{\mathcal{L}}_{\epsilon})^{-1}\right)^{-1}-\mathbf{I}\right\|_{\mathrm{op},\alpha}\\
    &\leq \left(\epsilon rC_1'+2C_2'\right)\cdot\left(2\epsilon C_1''+4\frac{C_2''}{r}\right)=2\epsilon^2 rC_1'C_1''+4\epsilon C_2'C_1''
+4\epsilon C_1'C_2''+8\frac{C_2'C_2''}{r}\\
&\leq \frac{1}{2r}\cdot\frac{1}{8C_1'C_1''}\cdot(2rC_1'C_1'')+\frac{1}{32C_2'C_1''}\cdot(4C_2'C_1'')\\&\phantom{=}+\frac{1}{32C'_1C''_2}\cdot(4C_1'C_2'')+\frac{1}{64C_2'C_2''}\cdot(8C_2'C_2'')=\frac{1}{2}.
\end{aligned}
\end{equation*}
Then for all $\alpha\geq \alpha_0$, $\epsilon\in(0,\epsilon_0]$ and $\xi\in \Gamma_{\epsilon r}$, we have
\[\left\|\left(\xi\mathbf{I}-\mathcal{L}_{\epsilon}\right)^{-1}-\left(\xi\mathbf{I}-\tilde{\mathcal{L}}_{\epsilon}\right)^{-1}\right\|_{\mathrm{op},\alpha}\leq\frac{1}{2\epsilon r},\]
\begin{comment}
We easily verify that $\tilde{\mathcal{P}}_{\epsilon,\alpha}$ is the spectral projection of $\tilde{\mathcal{L}}_{\epsilon}$ to the eigenvalue $0$ and
\[\mathrm{Im}\left(\tilde{\mathcal{P}}_{\epsilon,\alpha}\right)=\mathrm{span}\left(\{\ket{\alpha}\bra{\alpha},\ket{-\alpha}\bra{-\alpha}\}\right)\]
\end{comment}
Finally, we conclude that
\begin{equation*}
\begin{aligned}
    &\left\|\mathcal{P}_{\epsilon,\alpha}-\tilde{\mathcal{P}}_{\epsilon,\alpha}\right\|_{\mathrm{op},\alpha}\le\left\|\frac{1}{2\pi i}\oint_{\Gamma_{\epsilon r}}\left(\left(\xi\mathbf{I}-\mathcal{L}_{\epsilon}\right)^{-1}-\left(\xi\mathbf{I}-\tilde{\mathcal{L}}_{\epsilon}\right)^{-1}\right)\text{d}\xi\right\|_{\mathrm{op},\alpha}\\&\leq\frac{1}{2\pi}\oint_{\Gamma_{\epsilon r}}\left\|\left(\xi\mathbf{I}-\mathcal{L}_{\epsilon}\right)^{-1}-\left(\xi\mathbf{I}-\tilde{\mathcal{L}}_{\epsilon}\right)^{-1}\right\|_{\mathrm{op},\alpha}|\text{d}\xi|\leq \frac{1}{2\pi}\oint_{\Gamma_{\epsilon r}}\frac{1}{2\epsilon r}|\text{d}\xi|=\frac{1}{2}.
\end{aligned}
\end{equation*}
\end{proof}

\begin{proof}[Proof of Theorem~\ref{theorem:main_2}]
Take $\alpha_0$, $\epsilon_0$ and $r$ as given in Proposition~\ref{prop:projection}. Then for all $\alpha\geq\alpha_0$, $\epsilon\in(0,\epsilon_0]$, the two spectral projections
\begin{equation*}
    \mathcal{P}_{\epsilon,\alpha}=\frac{1}{2\pi i}\oint_{\Gamma_{\epsilon r}}(\xi\mathbf{I}-\mathcal{L}_{\epsilon})^{-1}\mathrm{d}\xi,\quad \tilde{\mathcal{P}}_{\epsilon,\alpha}=\frac{1}{2\pi i}\oint_{\Gamma_{\epsilon r}}(\xi\mathbf{I}-\tilde{\mathcal{L}}_{\epsilon})^{-1}\mathrm{d}\xi,
\end{equation*}
are well defined and 
    \[\left\|\mathcal{P}_{\epsilon,\alpha}-\tilde{\mathcal{P}}_{\epsilon,\alpha}\right\|_{\mathrm{op},\alpha}\leq \frac{1}{2}.\]
Moreover, the contour $\Gamma_{\epsilon r}$ separates the two excluded disks in \eqref{eq:theorem_1}. By Proposition~\ref{prop:resolvent_tilde}, $\tilde{\mathcal{P}}_{\epsilon,\alpha}$ is the spectral projection of
$\tilde{\mathcal{L}}_{\epsilon}$ associated with the eigenvalue $0$, and
\[\mathrm{Im}\left(\tilde{\mathcal{P}}_{\epsilon,\alpha}\right)=\mathrm{span}\left(\{\ket{\alpha}\bra{\alpha},\ket{-\alpha}\bra{-\alpha}\}\right).\]
Moreover, since $\mathcal{P}_{\epsilon,\alpha}$ and $\tilde{\mathcal{P}}_{\epsilon,\alpha}$ are both projections and
\[\left\|\mathcal{P}_{\epsilon,\alpha}-\tilde{\mathcal{P}}_{\epsilon,\alpha}\right\|_{\mathrm{op},\alpha}\leq \frac{1}{2},\]
their ranges are isomorphic (See \cite{kato1995}).Therefore, \[\mathrm{Rank}(\mathcal{P}_{\epsilon,\alpha})=\mathrm{Rank}(\tilde{\mathcal{P}}_{\epsilon,\alpha})=2.\]
\end{proof}

\begin{remark}
\label{remark:inverse_L_epsilon}
    Take $\alpha_0$, $\epsilon_0$ and $r>0$ as given in Theorem~\ref{theorem:main_2}. We can further prove that there exists $C>0$ such that for all $\alpha\geq \alpha_0$, $\epsilon\in(0,\epsilon_0]$, $\Gamma_{\epsilon r}$ is contained in the resolvent set of $\mathcal{L}_{\epsilon}$ as an operator on $\mathcal{K}_{L}$ and 
    \[\|(\lambda\mathbf{I}-\mathcal{L}_{\epsilon})^{-1}\|_{\mathrm{op},\alpha}\leq \frac{C}{\epsilon},\quad\forall\lambda\in\Gamma_{\epsilon r}.\]
\end{remark}

\section{Eigenvectors of $\mathcal{L}_{\epsilon}$}
Take $\bar{\alpha}>0$, $\bar{\epsilon}>0$ and $\bar{r}>0$ as given in Theorem~\ref{theorem:main_2} and Remark~\ref{remark:inverse_L_epsilon}. For all $\alpha\geq\bar{\alpha}$ and $\epsilon\in(0,\bar{\epsilon}]$, the spectral projection $\mathcal{P}_{\epsilon,\alpha}$ given in~\eqref{eq:spectral_projection} is well defined and has rank $2$. Let us define
\begin{equation*}
\begin{aligned}
\rho_{\epsilon,\alpha}^{+}&=\mathcal{P}_{\epsilon,\alpha}(\ket{\alpha}\bra{\alpha}),
&\rho_{\epsilon,\alpha}^{-}&=\mathcal{P}_{\epsilon,\alpha}(\ket{-\alpha}\bra{-\alpha}),\\
f_{\epsilon,\alpha}^{+}&=\mathcal{L}_{\epsilon}(\rho_{\epsilon,\alpha}^{+}),
&f_{\epsilon,\alpha}^{-}&=\mathcal{L}_{\epsilon}(\rho_{\epsilon,\alpha}^{-}).
\end{aligned}
\end{equation*}

\subsection{Main result}
\begin{theorem}
\label{theorem:main_3}
There exist $\alpha_0\geq\bar{\alpha}$, $\epsilon_0\in(0,\bar{\epsilon}]$ and $C>0$ such that for all $\alpha\geq\alpha_0$ and $\epsilon\in(0,\epsilon_0]$, $\rho_{\epsilon,\alpha}^{+}$ and $\rho_{\epsilon,\alpha}^{-}$ are well defined and linearly independent. Moreover,
\[
\frac12\rho_{\epsilon,\alpha}^{+}+\frac12\rho_{\epsilon,\alpha}^{-},
\qquad
\frac12\rho_{\epsilon,\alpha}^{+}-\frac12\rho_{\epsilon,\alpha}^{-}
\]
are eigenvectors of $\mathcal{L}_{\epsilon}$ associated with the eigenvalues $0$ and $-\lambda$, respectively, where
\[
0\leq\lambda\leq C\alpha^6\exp\left(-\frac{\log2}{4}\alpha^{2}\right).
\]
\end{theorem}
In order to prove Theorem~\ref{theorem:main_3}, we write $\rho_{\epsilon,\alpha}^{+}$ and $\rho_{\epsilon,\alpha}^{-}$ in the corresponding shifted Fock bases. We then use a finite-dimensional truncation and an exponential weight to estimate the truncation error. 
\begin{comment}
Throughout this section, constants denoted by $C$ are independent of $\alpha$ and $\epsilon$ in the stated parameter ranges and may change from one occurrence to the next.
\end{comment}
\subsection{Shifted Lindbladian and finite-dimensional truncation}
For $\alpha\in\mathbb{R}$, its associated displacement operator is
\[
D(\alpha)=\exp(\alpha a^{\dagger}-\alpha a).
\]
Define the conjugation map
\[
\mathcal{U}_{\alpha}(\rho)=D(-\alpha)\rho D(\alpha).
\]
The shifted Lindbladian and its spectral projection are
\begin{equation*}
\begin{aligned}
\mathcal{L}_{\epsilon}^{s}
&=\mathcal{U}_{\alpha}\circ\mathcal{L}_{\epsilon}\circ\mathcal{U}_{\alpha}^{-1}
=D[a^2+2\alpha a]+\epsilon D[a+\alpha],\\
\mathcal{P}_{\epsilon,\alpha}^{s}
&=\mathcal{U}_{\alpha}\circ\mathcal{P}_{\epsilon,\alpha}\circ\mathcal{U}_{\alpha}^{-1}
=\frac{1}{2\pi i}\oint_{\Gamma_{\epsilon\bar r}}
(\lambda\mathbf{I}-\mathcal{L}_{\epsilon}^{s})^{-1}\,\mathrm{d}\lambda.
\end{aligned}
\end{equation*}
To retain the norms of Sections~1 and~2, set
\[
L^{s}=a^2+2\alpha a,\qquad
P^{s}=D(-\alpha){P}D(\alpha).
\]
\[
S^{s}=D(-\alpha)SD(\alpha)
=\bigl((L^{s})^{\dagger}L^{s}+P^{s}\bigr)^{\frac12}.
\]
Here $P^{s}$ is the orthogonal projection onto
$\mathrm{span}\{\ket0,\ket{-2\alpha}\}$. On
$\mathcal{K}_{L}^{s}=\mathcal{U}_{\alpha}(\mathcal{K}_{L})$, define
\[
\|\rho\|_{L,s}=\|S^{s}\rho S^{s}\|_1,\qquad
\|\rho\|_{\alpha,s}=\|\rho\|_{L,s}
+\alpha^2\|P^{s}\rho P^{s}\|_1.
\]
These are exactly the transported norms $\|\cdot\|_L$ and $\|\cdot\|_{\alpha}$. In particular, Lemma~\ref{lemma:gap_1} implies $S^{s}\geq I_{\mathcal H}$, and hence
\[
\|\rho\|_1\leq\|\rho\|_{L,s}\leq\|\rho\|_{\alpha,s}.
\]
Writing $R_{\lambda}^{s}=(\lambda\mathbf{I}-\mathcal{L}_{\epsilon}^{s})^{-1}$, Remark~\ref{remark:inverse_L_epsilon} gives
\begin{equation}
\label{eq:shifted_contour_bound}
\|R_{\lambda}^{s}\|_{\mathrm{op},\alpha,s}\leq\frac{C_0}{\epsilon},
\qquad \lambda\in\Gamma_{\epsilon\bar r}.
\end{equation}
{Define the space of finite Fock matrices:
\[\mathcal{K}_f=\textbf{span}_{\mathbb{C}}\left\{\ket{n}\bra{m}\mid n,m\in\mathbb{N}\right\}.\]}
For $M\geq2$, let $P_M$ be the orthogonal projection onto
\[
\mathcal H_M=\mathrm{span}\{\ket0,\ldots,\ket M\}.
\]
Thus $M$ is the highest retained Fock level and $\dim\mathcal H_M=M+1$. Set
\[
a_M=P_MaP_M,\qquad I_M=I_{\mathcal H_M},\qquad
\mathcal P_M(\rho)=P_M\rho P_M,
\]
and denote by $\mathbf I_M$ the identity superoperator on
$\mathcal P_M\mathcal K_f$. Define
\begin{equation*}
\begin{aligned}
\mathcal L_{0,M}&=D[a_M^2+2\alpha a_M]+\epsilon D[a_M],\\
\mathcal L_{\epsilon,M}&=D[a_M^2+2\alpha a_M]+\epsilon D[a_M+\alpha I_M].
\end{aligned}
\end{equation*}
Both operators generate completely positive, trace-preserving semigroups on $\mathcal P_M\mathcal K_f$. Finally, set
\[
X_M=P_M2^NP_M,\qquad F_M=X_M-P_0,\qquad Q_M=P_M-P_0,
\]
and define
\[
\|\rho\|_{X_M}=\|X_M^{\frac12}\rho X_M^{\frac12}\|_1.
\]
All inverse powers of $X_M$ below are taken on $\mathcal H_M$ and extended by zero on $\mathcal H_M^{\perp}$.

\subsection{Exponential weight and truncated resolvent}
\begin{lemma}
\label{lemma:truncated_lyapunov}
For all $\alpha\geq2$, $\epsilon\geq0$ and
\[
M=\left\lfloor\frac{\alpha^2}{2}+1\right\rfloor,
\qquad c=\frac{\alpha^2}{3}+\frac{\epsilon}{2},
\]
we have
\[
\mathcal L_{0,M}^{*}(F_M)\leq-cF_M.
\]
\end{lemma}
\begin{proof}
Let $\mathcal L_0^{s}=D[L^{s}]+\epsilon D[a]$. For a diagonal observable $h(N)$, a direct computation on finite Fock vectors gives
\begin{equation*}
\begin{aligned}
(\mathcal L_0^{s})^{*}(h(N))
&=N(N-1)\bigl(h(N-2)-h(N)\bigr)\\
&\quad +(4\alpha^2+\epsilon)N\bigl(h(N-1)-h(N)\bigr)\\
&\quad +\alpha a^{\dagger}N\bigl(2h(N-1)-h(N)-h(N+1)\bigr)\\
&\quad +\alpha N\bigl(2h(N-1)-h(N)-h(N+1)\bigr)a.
\end{aligned}
\end{equation*}
In particular,
\begin{equation*}
\begin{aligned}
(\mathcal L_0^{s})^{*}(2^N)
&=-\frac34N(N-1)2^N-\frac12(4\alpha^2+\epsilon)N2^N\\
&\quad-2\alpha\bigl(a^{\dagger}N2^N+N2^Na\bigr)\\
&=-\frac32\left(a+\frac{4\alpha}{3}\right)^{\dagger}
N2^N\left(a+\frac{4\alpha}{3}\right)
+\frac12\left(\frac{4\alpha^2}{3}-\epsilon\right)N2^N.
\end{aligned}
\end{equation*}
Since $L^{s}$ and $a$ leave $\mathcal H_M$ invariant,
\[
\mathcal L_{0,M}^{*}(X_M)=P_M(\mathcal L_0^{s})^{*}(2^N)P_M.
\]
Writing $V_M=P_MN2^NP_M$ and $A_M=\mathcal L_{0,M}^{*}(X_M)$, we obtain
\begin{equation*}
A_M=-\frac32\left(a_M+\frac{4\alpha}{3}I_M\right)^{\dagger}
V_M\left(a_M+\frac{4\alpha}{3}I_M\right)
+\frac12\left(\frac{4\alpha^2}{3}-\epsilon\right)V_M.
\end{equation*}
Since $A_M\ket0=0$, it suffices to work on
$\mathcal H'_M=Q_M\mathcal H_M$. On this subspace, define
\[
B_M=V_M^{\frac12}a_MV_M^{-\frac12}.
\]
Then $B_M\ket1=0$ and $B_M\ket n=\sqrt{(n-1)/2}\ket{n-1}$ for $2\leq n\leq M$, so
\[
\|B_M\|_{\infty}=\sqrt{\frac{M-1}{2}}\leq\frac{\alpha}{2}.
\]
Consequently, on $\mathcal H'_M$,
\begin{equation*}
\begin{aligned}
A_M
&=V_M^{\frac12}\left(
-\frac32\left(B_M+\frac{4\alpha}{3}I_{\mathcal H'_M}\right)^{\dagger}
\left(B_M+\frac{4\alpha}{3}I_{\mathcal H'_M}\right)
+\frac12\left(\frac{4\alpha^2}{3}-\epsilon\right)I_{\mathcal H'_M}
\right)V_M^{\frac12}\\
&\leq\left(-\frac32\left(\frac{4\alpha}{3}-\sqrt{\frac{M-1}{2}}\right)^2
+\frac12\left(\frac{4\alpha^2}{3}-\epsilon\right)\right)V_M\\
&\leq-\left(\frac{3\alpha^2}{8}+\frac{\epsilon}{2}\right)V_M
\leq-cF_M.
\end{aligned}
\end{equation*}
Moreover,
\[
\mathcal L_{0,M}^{*}(P_0)
=(\sqrt2\ket2+2\alpha\ket1)(\sqrt2\bra2+2\alpha\bra1)
+\epsilon\ket1\bra1\geq0.
\]
Therefore $\mathcal L_{0,M}^{*}(F_M)\leq A_M\leq-cF_M$.
\end{proof}

\begin{lemma}
\label{lemma:truncated_resolvent_0}
Under the assumptions of Lemma~\ref{lemma:truncated_lyapunov}, let
$T_M(t)=e^{t\mathcal L_{0,M}}$ and $\Pi_0(\rho)=\Tr(\rho)P_0$. Then, for all $t\geq0$,
\begin{equation}
\label{eq:truncated_decay}
\|T_M(t)\circ(\mathbf I_M-\Pi_0)\|_{\mathrm{op},X_M}
\leq2e^{-ct}+4e^{-ct/2}.
\end{equation}
For all $\lambda\neq0$ with $\mathrm{Re}\lambda>-c/2$, the operator
$\lambda\mathbf I_M-\mathcal L_{0,M}$ is invertible, and its inverse $R^0_{\lambda,M}$ satisfies
\begin{equation}
\label{eq:truncated_resolvent_bound_0}
\|R^0_{\lambda,M}\circ(\mathbf I_M-\Pi_0)\|_{\mathrm{op},X_M}
\leq\frac{2}{\mathrm{Re}\lambda+c}+\frac{8}{2\mathrm{Re}\lambda+c}.
\end{equation}
\end{lemma}
\begin{proof}
Take a positive $\rho\in\mathcal P_M\mathcal K_f$ and define
\[
p(t)=\Tr(F_MT_M(t)(\rho)).
\]
By Lemma~\ref{lemma:truncated_lyapunov},
\[
\dot p(t)=\Tr(\mathcal L_{0,M}^{*}(F_M)T_M(t)(\rho))\leq-cp(t),
\]
and hence $p(t)\leq e^{-ct}\|\rho\|_{X_M}$. In particular,
\[
\|Q_MT_M(t)(\rho)Q_M\|_{X_M}=p(t).
\]
Set $s(t)=\Tr(Q_MT_M(t)(\rho)Q_M)$. Since $Q_M\leq F_M$, we have
$0\leq s(t)\leq p(t)$. Trace preservation gives
\[
P_0T_M(t)(\rho)P_0=(\Tr(\rho)-s(t))P_0.
\]
Applying the Cauchy--Schwarz inequality to the blocks of
\[
X_M^{\frac12}T_M(t)(\rho)X_M^{\frac12}\geq0,
\]
we obtain
\begin{equation*}
\begin{aligned}
\|Q_MT_M(t)(\rho)P_0\|_{X_M}
&\leq\|Q_MT_M(t)(\rho)Q_M\|_{X_M}^{\frac12}
\|P_0T_M(t)(\rho)P_0\|_{X_M}^{\frac12}\\
&\leq e^{-ct/2}\|\rho\|_{X_M}.
\end{aligned}
\end{equation*}
The same estimate holds for the other off-diagonal block. {To extend these estimates to arbitrary $\rho$, we decompose   the real and imaginary parts of $X_{M}^{\frac{1}{2}}\rho X_{M}^{\frac{1}{2}}$ into positive and negative parts:
\[X_{M}^{\frac{1}{2}}\rho X_{M}^{\frac{1}{2}}=x_1-x_2+ix_3-ix_4,\quad x_{j}\geq 0,\quad \sum_{j=1}^4\|x_{j}\|_1\leq 2\|\rho\|_{X_{M}}.\]
Then we obtain a decomposition of $\rho$ into positive elements of $\mathcal{P}_{M}\mathcal{K}_{f}$:
\[\rho=\rho_1-\rho_2+i\rho_3-i\rho_4,\quad \rho_{j}=X_{M}^{-\frac{1}{2}}x_{j}X_{M}^{-\frac{1}{2}}\geq 0,\quad\sum_{j=1}^{4}\|\rho_j\|_{X_{M}}\leq 2\|\rho\|_{X_{M}}.\]}
This gives a factor $2$ for each off-diagonal block. Then by complete positivity, for all $\rho\in\mathcal{P}_{M}\mathcal{K}_{f}$,
\[
\begin{aligned}
&\|Q_MT_M(t)(\rho)Q_M\|_{X_M}\leq e^{-ct}\|\rho\|_{X_M},\\
&|s(t)|\leq\|Q_MT_M(t)(\rho)Q_M\|_{1}\leq e^{-ct}\|\rho\|_{X_M}\leq e^{-ct}\|\rho\|_{X_M}.
\end{aligned}
\]
Since $T_M(t)\circ\Pi_0=\Pi_0$, adding the four blocks proves~\eqref{eq:truncated_decay}.
{Then for all $\lambda\in \mathbb{C}\setminus\{0\}$ such that $\text{Re}\lambda> -\frac{c}{2}$, $\lambda\mathbf{I}_{M}-\mathcal{L}_{0,M}$ is invertible on $\mathcal{P}_{M}\mathcal{K}_f$ and} 
\[
R^0_{\lambda,M}
=\frac1\lambda\Pi_0+
\int_0^{\infty}e^{-\lambda t}T_M(t)\circ(\mathbf I_M-\Pi_0)\,\mathrm{d}t.
\]
Integrating~\eqref{eq:truncated_decay} proves~\eqref{eq:truncated_resolvent_bound_0}.
\end{proof}

We have $\mathcal L_{\epsilon,M}=\mathcal L_{0,M}+\epsilon\Delta_M$, where
\[
\Delta_M(\rho)=\frac{\alpha}{2}(a_M-a_M^{\dagger})\rho
+\frac{\alpha}{2}\rho(a_M^{\dagger}-a_M).
\]
Since
\[
\|X_M^{\frac12}a_MX_M^{-\frac12}\|_{\infty}=\sqrt{\frac M2},
\qquad
\|X_M^{\frac12}a_M^{\dagger}X_M^{-\frac12}\|_{\infty}=\sqrt{2M},
\]
we obtain
\begin{equation}
\label{eq:truncated_perturbation_bound}
\|\Delta_M\|_{\mathrm{op},X_M}
\leq\alpha\left(\sqrt{\frac M2}+\sqrt{2M}\right)
\leq\alpha\left(\sqrt{\frac{\alpha^2}{4}+1}+\sqrt{\alpha^2+2}\right).
\end{equation}
Moreover, $\Tr(\Delta_M(\rho))=0$, so
$\Delta_M=(\mathbf I_M-\Pi_0)\circ\Delta_M$.

\begin{lemma}
\label{lemma:truncated_resolvent}
There exist $\alpha_0\geq\max(2,\bar\alpha)$,
$\epsilon_0\in(0,\min(1,\bar\epsilon)]$ and $C>0$ such that, for all
$\alpha\geq\alpha_0$, $\epsilon\in(0,\epsilon_0]$ and
$\lambda\in\Gamma_{\epsilon\bar r}$, the inverse
$R_{\lambda,M}=(\lambda\mathbf I_M-\mathcal L_{\epsilon,M})^{-1}$ exists and
\begin{equation}
\label{eq:truncated_resolvent_bound}
\|R_{\lambda,M}\circ(\mathbf I_M-\Pi_0)\|_{\mathrm{op},X_M}
\leq\frac{4}{c-\epsilon\bar r}+\frac{16}{c-2\epsilon\bar r}\leq \frac{C}{\alpha^2}.
\end{equation}
\end{lemma}
\begin{proof}
By~\eqref{eq:truncated_resolvent_bound_0} and~\eqref{eq:truncated_perturbation_bound}, we may enlarge $\alpha_0$ and decrease $\epsilon_0$ so that
\[
\epsilon\bar r<c/4,\qquad
\epsilon\|R^0_{\lambda,M}\circ\Delta_M\|_{\mathrm{op},X_M}\leq\frac12
\]
for all $0<|\lambda|\leq\epsilon\bar r$. Thus
\[
R_{\lambda,M}
=\bigl(\mathbf I_M-\epsilon R^0_{\lambda,M}\circ\Delta_M\bigr)^{-1}
\circ R^0_{\lambda,M},
\]
and the inverse factor has norm at most $2$. Therefore,
\[\|R_{\lambda,M}\circ(\mathbf I_M-\Pi_0)\|_{\mathrm{op},X_M}
\leq\frac{4}{c-\epsilon\bar r}+\frac{16}{c-2\epsilon\bar r},\]
and there exists $C>0$ satisfying equation~\eqref{eq:truncated_resolvent_bound}.
\end{proof}

\subsection{Truncation error and localization}
Define
\[
\rho_{\epsilon,\alpha}^{s}=\mathcal P_{\epsilon,\alpha}^{s}(P_0),
\qquad
f_{\epsilon,\alpha}^{s}=\mathcal L_{\epsilon}^{s}(\rho_{\epsilon,\alpha}^{s}),
\qquad
d=\mathcal L_{\epsilon}^{s}(P_0)
=-\frac{\epsilon\alpha}{2}(\ket1\bra0+\ket0\bra1).
\]
Notice that
\[
\|d\|_{X_M}=\sqrt2\epsilon\alpha.
\]
For each $\lambda\in\Gamma_{\epsilon\bar{r}}$, applying the resolvent $R_{\lambda}^{s}$ to $\lambda P_0=\lambda P_0-\mathcal{L}_{\epsilon}^{s}(P_0)+d$, we obtain the following resolvent identity:
\[P_0+R_{\lambda}^{s}(d)=\lambda R_{\lambda}^{s}(P_0)\]
Moreover, since $\mathcal P_{\epsilon,\alpha}^{s}$ commutes with its generator,
\begin{equation}
\label{eq:shifted_contour_vectors}
\begin{aligned}
\rho_{\epsilon,\alpha}^{s}
&=P_0+\frac{1}{2\pi i}\oint_{\Gamma_{\epsilon\bar r}}
\frac1\lambda R_{\lambda}^{s}(d)\,\mathrm{d}\lambda,\\
f_{\epsilon,\alpha}^{s}
&=\frac{1}{2\pi i}\oint_{\Gamma_{\epsilon\bar r}}
R_{\lambda}^{s}(d)\,\mathrm{d}\lambda.
\end{aligned}
\end{equation}
For a finite Fock matrix $\rho$ supported on $\mathcal H_M$, set
\[
\delta_M(\rho)=\mathcal L_{\epsilon}^{s}(\rho)-\mathcal L_{\epsilon,M}(\rho).
\]
A direct computation gives
\begin{equation}
\label{eq:boundary_defect}
\delta_M(\rho)
=-\frac{\alpha\sqrt{M+1}}2(2M+\epsilon)
\bigl(\ket{M+1}\bra M\rho+\rho\ket M\bra{M+1}\bigr).
\end{equation}
All finite Fock matrices belong to the domain of $\mathcal L_{\epsilon}^{s}$, so the following resolvent identity is valid:
\begin{equation}
\label{eq:boundary_resolvent_identity}
R_{\lambda}^{s}(d)
=R_{\lambda,M}(d)
+R_{\lambda}^{s}\circ\delta_M\circ R_{\lambda,M}(d).
\end{equation}
Indeed, applying $\lambda\mathbf I-\mathcal L_{\epsilon}^{s}$ to $R_{\lambda,M}(d)$ gives $d-\delta_M(R_{\lambda,M}(d))$.

\begin{lemma}
\label{lemma:boundary_error}
After increasing $\alpha_0$ and decreasing $\epsilon_0$ if necessary, there exists $C>0$ such that, for all parameters as in Lemma~\ref{lemma:truncated_resolvent},
\begin{equation}
\label{eq:boundary_error_bound}
\sup_{\lambda\in\Gamma_{\epsilon\bar r}}
\|R_{\lambda}^{s}\circ\delta_M\circ R_{\lambda,M}(d)\|_{\alpha,s}
\leq C\alpha^6 2^{-M/2}.
\end{equation}
\end{lemma}
\begin{proof}
The Fock basis gives
\[
\|a2^{-N/2}\|_{\infty}=\frac1{\sqrt2},\qquad
\|a^22^{-N/2}\|_{\infty}=\frac{\sqrt3}{2}.
\]
Since $0\leq P^{s}\leq I_{\mathcal H}$,
\[
\|S^{s}2^{-N/2}\|_{\infty}^2
\leq\left(\sqrt2\alpha+\frac{\sqrt3}{2}\right)^2+1
\leq(\sqrt2\alpha+2)^2.
\]
Furthermore, writing $E_M=\ket{M+1}\bra M$, we have
\begin{equation*}
\begin{aligned}
\|S^{s}E_MX_M^{-\frac12}\|_{\infty}
&=2^{-M/2}\|S^{s}\ket{M+1}\|\\
&\leq\sqrt{(M+1)(M+4\alpha^2)+1}\,2^{-M/2}.
\end{aligned}
\end{equation*}
Consequently,
\begin{equation*}
\begin{aligned}
\|E_M\rho\|_{L,s}
&=\|S^{s}E_MX_M^{-\frac12}
X_M^{\frac12}\rho X_M^{\frac12}X_M^{-\frac12}S^{s}\|_1\\
&\leq(\sqrt2\alpha+2)\sqrt{(M+1)(M+4\alpha^2)+1}
\,2^{-M/2}\|\rho\|_{X_M}.
\end{aligned}
\end{equation*}
The same bound holds for $\rho E_M^{\dagger}$. For the additional term in the transported $\alpha$-norm,
\[
\alpha^2\|P^{s}E_M\rho P^{s}\|_1
\leq\alpha^2 2^{-M/2}\|\rho\|_{X_M},
\]
and similarly for $\rho E_M^{\dagger}$. Combining these estimates with~\eqref{eq:boundary_defect} yields
\begin{equation*}
\begin{aligned}
\|\delta_M(\rho)\|_{\alpha,s}
&\leq\alpha\sqrt{M+1}(2M+\epsilon)
\left[(\sqrt2\alpha+2)\sqrt{(M+1)(M+4\alpha^2)+1}+\alpha^2\right]\\
&\hspace{2cm}\cdot2^{-M/2}\|\rho\|_{X_M}.
\end{aligned}
\end{equation*}
Then there exists $C'>0$ such that 
\[\|\delta_M(\rho)\|_{\alpha,s}
\leq C'\alpha^7 2^{-M/2}\|\rho\|_{X_M}.\]
By Lemma~\ref{lemma:truncated_resolvent}, there exists $C''>0$ such that
\[
\|R_{\lambda,M}(d)\|_{X_M}
\leq\sqrt2\epsilon\frac{C''}{\alpha}.
\]
Applying~\eqref{eq:shifted_contour_bound}, we obtain
\[\begin{aligned}\|R_{\lambda}^s\circ\delta_M\circ R_{\lambda,M}(d)\|_{\alpha,s}&\leq\frac{C_0}{\epsilon}\cdot C'\alpha^7 2^{-M/2}\cdot \sqrt2\epsilon\frac{C''}{\alpha}\\&=\sqrt{2}C_0C'C''\alpha^6 2^{-M/2}.
\end{aligned}\]
We can conclude by setting $C=\sqrt{2}C_0C'C''$.
\end{proof}

\begin{lemma}
\label{lemma:shifted_localization}
There exists $C_1,C_2,C_3>0$ such that, under the preceding assumptions,
\begin{equation}
\label{eq:shifted_localization}
\begin{aligned}
\|\rho_{\epsilon,\alpha}^{s}-P_0\|_1
&\leq C_1\frac{\epsilon}{\alpha}+C_2\alpha^6 2^{-M/2},\\
\|\rho_{\epsilon,\alpha}^{s}-\mathcal P_M(\rho_{\epsilon,\alpha}^{s})\|_1
&\leq C_2\alpha^6 2^{-M/2},\\
\|f_{\epsilon,\alpha}^{s}\|_1
\leq\|f_{\epsilon,\alpha}^{s}\|_{\alpha,s}
&\leq C_3\epsilon\alpha^6 2^{-M/2}.
\end{aligned}
\end{equation}
Moreover, $\Tr(\rho_{\epsilon,\alpha}^{s})=1$.
\end{lemma}
\begin{proof}
Define the finite-dimensional approximation
\[
\sigma_M=P_0+\frac{1}{2\pi i}\oint_{\Gamma_{\epsilon\bar r}}
\frac1\lambda R_{\lambda,M}(d)\,\mathrm{d}\lambda.
\]
For all $\lambda\in\Gamma_{\epsilon\bar{r}}$, by Lemma~\ref{lemma:truncated_resolvent}, there exists $C_1>0$ such that
\[\|R_{\lambda,M}(d)\|_{X_{M}}\leq C_1\frac{\epsilon}{\alpha}.\]
Since $\mathcal{L}_{\epsilon,M}$ is trace preserving on $\mathcal{P}_{M}\mathcal{K}_{f}$, $\Tr(R_{\lambda,M}(d))=\frac{1}{\lambda}\Tr(d)=0$. Therefore, $\sigma_{M}$ is supported on $\mathcal H_M$, has trace $1$, and satisfies
\[
\|\sigma_M-P_0\|_{X_M}\leq C_1\frac{\epsilon}{\alpha}.
\]
Equations~\eqref{eq:shifted_contour_vectors} and~\eqref{eq:boundary_resolvent_identity}, together with Lemma~\ref{lemma:boundary_error}, there exists $C_2>0$ such that
\[
\|\rho_{\epsilon,\alpha}^{s}-\sigma_M\|_{\alpha,s}
\leq C_2\alpha^6 2^{-M/2}.
\]
Since $\|\cdot\|_1\leq\|\cdot\|_{\alpha,s}$ and compression by $P_M$ is a trace-norm contraction, this proves the first two estimates.

For the last estimate, since $d\in\mathcal{P}_M\mathcal{K}_f$,
\[\oint_{\Gamma_{\epsilon\bar r}}R_{\lambda,M}(d)\,\mathrm{d}\lambda=\Tr(d)P_0=0,\]
and
\[\begin{aligned}\|f_{\epsilon,\alpha}^{s}\|_{1}&\leq\|f_{\epsilon,\alpha}^s\|_{\alpha,s}
=\frac{1}{2\pi }\left\|\oint_{\Gamma_{\epsilon\bar r}}
R_{\lambda}^{s}\circ\delta_M\circ R_{\lambda,M}(d)\,\mathrm{d}\lambda\right\|_{\alpha,s}\\&\leq\frac{1}{2\pi}\oint_{\Gamma_{\epsilon\bar r}}\sup_{\lambda\in\Gamma_{\epsilon\bar r}}
\|R_{\lambda}^{s}\circ\delta_M\circ R_{\lambda,M}(d)\|_{\alpha,s}\mathrm{d}\lambda.\end{aligned}\]
By Lemma~\ref{lemma:boundary_error}, there exists $C_3>0$ satisfying the third estimate. The trace preserving also implies $\Tr(R_{\lambda}^s(d))=\frac{1}{\lambda}\Tr(d)=0$, the first identity in~\eqref{eq:shifted_contour_vectors} therefore gives $\Tr(\rho_{\epsilon,\alpha}^{s})=1$.
\end{proof}

\subsection{Eigenvectors and eigenvalues of $\mathcal L_{\epsilon}$}
Define
\[
P_M^{\pm\alpha}=D(\pm\alpha)P_MD(\mp\alpha),
\qquad
\mathcal P_M^{\pm\alpha}(\rho)=P_M^{\pm\alpha}\rho P_M^{\pm\alpha}.
\]
The parity operator $J=e^{i\pi N}$ satisfies $JaJ=-a$. Then for $\rho\in\mathcal{K}_1$, define the conjugation map
\[\mathcal{J}(\rho)=J\rho J.\]
$\mathcal{J}$ commutes with $\mathcal L_{\epsilon}$ and $\mathcal P_{\epsilon,\alpha}$. Consequently,
\[
\rho_{\epsilon,\alpha}^{-}=\mathcal{J}(\rho_{\epsilon,\alpha}^{+}),
\qquad
f_{\epsilon,\alpha}^{-}=\mathcal{J}(f_{\epsilon,\alpha}^{+}).
\]

\begin{lemma}[Localization of $\rho_{\epsilon,\alpha}^{\pm}$]
\label{lemma:exponential_1}
There exist $\alpha_0\geq\bar\alpha$, $\epsilon_0\in(0,\bar\epsilon]$ and $C>0$ such that, for all $\alpha\geq\alpha_0$ and $\epsilon\in(0,\epsilon_0]$,
\[
\Tr(\rho_{\epsilon,\alpha}^{\pm})=1,\qquad
\frac12\leq\|\rho_{\epsilon,\alpha}^{\pm}\|_1\leq\frac32,
\]
and
\begin{equation*}
\begin{aligned}
\|\rho_{\epsilon,\alpha}^{\pm}-\ket{\pm\alpha}\bra{\pm\alpha}\|_1
&\leq C\frac{\epsilon}{\alpha}+C\alpha^6 2^{-\alpha^2/4},\\
\|\rho_{\epsilon,\alpha}^{\pm}-\mathcal P_M^{\pm\alpha}(\rho_{\epsilon,\alpha}^{\pm})\|_1
&\leq C\alpha^6 2^{-\alpha^2/4}.
\end{aligned}
\end{equation*}
\end{lemma}
\begin{proof}
Apply Lemma~\ref{lemma:shifted_localization}, unitary invariance of the trace norm, and parity symmetry. The choice of $M$ gives $M\geq\alpha^2/2$, hence $2^{-M/2}\leq2^{-\alpha^2/4}$. Increasing $\alpha_0$ makes the first error at most $1/2$, which gives the norm bounds.
\end{proof}

\begin{lemma}[Estimate of $f_{\epsilon,\alpha}^{\pm}$]
\label{lemma:exponential_2}
Under the same assumptions, there exists $C>0$ such that
\[
\|f_{\epsilon,\alpha}^{\pm}\|_1\leq C\epsilon\alpha^6 2^{-\alpha^2/4}.
\]
\end{lemma}
\begin{proof}
The first estimate follows from Lemma~\ref{lemma:shifted_localization} by displacement and parity symmetry.
\end{proof}

\begin{proof}[Proof of Theorem~\ref{theorem:main_3}]
By Lemma~\ref{lemma:exponential_1} and
\[
\|\ket\alpha\bra\alpha-\ket{-\alpha}\bra{-\alpha}\|_1
=2\sqrt{1-e^{-4\alpha^2}},
\]
we may increase $\alpha_0$ so that
\[
\|\rho_{\epsilon,\alpha}^{+}-\rho_{\epsilon,\alpha}^{-}\|_1\geq1.
\]
Both operators have trace $1$, so they are linearly independent and form a basis of the rank-two range of $\mathcal P_{\epsilon,\alpha}$. Define 
\[
u_+=\frac12(\rho_{\epsilon,\alpha}^{+}+\rho_{\epsilon,\alpha}^{-}),
\qquad
u_-=\frac12(\rho_{\epsilon,\alpha}^{+}-\rho_{\epsilon,\alpha}^{-}).
\]
Since \[\mathcal{J}(u_+)=u_+,\quad \mathcal{J}(u_-)=-u_-,\]
they decompose the two-dimensional subspace $\text{Im}\left(\mathcal{P}_{\epsilon,\alpha}\right)$
into even and odd subspaces:
\[
E_+=\mathbb C u_+,\qquad E_-=\mathbb C u_-.
\]
Since \(\mathcal L_\epsilon\) preserves $\text{Im}\left(\mathcal{P}_{\epsilon,\alpha}\right)$ and commutes with $\mathcal{J}$,
\[
\mathcal{J}(\mathcal L_\epsilon u_\pm)
=\mathcal L_\epsilon(\mathcal{J}u_\pm)
=\pm\mathcal L_\epsilon u_\pm.
\]
We can easily deduce that $\mathcal{L}_{\epsilon}$ preserves $E_{\pm}$ and $u_{\pm}$ are eigenvectors of $\mathcal{L}_{\epsilon}$. Since $\Tr(u_+)=1$ and the generator preserves trace, the eigenvalue associated with $u_+$ is zero.
The contour is invariant under complex conjugation and the generator preserves adjoints, so $\rho_{\epsilon,\alpha}^{\pm}$ and $u_-$ are self-adjoint. The eigenvalue of $u_-$ is therefore real. Trace-norm contractivity of the trace-preserving Lindblad semigroup implies that this eigenvalue is nonpositive; write it as $-\lambda$, with $\lambda\geq0$.
Finally, by Lemma~\ref{lemma:exponential_2}, there exists $C>0$ such that 
\begin{equation}
\label{eq:improved_eigenvalue_bound}
0\leq\lambda
=\frac{\|f_{\epsilon,\alpha}^{+}-f_{\epsilon,\alpha}^{-}\|_1}
{\|\rho_{\epsilon,\alpha}^{+}-\rho_{\epsilon,\alpha}^{-}\|_1}
\leq C\epsilon\alpha^6 2^{-\alpha^2/4}.
\end{equation}
This implies the bound stated in Theorem~\ref{theorem:main_3}.
\end{proof}
\begin{remark}
Estimate~\eqref{eq:improved_eigenvalue_bound} is stronger than the stated theorem. For every $\kappa\in(0,(\log2)/4)$ there is $C_\kappa>0$ such that
\[
0\leq\lambda\leq C_\kappa\epsilon e^{-\kappa\alpha^2}.
\]
\end{remark}
\bibliography{references}

@article{rouchonazouit,
  author       = {Azouit, R{\'e}mi and
                  Sarlette, Alain and
                  Rouchon, Pierre},
  title        = {Well-posedness and convergence of the {Lindblad} master equation for a quantum harmonic oscillator with multi-photon drive and damping},
  journal      = {ESAIM: Control, Optimisation and Calculus of Variations},
  year         = {2016},
  volume       = {22},
  number       = {4},
  pages        = {1353--1369},
  doi          = {10.1051/cocv/2016050},
  url          = {https://doi.org/10.1051/cocv/2016050}
}

@misc{costarico2025,
  author       = {Costa Rico, Pablo and Gondolf, Paul and M{\"o}bus, Tim},
  title        = {Instantaneous {Sobolev} Regularization for Dissipative Bosonic Dynamics},
  year         = {2025},
  eprint       = {2512.04066},
  archivePrefix= {arXiv},
  primaryClass = {math-ph},
  note         = {arXiv:2512.04066 [math-ph]},
  doi          = {10.48550/arXiv.2512.04066},
  url          = {https://arxiv.org/abs/2512.04066}
}

@article{mirrahimi2014,
  author       = {Mirrahimi, Mazyar and
                  Leghtas, Zaki and
                  Albert, Victor V and
                  Touzard, Steven and
                  Schoelkopf, Robert J and
                  Jiang, Liang and
                  Devoret, Michel H},
  title        = {Dynamically protected cat-qubits: a new paradigm for universal quantum computation},
  journal      = {New Journal of Physics},
  year         = {2014},
  volume       = {16},
  number       = {4},
  pages        = {045014},
  doi          = {10.1088/1367-2630/16/4/045014},
  url          = {https://doi.org/10.1088/1367-2630/16/4/045014}
}

@article{leghtas2015,
  author       = {Leghtas, Z. and
                  Touzard, S. and
                  Pop, I. M. and
                  Kou, A. and
                  Vlastakis, B. and
                  Petrenko, A. and
                  Sliwa, K. M. and
                  Narla, A. and
                  Shankar, S. and
                  Hatridge, M. J. and
                  Reagor, M. and
                  Frunzio, L. and
                  Schoelkopf, R. J. and
                  Mirrahimi, M. and
                  Devoret, M. H.},
  title        = {Confining the state of light to a quantum manifold by engineered two-photon loss},
  journal      = {Science},
  year         = {2015},
  volume       = {347},
  number       = {6224},
  pages        = {853--857},
  doi          = {10.1126/science.aaa2085},
  url          = {https://doi.org/10.1126/science.aaa2085}
}

@article{lescanne2020,
  author       = {Lescanne, Rapha{\"e}l and
                  Villiers, Marius and
                  Peronnin, Th{\'e}au and
                  Sarlette, Alain and
                  Delbecq, Matthieu and
                  Huard, Benjamin and
                  Kontos, Takis and
                  Mirrahimi, Mazyar and
                  Leghtas, Zaki},
  title        = {Exponential suppression of bit-flips in a qubit encoded in an oscillator},
  journal      = {Nature Physics},
  year         = {2020},
  volume       = {16},
  number       = {5},
  pages        = {509--513},
  doi          = {10.1038/s41567-020-0824-x},
  url          = {https://doi.org/10.1038/s41567-020-0824-x}
}

@article{berdou2023,
  author       = {Berdou, C. and
                  Murani, A. and
                  R{\'e}glade, U. and
                  Smith, W. C. and
                  Villiers, M. and
                  Palomo, J. and
                  Rosticher, M. and
                  Denis, A. and
                  Morfin, P. and
                  Delbecq, M. and
                  Kontos, T. and
                  Pankratova, N. and
                  Rautschke, F. and
                  Peronnin, T. and
                  Sellem, L.-A. and
                  Rouchon, P. and
                  Sarlette, A. and
                  Mirrahimi, M. and
                  Campagne-Ibarcq, P. and
                  Jezouin, S. and
                  Lescanne, R. and
                  Leghtas, Z.},
  title        = {One Hundred Second Bit-Flip Time in a Two-Photon Dissipative Oscillator},
  journal      = {PRX Quantum},
  year         = {2023},
  volume       = {4},
  number       = {2},
  pages        = {020350},
  doi          = {10.1103/PRXQuantum.4.020350},
  url          = {https://doi.org/10.1103/PRXQuantum.4.020350}
}

@article{reglade2024,
  author       = {R{\'e}glade, U. and
                  Bocquet, A. and
                  Gautier, R. and
                  Cohen, J. and
                  Marquet, A. and
                  Albertinale, E. and
                  Pankratova, N. and
                  Hall{\'e}n, M. and
                  Rautschke, F. and
                  Sellem, L.-A. and
                  Rouchon, P. and
                  Sarlette, A. and
                  Mirrahimi, M. and
                  Campagne-Ibarcq, P. and
                  Lescanne, R. and
                  Jezouin, S. and
                  Leghtas, Z.},
  title        = {Quantum control of a cat qubit with bit-flip times exceeding ten seconds},
  journal      = {Nature},
  year         = {2024},
  volume       = {629},
  number       = {8013},
  pages        = {778--783},
  doi          = {10.1038/s41586-024-07294-3},
  url          = {https://doi.org/10.1038/s41586-024-07294-3}
}

@article{putterman2025,
  author       = {Putterman, Harald and
                  Noh, Kyungjoo and
                  Patel, Rishi N. and
                  Peairs, Gregory A. and
                  MacCabe, Gregory S. and
                  Lee, Menyoung and
                  Aghaeimeibodi, Shahriar and
                  Hann, Connor T. and
                  Jarrige, Ignace and
                  Marcaud, Guillaume and
                  He, Yuan and
                  Moradinejad, Hesam and
                  Owens, John Clai and
                  Scaffidi, Thomas and
                  Arrangoiz-Arriola, Patricio and
                  Iverson, Joe and
                  Levine, Harry and
                  Brand{\~a}o, Fernando G. S. L. and
                  Matheny, Matthew H. and
                  Painter, Oskar},
  title        = {Preserving Phase Coherence and Linearity in Cat Qubits with Exponential Bit-Flip Suppression},
  journal      = {Physical Review X},
  year         = {2025},
  volume       = {15},
  number       = {1},
  pages        = {011070},
  doi          = {10.1103/PhysRevX.15.011070},
  url          = {https://doi.org/10.1103/PhysRevX.15.011070}
}

@article{guillaud2019,
  author       = {Guillaud, J{\'e}r{\'e}mie and
                  Mirrahimi, Mazyar},
  title        = {Repetition Cat Qubits for Fault-Tolerant Quantum Computation},
  journal      = {Physical Review X},
  year         = {2019},
  volume       = {9},
  number       = {4},
  pages        = {041053},
  doi          = {10.1103/PhysRevX.9.041053},
  url          = {https://doi.org/10.1103/PhysRevX.9.041053}
}

@article{guillaud2023,
  author       = {Guillaud, J{\'e}r{\'e}mie and
                  Cohen, Joachim and
                  Mirrahimi, Mazyar},
  title        = {Quantum computation with cat qubits},
  journal      = {SciPost Physics Lecture Notes},
  year         = {2023},
  pages        = {72},
  doi          = {10.21468/SciPostPhysLectNotes.72},
  url          = {https://doi.org/10.21468/SciPostPhysLectNotes.72}
}

@article{macieszczak2016,
  author       = {Macieszczak, Katarzyna and
                  Gu{\c t}{\u a}, M{\u a}d{\u a}lin and
                  Lesanovsky, Igor and
                  Garrahan, Juan P.},
  title        = {Towards a Theory of Metastability in Open Quantum Dynamics},
  journal      = {Physical Review Letters},
  year         = {2016},
  volume       = {116},
  number       = {24},
  pages        = {240404},
  doi          = {10.1103/PhysRevLett.116.240404},
  url          = {https://doi.org/10.1103/PhysRevLett.116.240404}
}

@inproceedings{azouit2015,
  author       = {Azouit, R{\'e}mi and
                  Sarlette, Alain and
                  Rouchon, Pierre},
  title        = {Convergence and adiabatic elimination for a driven dissipative quantum harmonic oscillator},
  booktitle    = {2015 54th IEEE Conference on Decision and Control (CDC)},
  year         = {2015},
  pages        = {6447--6453},
  publisher    = {IEEE},
  doi          = {10.1109/CDC.2015.7403235},
  url          = {https://doi.org/10.1109/CDC.2015.7403235}
}

@article{dubovitskii,
  author       = {Dubovitskii, Kirill S.},
  title        = {Bit-flip errors in dissipative cat qubits: Second-order perturbation theory},
  journal      = {Physical Review A},
  year         = {2025},
  volume       = {111},
  number       = {1},
  pages        = {012617},
  doi          = {10.1103/PhysRevA.111.012617},
  url          = {https://doi.org/10.1103/PhysRevA.111.012617}
}

@article{gautier2022,
  author       = {Gautier, Ronan and
                  Sarlette, Alain and
                  Mirrahimi, Mazyar},
  title        = {Combined Dissipative and {Hamiltonian} Confinement of Cat Qubits},
  journal      = {PRX Quantum},
  year         = {2022},
  volume       = {3},
  number       = {2},
  pages        = {020339},
  doi          = {10.1103/PRXQuantum.3.020339},
  url          = {https://doi.org/10.1103/PRXQuantum.3.020339}
}

@article{thompson2022,
  author       = {Thompson, Foster and
                  Kamenev, Alex},
  title        = {Qubit decoherence and symmetry restoration through real-time instantons},
  journal      = {Physical Review Research},
  year         = {2022},
  volume       = {4},
  number       = {2},
  pages        = {023020},
  doi          = {10.1103/PhysRevResearch.4.023020},
  url          = {https://doi.org/10.1103/PhysRevResearch.4.023020}
}

@book{kato1995,
  author       = {Kato, Tosio},
  title        = {Perturbation Theory for Linear Operators},
  edition      = {Second},
  series       = {Classics in Mathematics},
  publisher    = {Springer},
  address      = {Berlin, Heidelberg},
  year         = {1995},
  isbn         = {978-3-540-58661-6},
  doi          = {10.1007/978-3-642-66282-9},
  url          = {https://doi.org/10.1007/978-3-642-66282-9}
}
\end{document}